\documentclass[letterpaper]{article} 

\usepackage{amsmath,amsfonts}  
\usepackage{adjustbox}
\usepackage{booktabs}
\usepackage{multirow}
\usepackage{appendix}

\usepackage{graphicx}
\usepackage{multicol}
\usepackage{subcaption}

\usepackage{cleveref}
\crefname{equation}{Eq.}{Eqs.}

\usepackage{aaai24}  
\usepackage{times}  
\usepackage{helvet}  
\usepackage{courier}  
\usepackage[hyphens]{url}  
\usepackage{graphicx} 
\usepackage{natbib}  
\usepackage{caption} 
\usepackage{algorithm}
\usepackage{algorithmic}

\usepackage{newfloat}
\usepackage{listings}
\DeclareCaptionStyle{ruled}{labelfont=normalfont,labelsep=colon,strut=off} 
\floatstyle{ruled}
\newfloat{listing}{tb}{lst}{}
\floatname{listing}{Listing}
\title{Beyond Traditional Threats: A Persistent Backdoor Attack on Federated Learning}
\author{
    Tao Liu\textsuperscript{\rm 1},
    Yuhang Zhang\textsuperscript{\rm 1},
    Zhu Feng\textsuperscript{\rm 1},
    Zhiqin Yang\textsuperscript{\rm 2},
    Chen Xu\textsuperscript{\rm 1},
    Dapeng Man\textsuperscript{\rm 1}\thanks{Corresponding author.},
    Wu Yang\textsuperscript{\rm 1}\footnotemark[1]
}
\affiliations{

    \textsuperscript{\rm 1}College of Computer Science and Technology, Harbin Engineering University, China\\
    \textsuperscript{\rm 2}Southampton Ocean Engineering Joint Institute, Harbin Engineering University, China\\
    ltaoheu@163.com, \{yangwu, mandapeng, chen.xu\}@hrbeu.edu.cn
}

\usepackage{bibentry}

\begin{document}

\maketitle 

\begin{abstract}
Backdoors on federated learning will be diluted by subsequent benign updates. This is reflected in the significant reduction of attack success rate as iterations increase, ultimately failing. We use a new metric to quantify the degree of this weakened backdoor effect, called attack persistence. Given that research to improve this performance has not been widely noted, we propose a Full Combination Backdoor Attack (FCBA) method. It aggregates more combined trigger information for a more complete backdoor pattern in the global model. Trained backdoored global model is more resilient to benign updates, leading to a higher attack success rate on the test set. We test on three datasets and evaluate with two models across various settings. FCBA's persistence outperforms SOTA federated learning backdoor attacks. On GTSRB, post-attack 120 rounds, our attack success rate rose over 50\% from baseline. The core code of our method is available at https://github.com/PhD-TaoLiu/FCBA.
\end{abstract}

\section{Introduction}
Federated Learning (FL) is a novel machine learning paradigm that allows model training across multiple devices while preserving data privacy at its source~\cite{mcmahan2017communication}. However, its distributed framework and the non-i.i.d. data heterogeneity can inadvertently facilitate backdoor attacks~\cite{zhao2018federated}. The concept of backdoor attacks in FL involves embedding a unique trigger in the training dataset~\cite{pixel}. The resulting global model behaves typically, but when exposed to this trigger in an input, it deliberately misclassifies to an attacker-specified category~\cite{HB}.

Presently, backdoor attacks in federated learning are emerging and largely unexplored. In FL, traditional backdoor attacks are less effective due to aggregation diminishing malicious impacts. The \textit{Model Replacement} method scales malicious updates pre-submission to ensure resilience during model averaging~\cite{CBA}. \textit{Distributed Backdoor Attack} (DBA) exploits decentralization of FL by spliting global triggers into multiple local ones, each embedded by separate adversaries~\cite{DBA}. However, these methods boost attack success rate (ASR) only for a brief period post-poison injection, questioning their long-term efficacy. Enhancing the durability of backdoor attacks in FL presents a contemporary research bottleneck and challenge.

Federated Learning inherently embodies Online Learning traits with continuous global model training~\cite{quanrud2015online,veness2017online}. Subsequent benign updates readily dilute the global model's backdoor, with the backdoor efficacy declining markedly over iterations. This mirrors catastrophic forgetting in multi-task learning~\cite{li2022multi}, marked by a sharp drop in ASR. Catastrophic forgetting of backdoors intuitively explains the limited persistence of backdoor attacks in federated learning~\cite{kemker2018measuring}.

To enhance the persistence of backdoor attacks in FL, we propose a \textit{Full Combination Backdoor Attack} (FCBA) method. In convolutional neural networks, the mechanism of hierarchical feature extraction has been empirically demonstrated to produce pronounced responses to particular stimulus patterns or triggers~\cite{wang2022hfenet}. Leveraging this observation, we propose a novel methodology that constructs an expanded combinatorial set of local triggers to amplify this inherent response. Subsequent to this local enhancement, we introduce a central aggregation mechanism that compiles these decentralized responses. Our primary motivation is to enhance the global model's capacity to learn and recognize backdoors, thereby offering robustness against backdoor forgetting~\cite{li2023permanence}.


Our main contributions can be summarized as follows:

(1)We propose a new backdoor attack, FCBA, with persistence beyond SOTA attack methods in FL. In three categorization tasks, ASR after 120 injection rounds surpasses the baseline by 34.9\%, 8.8\%, and 56.8\% respectively.

(2)Using combinatorics, we innovatively design trigger strategies and identify malicious participants.

(3)We verify that FCBA exhibits strong robustness across various environments. Ablation studies indicate that the majority of factors exert limited influence on this attack, while most existing defense strategies fail to effectively counter our assault.

\section{Related Work}
\subsection{Backdoor Attack on Federated Learning}
The research on backdoor attack in federated learning is in its infancy. ~\citet{CBA} first proposed backdoor attack in federated learning to achieve model replacement by amplifying malicious updates. However, the success rate of this attack decreases significantly with iteration increase in single-shot attack setting. ~\citet{bhagoji2019analyzing} proposed increasing the attacker's local learning rate to achieve an attack when the model does not converge and proposed the alternate minimization strategy to enhance the stealthiness of the attack. The ASR decay problem is still not solved rather it is more serious.~\citet{DBA} proposed for the first time the distributed backdoor attack, which decomposes global triggers into multiple local triggers trained separately, and then aggregates the dispersed backdoor information to improve the durability and covertness of the attack. However, the attack persistence is not satisfactory in a single-shot attack setting.~\citet{wang2020attack} theoretically proved that backdoor attacks cannot be avoided in federated learning and proposed an edge-case backdoor that forces the model to perform poorly on long-tailed samples. However, they only used projected gradient descent to improve the stealthiness of the attack and did not improve the persistence improvement. In our study, we introduce FCBA, a novel backdoor attack showcasing unparalleled attack persistence, even within the constraints of a single-shot attack setting.

\subsection{Defenses against Backdoor Attack}
Backdoor defense research remains rooted in traditional computing, employing methods like \textit{Neuron Pruning}~\cite{liu2018fine}, \textit{STRIP}~\cite{gao2019strip}, \textit{AC}~\cite{chen2018detecting}, \textit{Neural Cleanse}~\cite{wang2019neural}, and \textit{FLguard}~\cite{nguyen2021flguard}. These anomaly detection-based strategies, often rely on raw data or model updates, which conflict with the principles of federated learning or secure aggregation mechanisms, limiting their deployment in federated learning context.

In response to backdoor challenges, robust federated learning defenses have surfaced. Unlike anomaly detection, robust federation learning aims to directly mitigate backdoor attacks during training. Notably, incorporating \textit{Byzantine Tolerance Distributions} into robust aggregation ~\cite{blanchard2017machine, chen2017distributed, damaskinos2018asynchronous, xie2018zeno, yin2018byzantine} seeks to combat federated learning attacks. However, based on flawed assumptions about data distribution and attack objectives, their defense can be compromised. \textit{Differential Private Federated Learning}~\cite{geyer2017differentially, mcmahan2017learning} achieves low-complexity elimination of backdoors by trimming model weights and injecting noise to limit each participant's influence on the global model. The method comes at the expense of the main task accuracy and requires a good tradeoff. Recently, a feedback-based federated learning \textit{BaFFle}~\cite{andreina2021baffle} 
 utilizes participants' validation results of the global model to eliminate backdoors. The computational complexity of this method is limited by the complexity of the master task and the total number of clients. In this work, these defense methods are difficult to practically and effectively defend against our attacks, proving that FCBA is strongly robust.

\section{Full Combination Backdoor Attack on Federated Learning}
\subsection{Federated Learning}

In this study, we utilize the horizontal federated learning approach~\cite{kairouz2021advances}, aiming for a global model with enhanced generalization from aggregating local participants' training outcomes, as shown in \cref{eq1}:

\begin{equation}
G^{t+1} =G^{t}+\frac{n}{\eta}\sum_{i=1}^{m} (L_{i}^{t+1} -G^{t}) 
\label{eq1}
\end{equation}
Given $\eta=n/m$, the global model is substituted by the mean of local models.

\subsection{Threat Model} 
\subsubsection{Attack scenario.} In FL, some participants may be malicious with a common backdoor task~\cite{bonawitz2019towards}. This can arise from colluding clients~\cite{conti2018internet} or a powerful attacker exploiting weak-security clients~\cite{wu2020federated}. Our method covers both, but we detail the latter for brevity.

\subsubsection{Attacker’s knowledge and capabilities.} 
Based on Kerckhoffs’s theory~\cite{shannon1949communication}, we make the same assumptions about the knowledge and capabilities of the attacker as ~\citet{DBA}. The attacker is a fully informed adversary and can completely control the local training process of the client. He can control the local data and the model updates, and possesses the ability to adaptively fine-tune the local training hyperparameters with each iteration. This assumption does not have the ability to directly affect other participants and the central server, and is very practical in FL scenarios.

\subsubsection{Attack workflow.} We use small white pixel blocks, about 2\% of the image, as fixation triggers in the upper left corner. Combined with label flipping, we poison a fraction of the training data. This poisoned data is mixed with clean data for each malicious participant. During training, we employ the model replacement method~\cite{CBA}, optimizing the local epoch and learning rate for enhanced backdoor efficacy.

After local training, the update is amplified with a scale factor to ensure that the backdoor survives the average aggregation. The scale factor, denoted as $\gamma$, is defined from \cref{eq2} in the model replacement.
\begin{equation}
\begin{split}
    \widetilde{L}_{m}^{t+1}&=\frac{n}{\eta} X-(\frac{n}{\eta}-1)G^{t}-\sum_{i=1}^{m-1} (L_{i}^{t+1} -G^{t})\\&\approx\frac{n}{\eta}(X-G^t)+G^t
\end{split}     
\label{eq2}
\end{equation}
where $L$ represents the local model, $G$ the global model, $X$ the malicious model, and $t$ the current iteration round. $\gamma$ is initially set as $n/\eta$, but can be adjusted to determine model replacement degree. It's presumed the $m$-th client is under the attacker's control.

After aggregation, a backdoored global model emerges. In testing, it functions normally on clean samples but misclassifies triggered ones to target classes.


\subsection{Full Combination Backdoor Attack} 

Remember, DBA~\cite{DBA} is an advanced FL backdoor attack that employs a distributed trigger strategy, capitalizing on FL's decentralized aggregation. During training, a global trigger is divided into $m$ distinct parts for decentralized backdoor pattern learning. In inference, the full global trigger evaluates the backdoor model, enhancing its persistence and stealthiness.

We propose a new \textit{Full Combination Backdoor Attack} (FCBA) method using combinatorics theory, which consists of the following three main works.
\begin{itemize}
    \item Generate Full Combination Triggers.
    \item Identify Malicious Clients.
    \item Designing Attack Objective Functions.
\end{itemize}
\begin{figure}[H] 
\centering 
\includegraphics[width=0.47\textwidth]{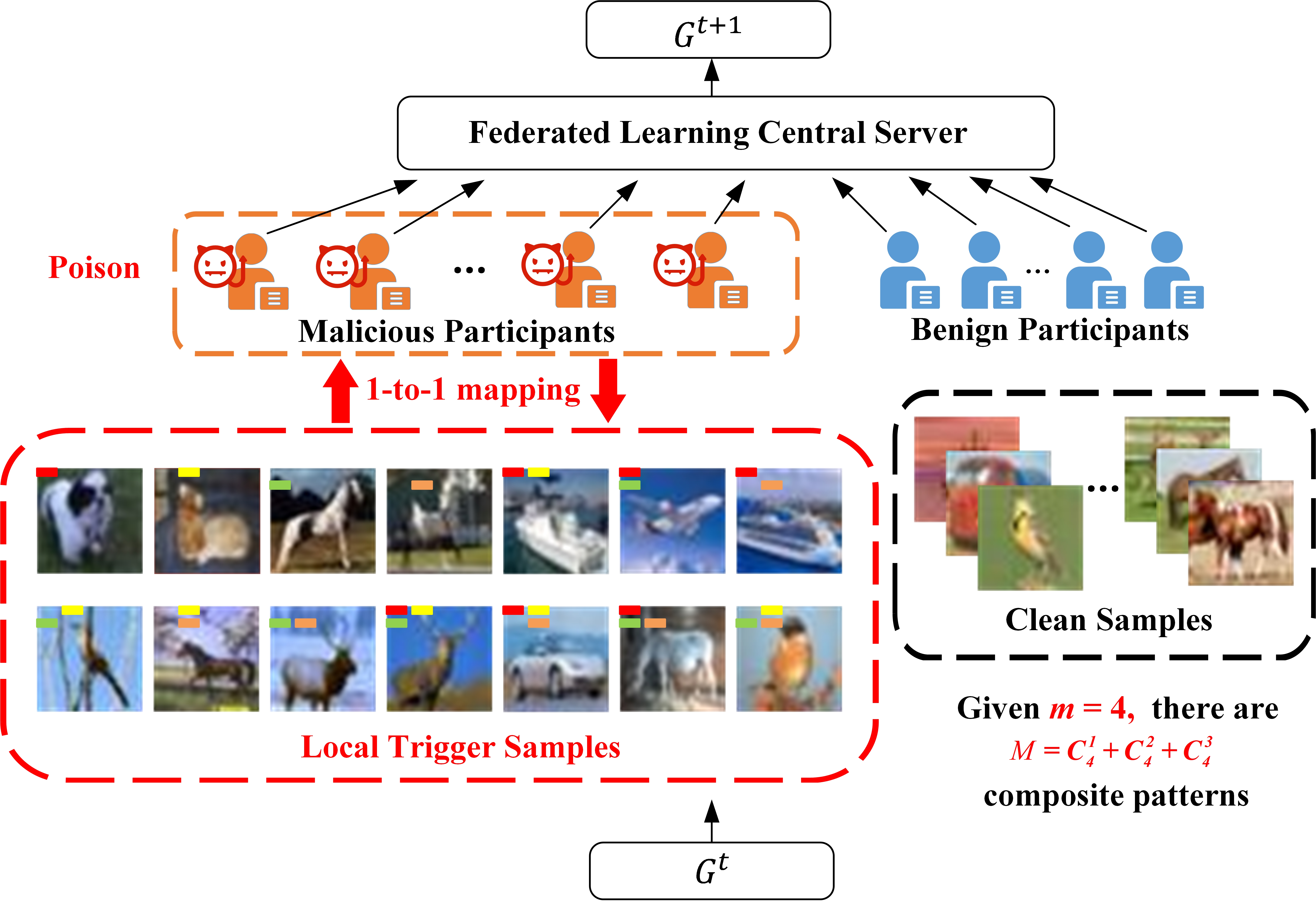} 
\caption{Overview of full combination backdoor attack (FCBA) in FL. At round $t+1$, the aggregator merges local data (both benign and adversarial) from $t$ to update $G_{t+1}$. During a backdoor attack, the attacker uses trigger partition $m$ to create local trigger patterns and identifies $M$ malicious clients, each with a unique trigger pattern.} 
\label{FCBA-describe} 
\end{figure}

\subsubsection{Generate Full Combination Trigger.} We introduce a novel local trigger generation strategy, $O_{FC}(i)$. Given a trigger partition number $m$, the global trigger is divided into $m$ distinct parts. These parts are treated as units, and we generate local triggers by combining them in various styles, as depicted in Fig. 1. For $m=4$, an attacker might select one, two, or three differently colored pixel blocks as local triggers. This approach equates to solving the sum of combinations (see \cref{eq3}), with $LT$ being the total trigger count. We term this the \textit{Full Combination} Problem, from which \textit{Full Combination} derives.
\begin{equation}
    LT=C_{m}^{1}+C_{m}^{2}+\cdots+C_{m}^{m-1}
    \label{eq3}
\end{equation}

Four notes are needed here: (1) $m$ should be $\ge 2$ but not exceedingly large. $m=1$ signifies a centralized attack; overly large $m$ leads to tiny local triggers, affecting backdoor efficacy and increasing strategy computation. (2)  Fig. 1's local trigger samples exclude 0-block and 4-block cases. Samples with 0 blocks are clean, while those with 4 blocks are global triggers used solely in the testing phase. (3) Attackers with specific local triggers only poison data using patterns from the related region.
(4) To maintain fairness, we ensure a comparable count of total injected triggers (e.g., altered pixels) between FCBA and DBA (refer to Appendix A.1).
\begin{equation}
\begin{split}
    (a+b)^n&=C_{n}^{0}a^n+ C_{n}^{1}a^{n-1}b^1+C_{n}^{2}a^{n-2}b^2+\cdots\\&+C_{n}^{n-1}a^{1}b^{n-1}+C_{n}^{n}b^n
\end{split}
\label{eq4}
\end{equation}

\subsubsection{Identify Malicious Clients.} To ensure the efficacy of the backdoor attack, each malicious client receives a unique local trigger. The total number of malicious clients, $M$, corresponds to the total number of local triggers, $LT$. This relationship is elucidated using combinatorics, drawing inspiration from \textit{Newton's Binomial Theorem} \cite{newton1732arithmetica}(\cref{eq4}). When both variables in \cref{eq4} are 1, it reduces to \cref{eq5}~\cite{knuth1997art}. Inverting this equation provides the sum of combinatorial numbers. Employing a variant of \cref{eq5}, as illustrated above, we determine $M$. For instance, in Fig. 1 where $m=4$, \cref{eq6} yields $M=LT=14$, signifying 14 unique local triggers and their respective malicious clients, with the remainder being benign. For fairness, $M$ clients are randomly designated as malicious from the entire client pool.
\begin{equation}
2^n=C_{n}^{0}+C_{n}^{1}+\cdots+C_{n}^{n-1}+C_{n}^{n}
\label{eq5}
\end{equation}
\begin{equation}
M=LT=2^m-2
\label{eq6}
\end{equation}
\subsubsection{Designing Attack Objective Functions.} Different from DBA, FCBA considers all the combination styles of sub-pixel blocks in depth after dividing the global triggers, capturing correlations between these blocks and their environment. It helps the global model to learn a more complete backdoor pattern and improves the performance of backdoor attacks. Given the direct mapping between malicious clients and local triggers, each local model can be targeted by unique backdoor attacks. We segment FCBA into $M$ sub-attack problems. Each aims to manipulate the local model to fit both the main and backdoor tasks, ensuring correct operation on clean inputs but misclassification on backdoor ones. For round $t$, the adversarial objective of attacker $i$ with local dataset $D_i$ and target label $\tau$ is described as:
\begin{equation}
\begin{split}
    \omega _i^*&=\arg\underset{\omega _i}{\max}(\sum_{j\in D_{i}^{poi} }^{}P[G^{t+1}(B(x_{i}^{j},\phi_{i}^*))=\tau;\gamma;I]\\&+\sum_{j'\in D_{i}^{cln} }^{}P[G^{t+1}(x_{i}^{j'})=y_{i}^{j'}]), \forall i\in M
\end{split}
\end{equation}
\begin{table*}[h]
\resizebox{\textwidth}{!}{
\begin{adjustbox}{scale={0.5}{0.5}}
\begin{tabular}{@{}c|cccccc@{}}
\toprule
Trigger ID   &  & Red $\to$ 1        & Yellow $\to$ 2     & Green $\to$ 3     & Orange $\to$ 4    &  \\ \midrule
$O_{SD}(i)$ &  & $O_{SD}(1)=1$ & $O_{SD}(2)=2$ & $O_{SD}(3)=3$ & $O_{SD}(4)=4$ &  \\ \midrule
&  & $O_{FC}(1)=1$ & $O_{FC}(2)=2$ & $O_{FC}(3)=3$ & $O_{FC}(4)=4$ &  \\
$O_{FC}(i)$ & $O_{FC}(5)=1,2$ & $O_{FC}(6)=1,3$    & $O_{FC}(7)=1,4$    & $O_{FC}(8)=2,3$    & $O_{FC}(9)=2,4$    & $O_{FC}(10)=3,4$ \\
&                 & $O_{FC}(11)=1,2,3$ & $O_{FC}(12)=1,2,4$ & $O_{FC}(13)=1,3,4$ & $O_{FC}(14)=2,3,4$ &                  \\ \bottomrule
\end{tabular}
\end{adjustbox}
}
\caption{DBA and FCBA's local trigger generation strategies $O_{SD}(i)$ and $O_{FC}(i)$. Here, $m$ is set to 4, and the 4 different colored sub-pixel blocks are marked as 1, 2, 3, and 4.}
\label{tab1}
\end{table*}

Here, the poisoned dataset $D_{i}^{poi}$ and the clean dataset $D_{i}^{cln}$ satisfy $D_{i}^{poi}\cap D_{i}^{cln}=\phi$ and $D_{i}^{poi}\cup D_{i}^{cln}= D_{i}$. Function $B$ uses the parameter $\phi_{i}^*$ to convert the clean data in any class into backdoor data with a local trigger pattern of the attacker's choosing. $\phi_{i}^*=\{\phi,m,O(i)\}$ is the local trigger of the malicious client $M_i$. $\phi$ is the global trigger used to control the global trigger style. For image data, it can be decomposed into factors such as Trigger Size $TS$, Trigger Location $TL$ and Trigger Gap $TG$ ($\phi=\{TS,TL,TG\}$) as shown in Fig. 2. $O(i)=\{O_{SD}(i), O_{FC}(i)\}$ is an optional local trigger generation policy that generates a set of local trigger styles based on the first two parameters. $SD$ stands for the simple division strategy used by DBA and $FC$ stands for the full-combinatorial strategy used by FCBA, here we choose the latter. The attacker performs backdoor injection using a poisoning round interval $I$, manipulates its updates using a scale factor $\gamma$ before submitting them to the aggregator, and chooses the optimal poisoning ratio $r$ to produce a better model parameter $\omega _i^*$, $G_{t+1}$ that simultaneously assigns with highest probability a target label $\tau $ for the backdoor data $B(x_{i}^{j},\phi_{i}^*)$ and a ground true label $y_{i}^{j'}$ for the clean data $x_{i}^{j'}$. We will present the important factors affecting the performance of the attack in the experimental section.
\begin{figure}[H] 
\centering 
\includegraphics[width=0.43\textwidth]{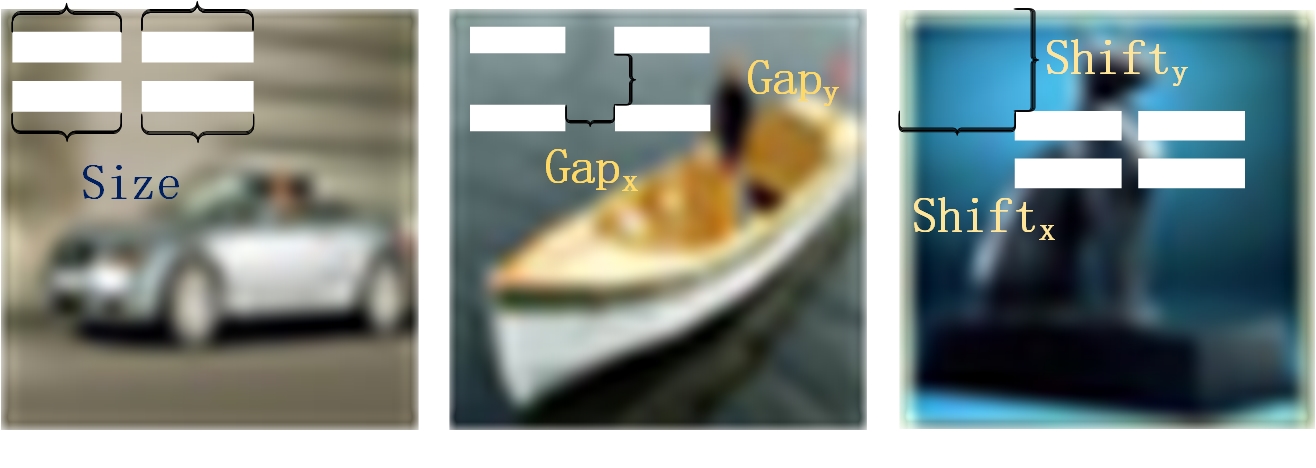} 
\caption{Trigger factors (size, gap and location) in backdoored images.} 
\label{Size-gap} 
\end{figure}
The local trigger generation strategies $O_{SD}(i)$ and $O_{FC}(i)$ for $m=4$ are shown in Tab. 1, where the $O_{FC}(i)$ strategy is shown in detail in Fig. 1. Under the premise where the parameters and the total amount of poison are almost aligned, the two attacks were evaluated using the same global triggers. The results show that FCBA's performance is significantly better than DBA, especially in terms of attack persistence. 

Note that the above procedure, which calculates the total number of malicious clients $M$ based on the number of trigger divisions $m$, can also be solved in reverse in real deployments. This is because attackers in FL can flexibly adjust their attack strategies according to their capabilities. Our work applies combinatorics theory to provide an explicit mapping relationship between these two variables.


\section{Experiment \& Analysis} 
In this section, we detail our experimental setup and compare the performance of the FCBA to leading backdoor attacks in FL across three datasets and two model architectures, highlighting the superiority of the FCBA attack in terms of its efficiency and durability. We use data distribution plots to depict ASR decay over iterations and t-SNE distance plots to illuminate the persistence of FCBA attacks. We also analyze the effects brought by different factors, demonstrating that FCBA has a wide range of attack persistence. Finally, our analysis and experiments show that it is difficult for existing defense methods to effectively defend FCBA, proving its robustness.
\subsection{Experiment Setup} 
\subsubsection{Datasets \& Model Architecture.} We provide a brief overview of tasks for each dataset in Appendix A.2. Tab. 2 showcases the model architecture and other specifics for these datasets.
\begin{table}[!h]
\centering
\resizebox{\columnwidth}{!}{
\begin{tabular}{c|c|c|c|c|c}
\toprule
\multirow{2}*{Dataset} & \multirow{2}*{Lables}& \multirow{2}*{Image Size} & \multicolumn{2}{c|}{{\footnotesize Images}} & \multirow{2}*{Model Architecture} \\
 \cmidrule(lr){4-5}
& & &{\footnotesize Training} & {\footnotesize Testing} \\
\midrule
MNIST & 10 & 28 x 28 x 1 & 50000 & 10000 & 2Conv + 2fc \\  
CIFAR-10 & 10 & 32 x 32 x 3 & 50000 & 10000 & lightweight Resnet-18 \\
GTSRB & 16 & 32 x 32 x3 & 23050 & 4310 & lightweight Resnet-18 \\
\bottomrule
\end{tabular}
}
\caption{ Dataset and model aechitecture.}
\label{tab2}  
\end{table}

\begin{figure*}[htbp]
    \centering
    \begin{subfigure}[b]{0.30\textwidth}
        \includegraphics[width=\linewidth]{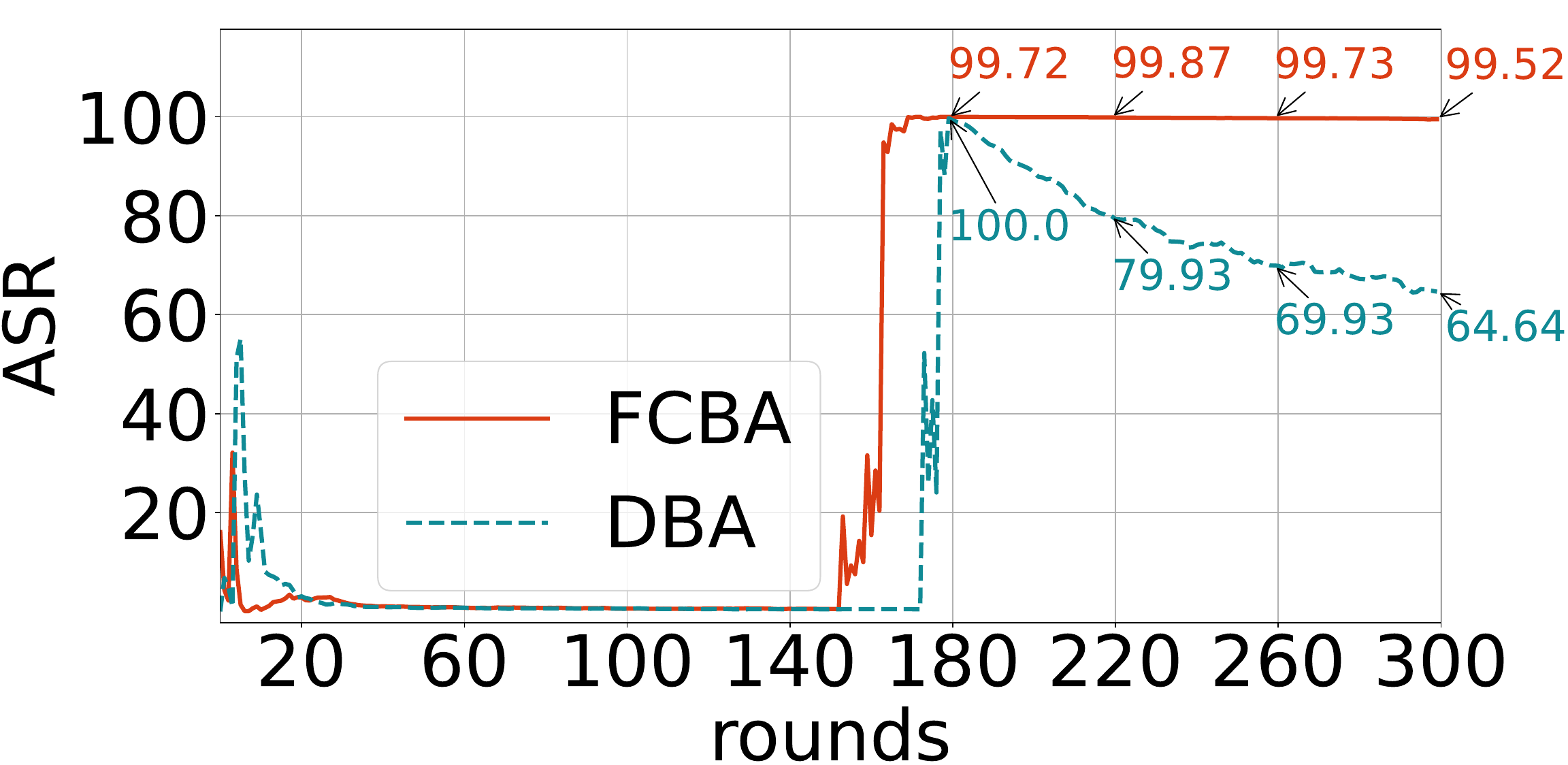}
        \caption{MNIST}
    \end{subfigure}
    \begin{subfigure}[b]{0.30\textwidth}
        \includegraphics[width=\linewidth]{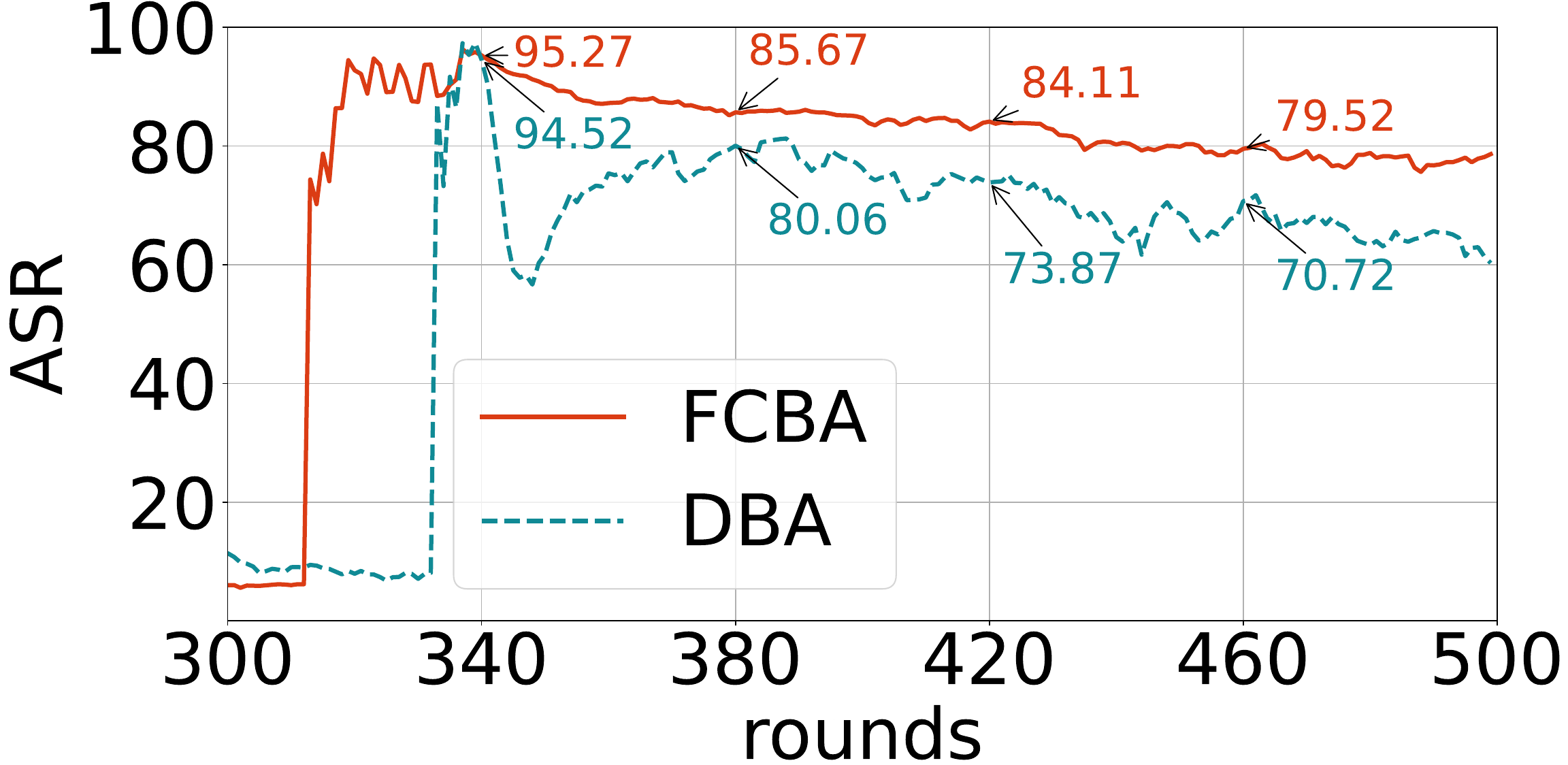}
        \caption{CIFAR-10}
    \end{subfigure}
    \begin{subfigure}[b]{0.30\textwidth}
        
        \includegraphics[width=\linewidth]{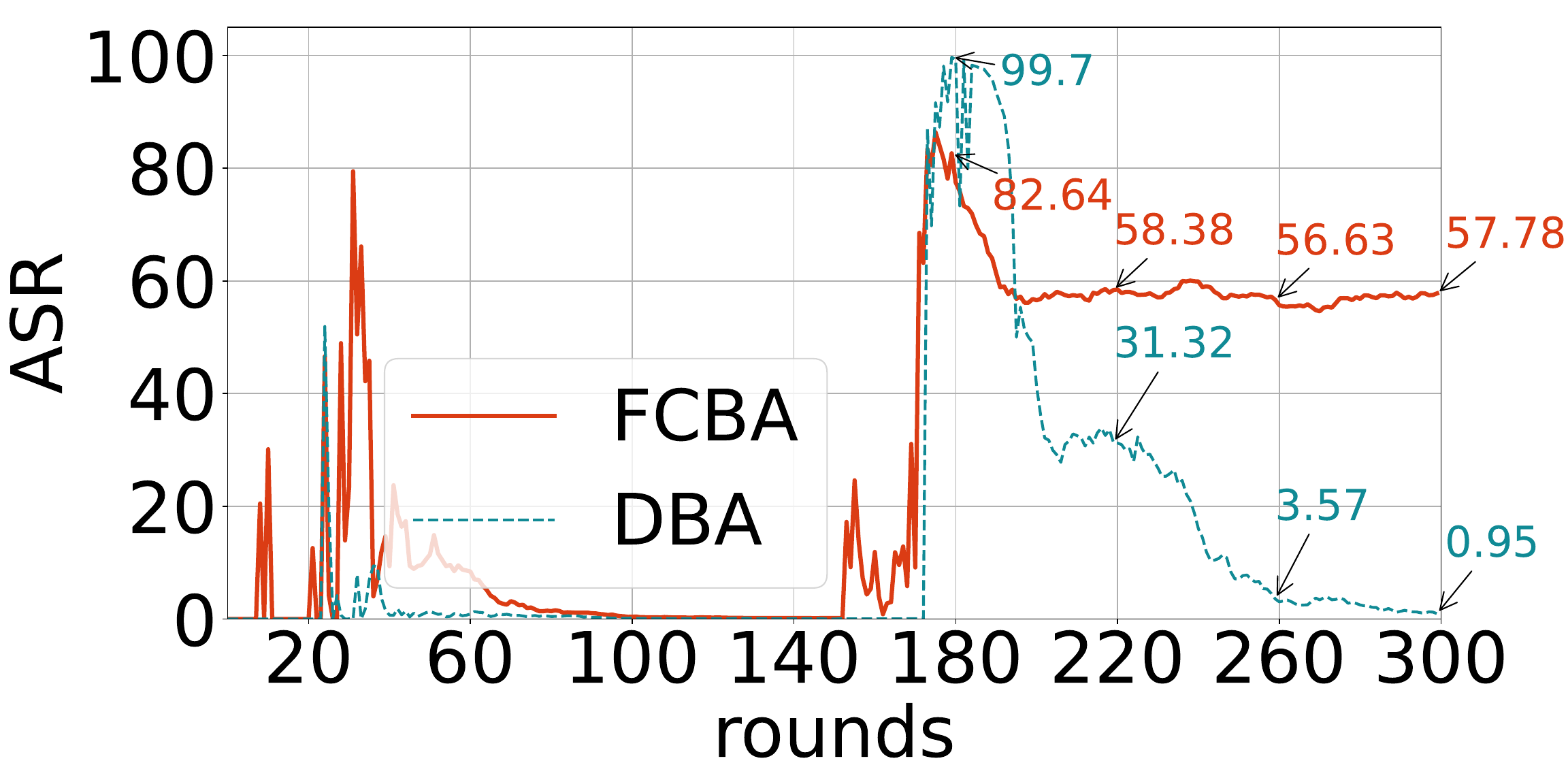}
        \caption{GTSRB}
    \end{subfigure}
    \caption{ASR of FCBA and DBA. FCBA is more effective and persistent than DBA.}
    \label{ASR-DBA-FCBA}
\end{figure*} 
\begin{figure*}[htbp]
    \centering    
    \begin{subfigure}[b]{0.30\textwidth}
        \includegraphics[width=\linewidth]{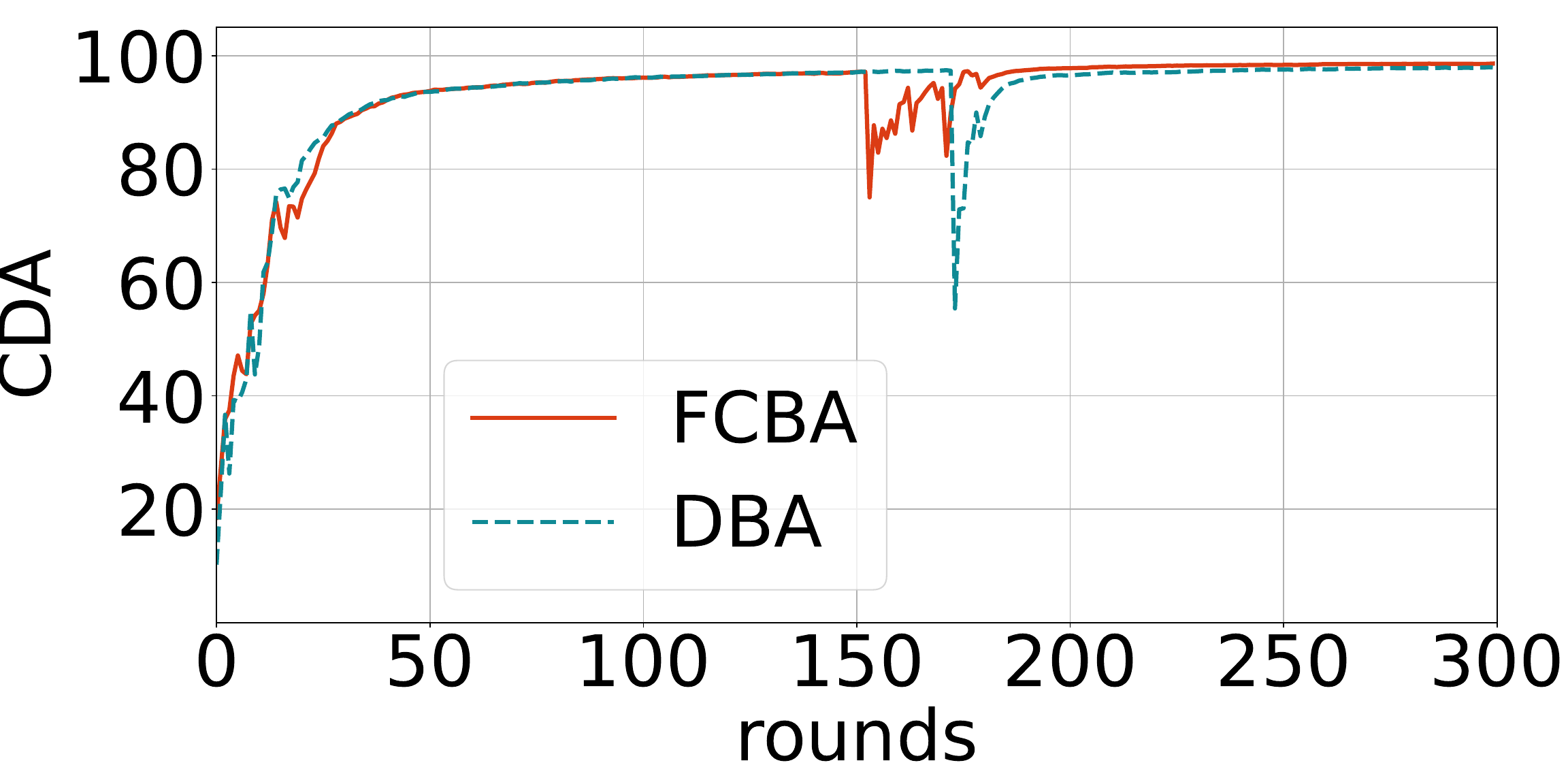}
        \caption{MNIST}
    \end{subfigure}
    \begin{subfigure}[b]{0.30\textwidth}
        \includegraphics[width=\linewidth]{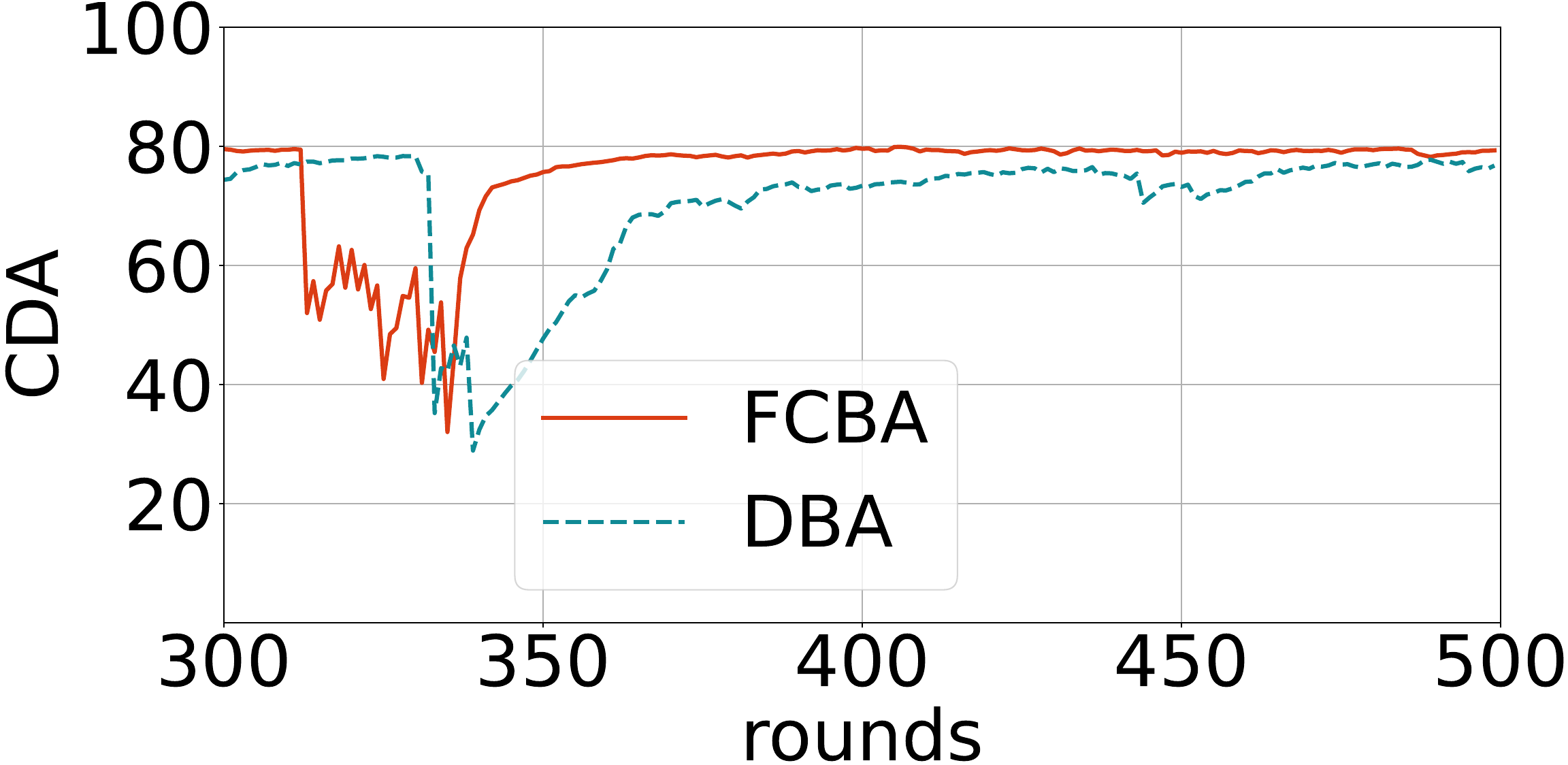}
        \caption{CIFAR-10}
    \end{subfigure}
    \begin{subfigure}[b]{0.30\textwidth}
        
        \includegraphics[width=\linewidth]{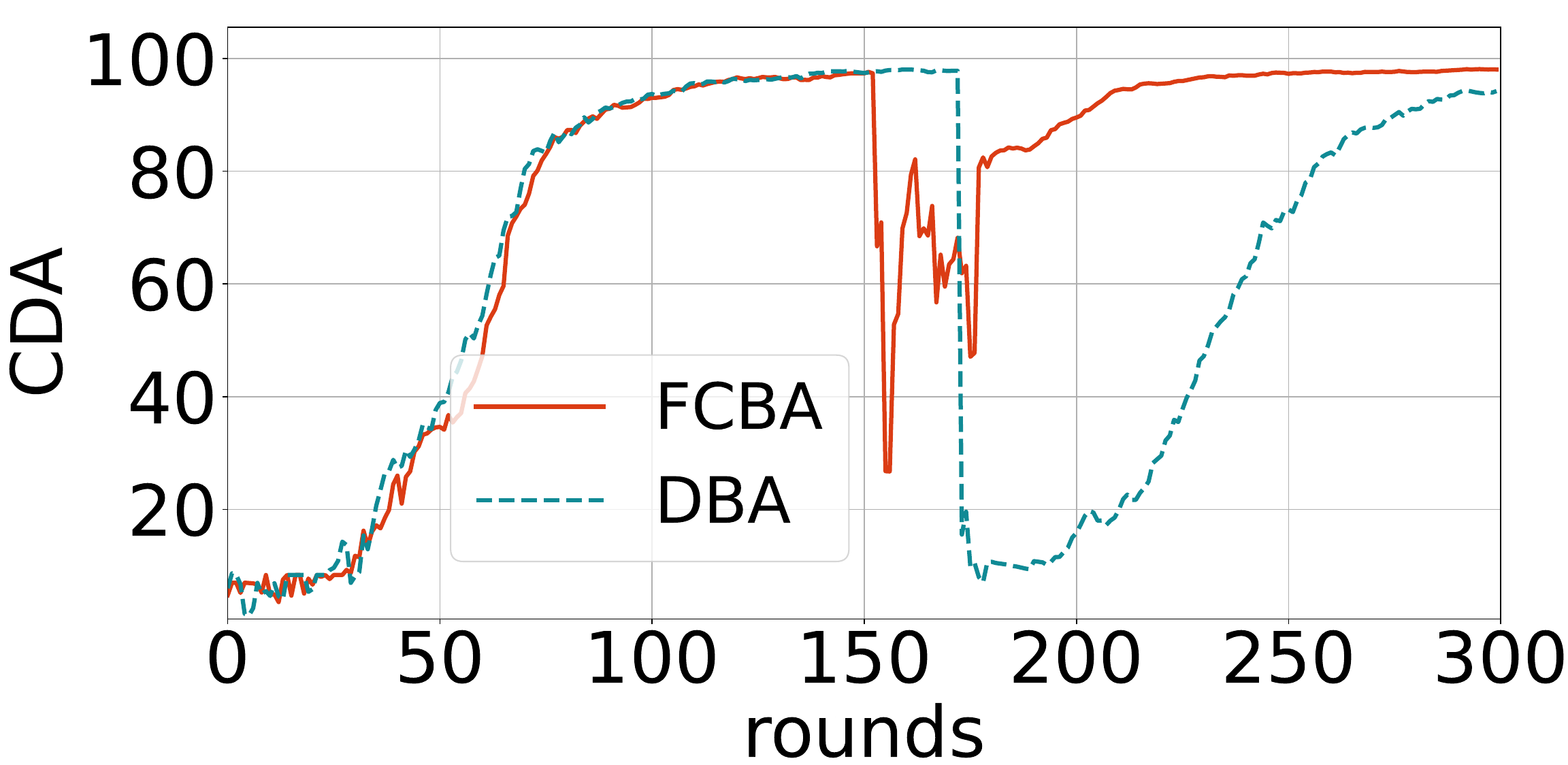}
        \caption{GTSRB}
    \end{subfigure}
    \caption{CDA of FCBA and DBA. FCBA is more hidden than DBA.}
    \label{CDA-DBA-FCBA}
\end{figure*}
\subsubsection{Parameters for training.} For three datasets, training images were allocated to 100 participants using a Dirichlet distribution with a hyperparameter of 0.5. During training with \textit{Stochastic Gradient Descent} (SGD) optimizer, each participant trained for $E$ local epochs with a specific local learning rate $lr$  and batch size of 64. In each round, 10 clients were chosen to submit their local updates for aggregation. The target labels for the backdoor are ``2'' in MNIST, ``Bird'' in CIFAR-10, and ``Pass by on right'' in GTSRB. Excluding analysis of crucial factors, the trigger factors were $\phi = \{4, 2, 0\}$ for MNIST, $\phi = \{6, 3, 0\}$ for CIFAR-10 and GTSRB, all involving 14 malicious parties. When the attack commence, malicious parties' batches comprise both clean and backdoor data, with a specific poisoning ratio $r$. Malicious participants have their own local poisoning $lr'$ and poisoning $E'$ (see Tab. 3) to maximize their backdoor performance and remain stealthy. Hardware details for the experiment are provided in Appendix A.3.
\begin{table}[!h]
  \centering
  \footnotesize
  \resizebox{\columnwidth}{!}{
  \begin{adjustbox}{scale={0.1}{0.1}}
  \begin{tabular}{c|c|c|c}
    \toprule
    {\footnotesize Dataset} & {\footnotesize $\text { Benign } l_{r} / E$} & {\footnotesize $\text { Poison } l_{r} ^{'}/E^{'}$}& {\footnotesize Poison Ratio $r$} \\
    \midrule
    MNIST  & 0.1 / 1 & 0.05 / 10 & 3 / 64 \\        
    CIFAR-10  & 0.1 / 2 & 0.05 / 6 & 2 / 64 \\        
    GTSRB  & 0.1 / 1 & 0.05 / 10 & 4 / 64 \\        
    \bottomrule
  \end{tabular}
  \end{adjustbox}
  }
  \caption{Parameters for training.}
  \label{tab3}
\end{table}
\subsubsection{Evaluation Metric.} We evaluate the performance of the new backdoor attack by three metrics.
\begin{itemize}
    \item \textbf{Clean Data Accuracy (CDA)} is the classification accuracy of backdoored model for clean samples that are with no triggers.
    \item \textbf{Attack Success Rate (ASR)} is the probability that trigger inputs are misclassified into the attacker targeted labels.
    \item \textbf{Attack Success Rate after $t$ rounds (ASR-$t$)} is the attack success rate of $t$ rounds after a complete FCBA is performed, used to quantify durability. Here, he larger the value of $t$, the greater the ASR-$t$, indicating higher persistence.
\end{itemize}

As for CDA, a backdoor attacker should retain it similar to the clean model counterpart. As for ASR and ASR-$t$, the attacker should maximize them.

\subsection{Full Combination  Backdoor Attack vs. Distributed Backdoor Attack}  
Single-shot and multiple-shot attacks are two main attack setups. Multiple-shot attacks rely on the continual selection of malicious clients for aggregation; otherwise, benign updates could neutralize the backdoor in the global model. In their tests~\cite{DBA}, malicious clients were prioritized for aggregation, with benign ones chosen at random. This method, however, diverges from real-world random aggregation. The threat model states attackers can't change the server's aggregation rules. The threat model states attackers can't change the server's aggregation rules. The likelihood of selecting few malicious clients consecutively in a random setup is low. Moreover, frequent inclusion of malicious clients might alert the server, compromising the backdoor's efficacy.

We opt for the more pragmatic single-shot attack in our experiments. In this setup, each malicious participant submits a single update, enabling the global model to quickly show a high backdoor success rate over several iterations. Traditional data poisoning struggles to achieve this. We utilize the concept of model replacement~\cite{CBA} to amplify malicious updates, ensuring the backdoor survives average aggregation without rapid deterioration.
\begin{figure*}[htbp]
    \centering
    \begin{subfigure}[b]{0.18\textwidth}
        \includegraphics[width=\linewidth]{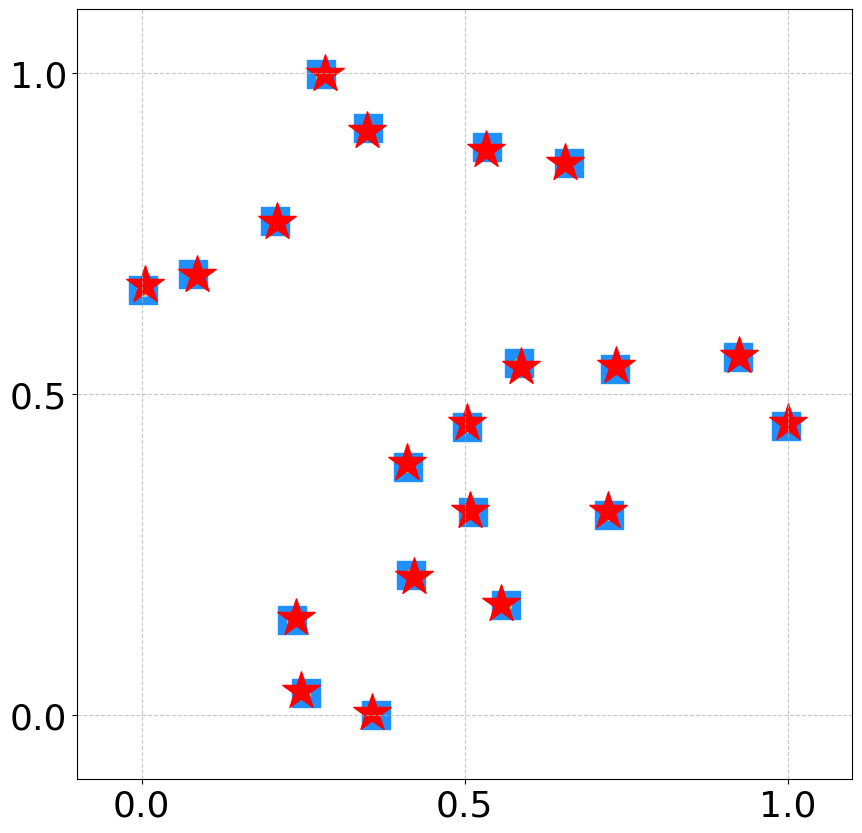}
        \caption{Pre-attack}
    \end{subfigure}
    \begin{subfigure}[b]{0.18\textwidth}
        \includegraphics[width=\linewidth]{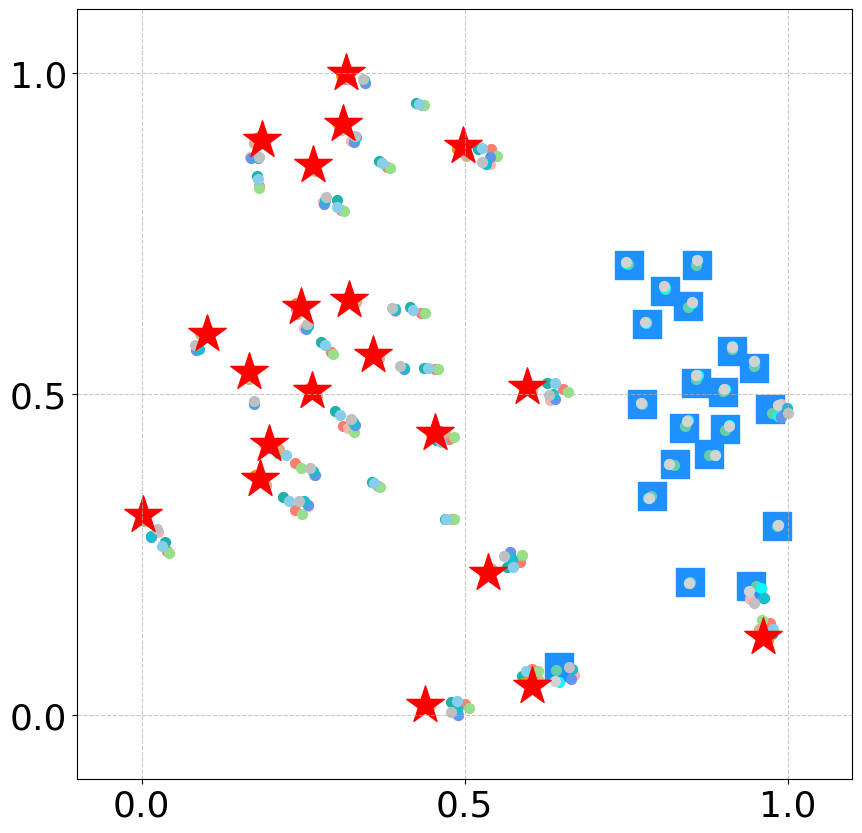}
        \caption{Round 0 Post-attack}
    \end{subfigure}
    \begin{subfigure}[b]{0.18\textwidth}
        \includegraphics[width=\linewidth]{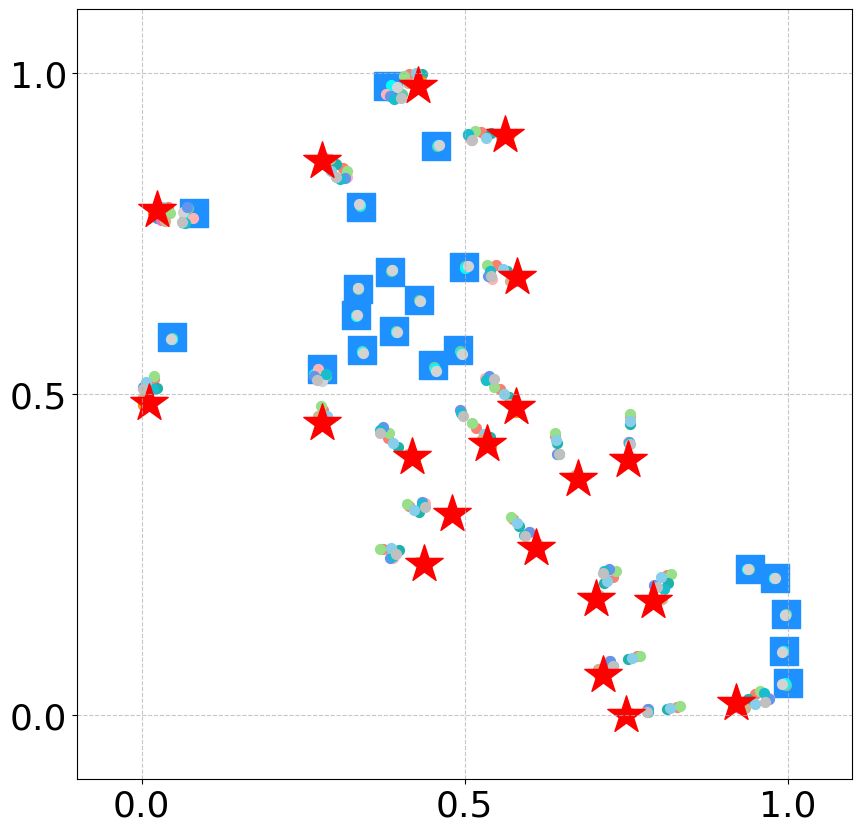}
        \caption{Round 40 Post-attack}
    \end{subfigure}
    \begin{subfigure}[b]{0.18\textwidth}
        \includegraphics[width=\linewidth]{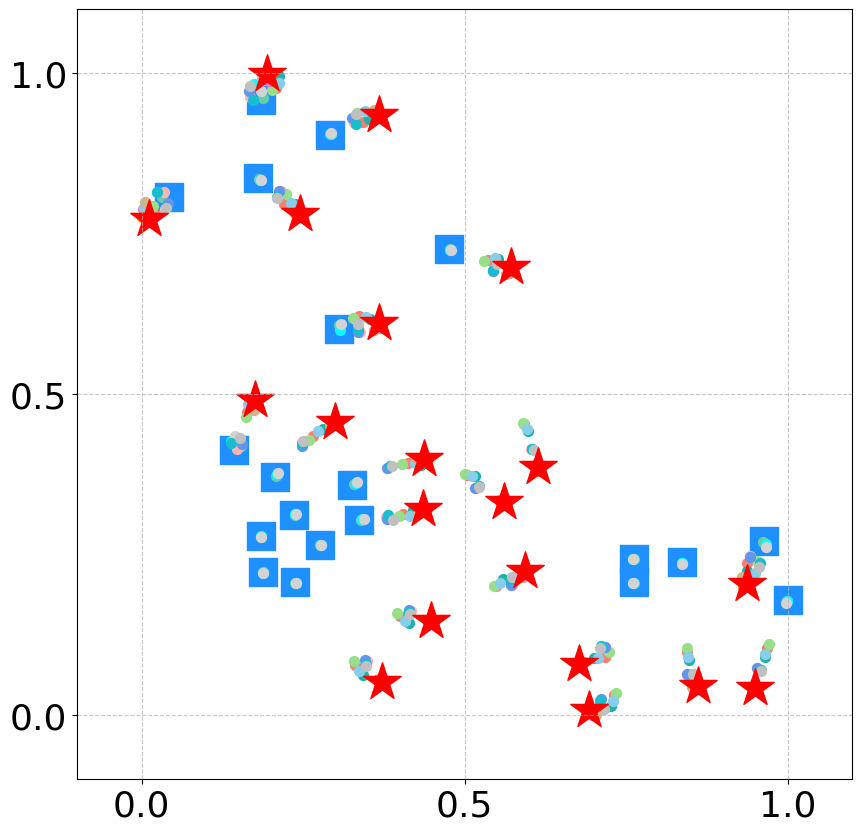}
        \caption{Round 80 Post-attack}
    \end{subfigure}
    \begin{subfigure}[b]{0.2484\textwidth}
        \includegraphics[width=\linewidth]{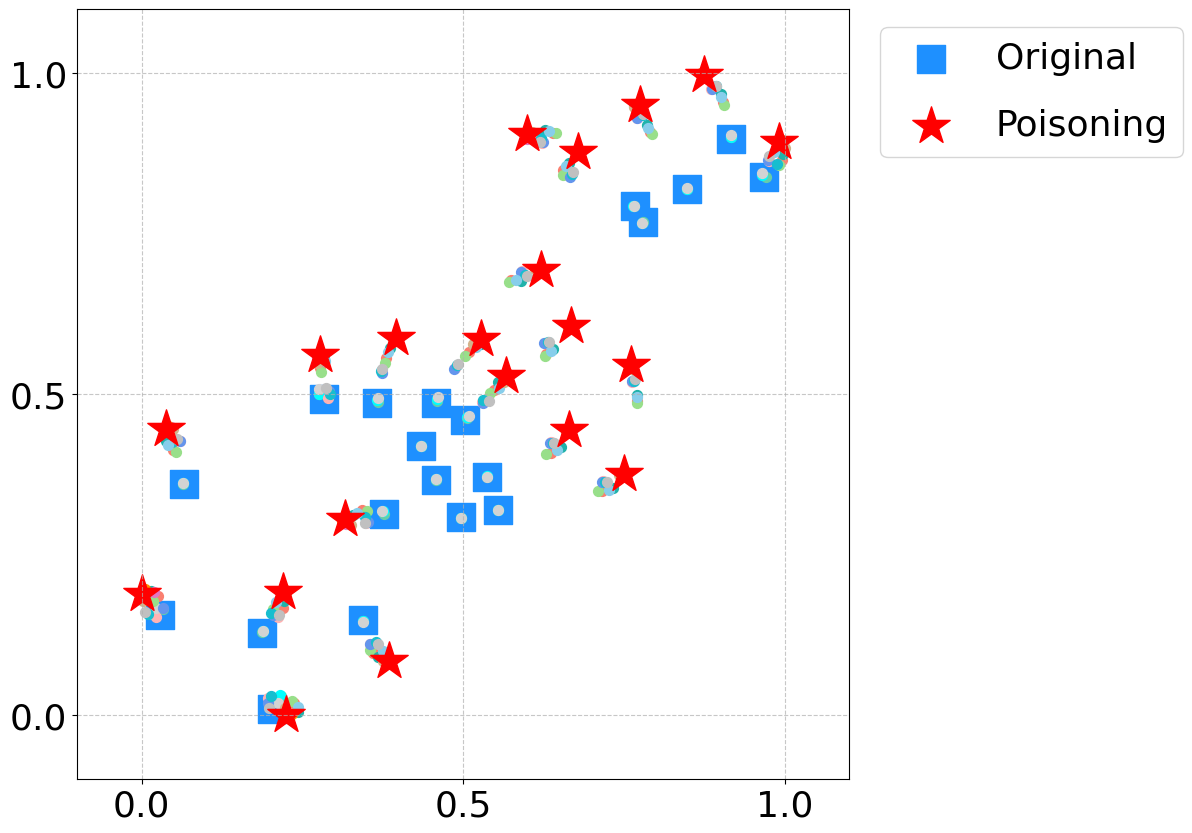}
        
        \caption{Round 120 Post-attack}

    \end{subfigure}
    
    \caption{Visualization of clean samples and trigger samples before and after DBA. After the attack, the global model can clearly distinguish between the two types of samples. However, as the rounds increase, the separation between the sample distributions becomes increasingly blurred.}
    \label{tsne-DBA}
\end{figure*}
\begin{figure*}[htbp]
    \centering
    
    \begin{subfigure}[b]{0.18\textwidth}
        \includegraphics[width=\linewidth]{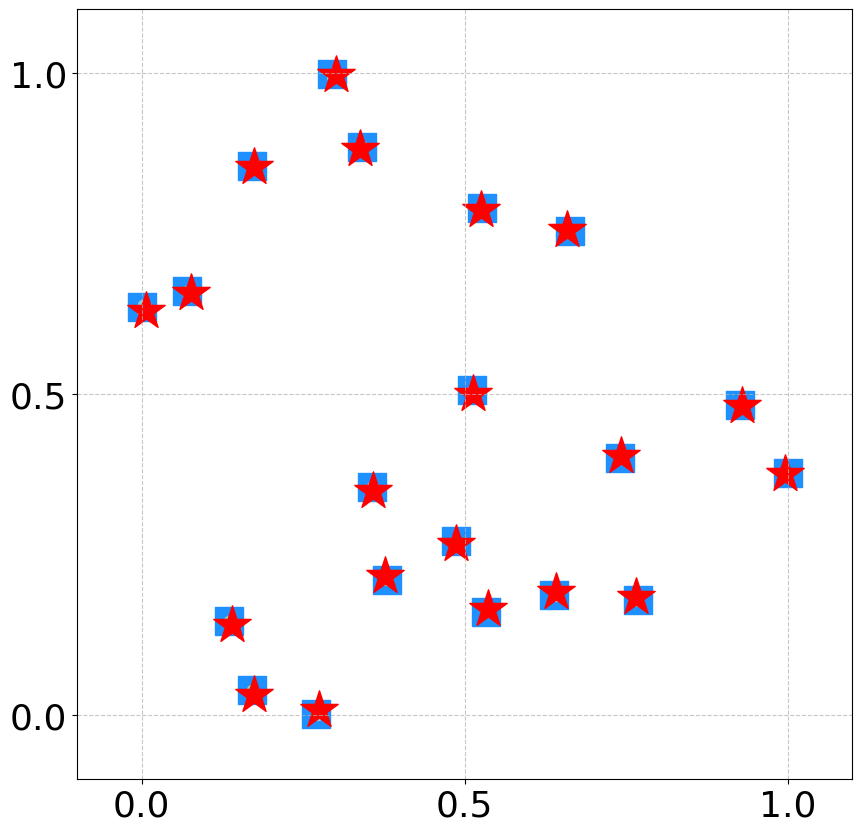}
        \caption{Pre-attack}
    \end{subfigure}
    \begin{subfigure}[b]{0.18\textwidth}
        \includegraphics[width=\linewidth]{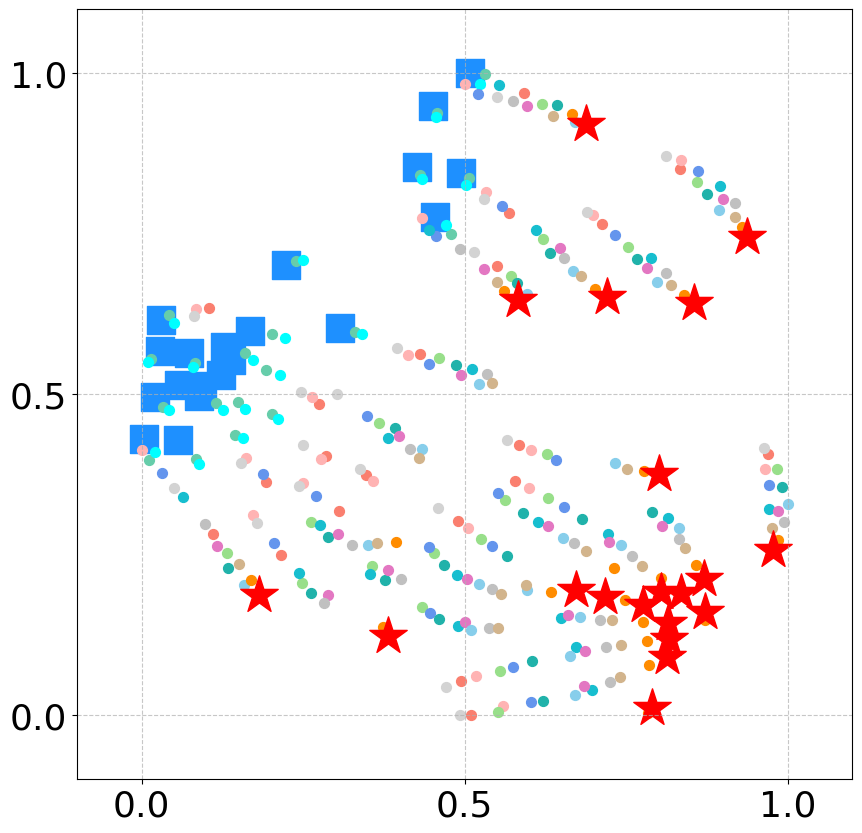}
        \caption{Round 0 Post-attack}
    \end{subfigure}
    \begin{subfigure}[b]{0.18\textwidth}
        \includegraphics[width=\linewidth]{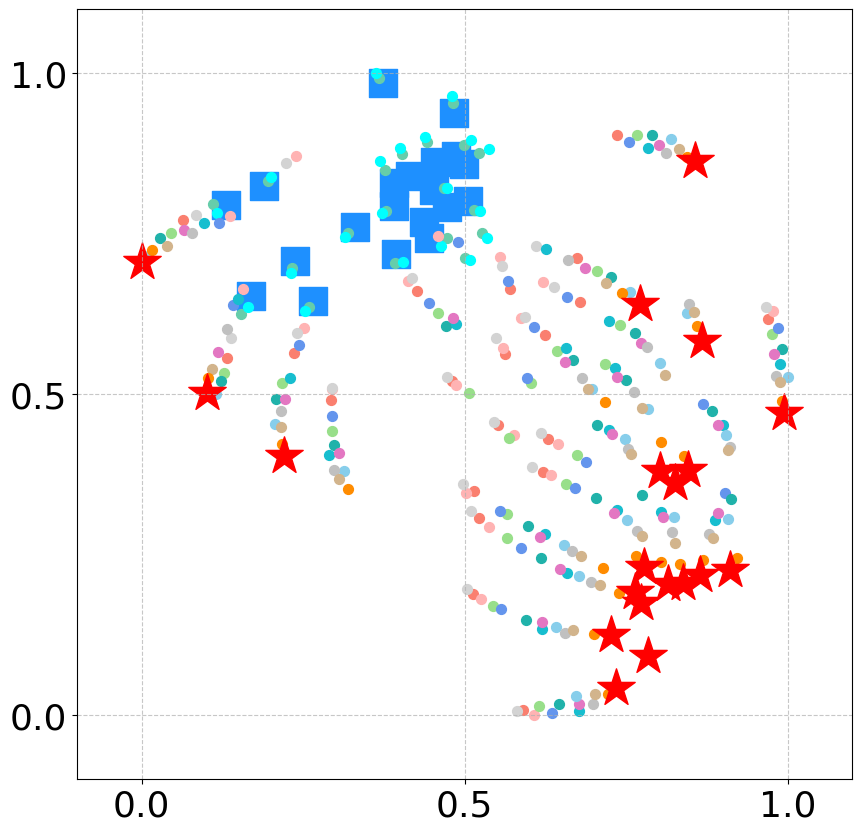}
        \caption{Round 40 Post-attack}
    \end{subfigure}
    \begin{subfigure}[b]{0.18\textwidth}
        \includegraphics[width=\linewidth]{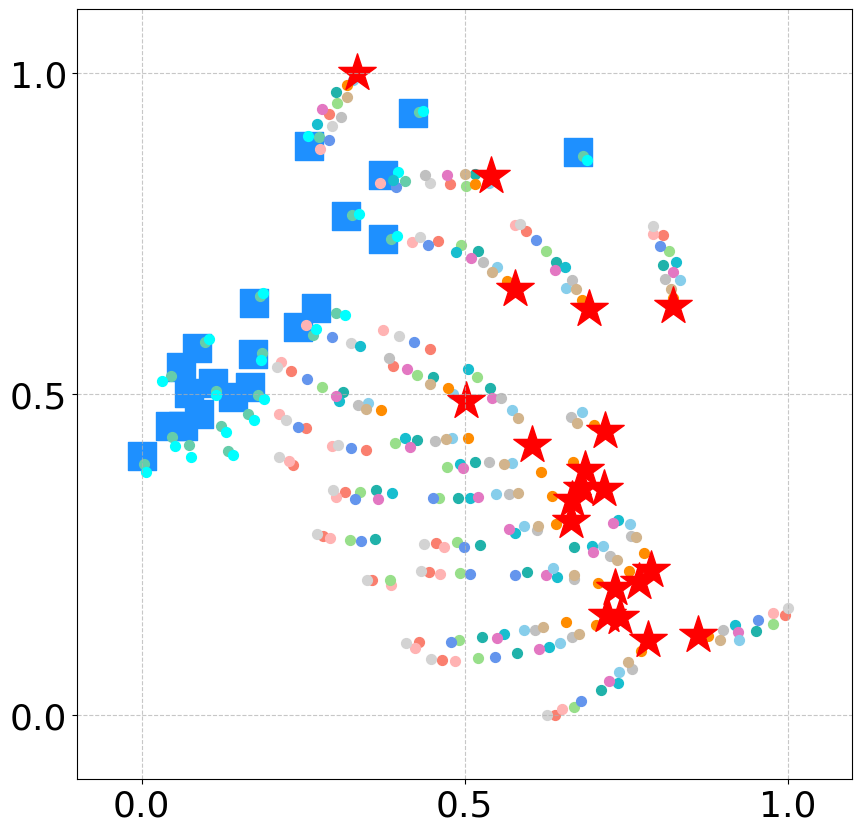}
        \caption{Round 80 Post-attack}
    \end{subfigure}
    \begin{subfigure}[b]{0.2484\textwidth}
        \includegraphics[width=\linewidth]{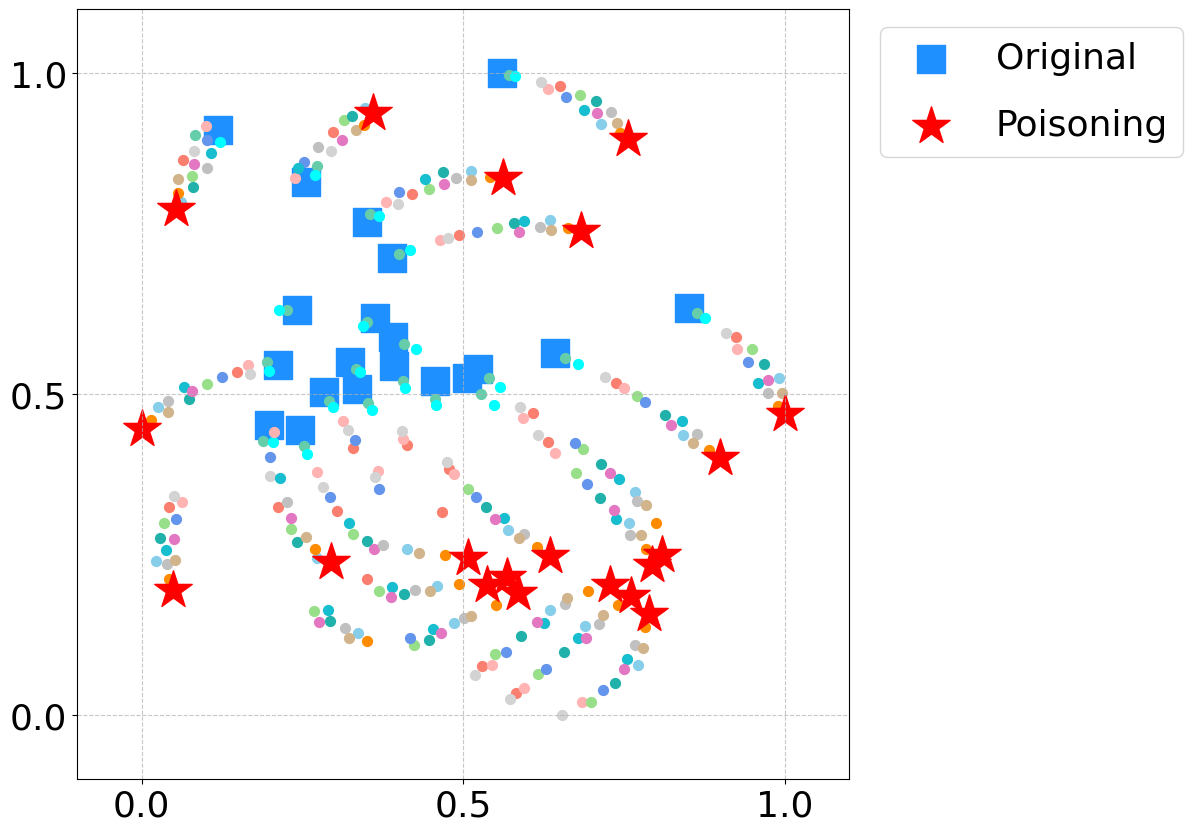}
        
        \caption{Round 120 Post-attack}
    \end{subfigure}
    
    \caption{Visualization of clean samples and trigger samples before and after FCBA. After the attack, the global model can clearly distinguish the two types of samples. As the rounds progress, a noticeable separation between the sample distributions still remains.}
    \label{tsne-FCBA}
\end{figure*}
To ensure fairness, we aim to keep the total poisoned pixels of FCBA similar to or less than that of DBA . Both FCBA and DBA complete backdoor injection in the same round, e.g., round 340 for CIFAR-10. We use a consistent global trigger for evaluation and omit the target class test data to prevent bias. We initiate the attack post global model accuracy convergence (The reasons are detailed in Appendix A.4). The global learning rate, $\eta$, is set to 0.1 for all datasets.

Here we focus on the persistence of backdoor attacks and use ASR-$t$ to portray it indirectly. Displaying ASR after 0, 40, 80, and 120 rounds post-poison injection, we use the ASR curve trend to indirectly illustrate attack persistence.

As shown in  Fig. 3, FCBA registers nearly 100\% ASR in MNIST and CIFAR-10 post full-attack ($\gamma=100$). Though benign updates can dilute ASR, FCBA's decay rate is slower than DBA. After 120 rounds, FCBA's ASR-t for MNIST, CIFAR-10, and GTSRB stands at 99.52\%, 79.52\%, and 57.78\%, versus 64.64\%, 70.72\%, and 0.95\% for DBA, highlighting FCBA's enhanced persistence. Fig. 4 reveals that while the model's primary accuracy dips during the backdoor injection, it rebounds with additional rounds, signifying minimal adverse impact from FCBA on main task performance. Moreover, the quicker recovery of FCBA's CDA post-attack further underscores its lesser impact on main tasks.

Fig. 4 reveals that while the model's main accuracy dips during the backdoor injection, it rebounds with additional rounds, signifying minimal adverse impact from FCBA on main task performance. Moreover, the quicker recovery of FCBA's CDA post-attack further underscores its lesser impact on main tasks.

\subsection{Why FCBA Attack Persistence Is High?}  

Using t-SNE, we visualize data distributions for 20 clean and 20 backdoor samples of the MNIST ``3'' class to study benign updates' impact and FCBA's persistence. In Fig. 5 , increasing rounds blur the distinction between these distributions with a DBA-implanted backdoor. However, Fig. 6 shows a clearer split for FCBA-implanted models. This demonstrates: (1) Benign updates dilute the backdoor effect, causing the model to favor the true labels of backdoor samples with more iterations. (2) FCBA's training more effectively bridges the training-inference gap, enabling this discrimination.

We analyze the t-SNE distance between clean samples (excluding the target class) and their backdoor sample representations in the test set, with results shown in Fig. 7. The larger t-SNE distance for FCBA compared to DBA indicates FCBA's superior capability to differentiate clean from backdoor samples. The red curve's gentle decline, compared to the blue curve, suggests FCBA maintains this discernment over a more extended period, explaining its higher attack persistence. (For a deeper dive into FCBA's heightened persistence, see Appendix A.5).
\begin{figure}[H] 
 \centering 
 \includegraphics[width=0.4\textwidth]{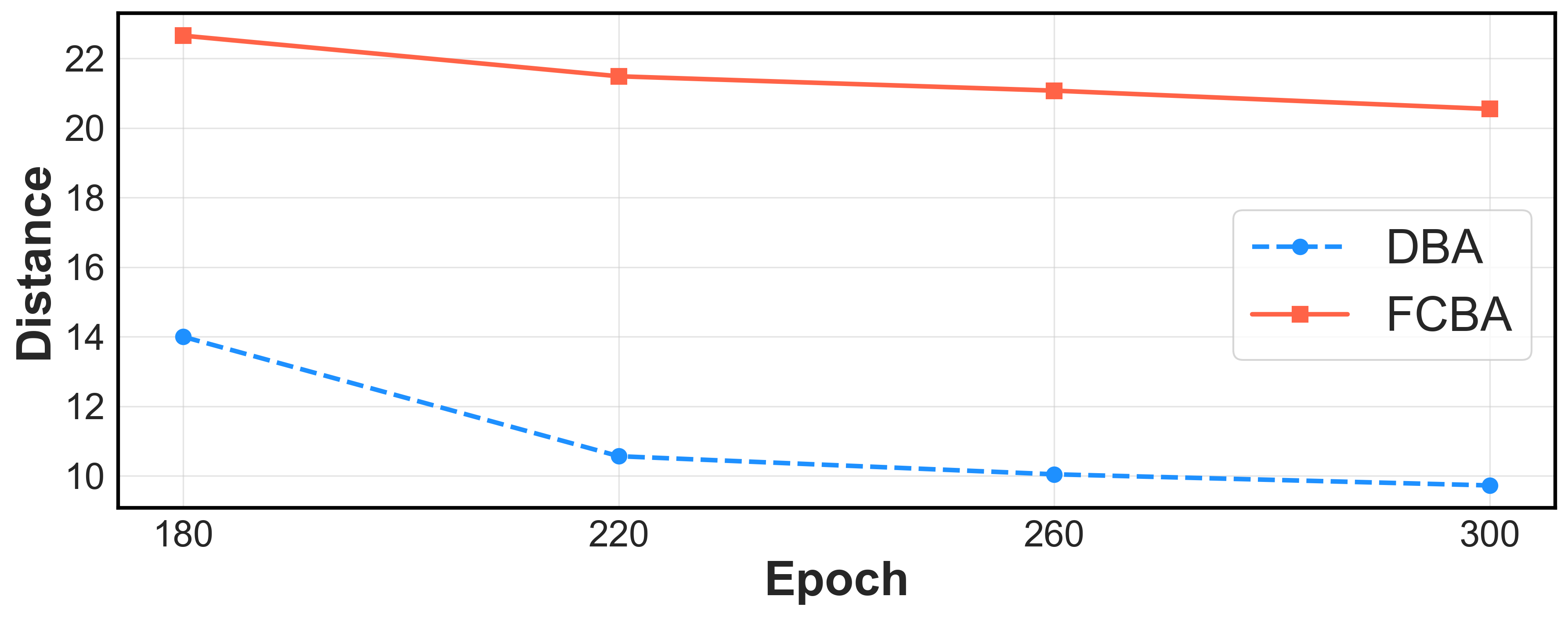} 
 \caption{Average distance between clean and triggered data} 
 \label{Average-distance} 
 \end{figure}

\subsection{Analysis of Crucial Factors in FCBA}  

Several factors in FCBA impact attack performance. Key ones are highlighted here. Fig. 2 elucidates the $TS$, $TL$, and $TG$ attributes in image datasets. For clarity, we represent sub-pixel blocks post-global trigger division as uniformly sized rectangles. We then investigate and analyze these factors on MNIST and CIFAR-10 within a reasonable value range.

\subsubsection{Effects of $\gamma$.} $\gamma$, as defined by~\citet{CBA}, is employed by attackers to amplify malicious updates. 

In Fig. 8, as $\gamma$ increases, ASR and ASR-$t$ initially rise before stabilizing. This is attributed to amplified malicious updates enhancing the backdoor effect up to its performance limit. Yet, when $\gamma$ surges further, ASR drops to 0\% because of model destabilization from large updates (see Appendix A.6). Selecting $\gamma$ involves tradeoffs, with extreme values compromising attack efficacy. 

\begin{figure}[htbp]
    \centering    
    \begin{subfigure}[b]{0.23\textwidth}
        \includegraphics[width=\linewidth]{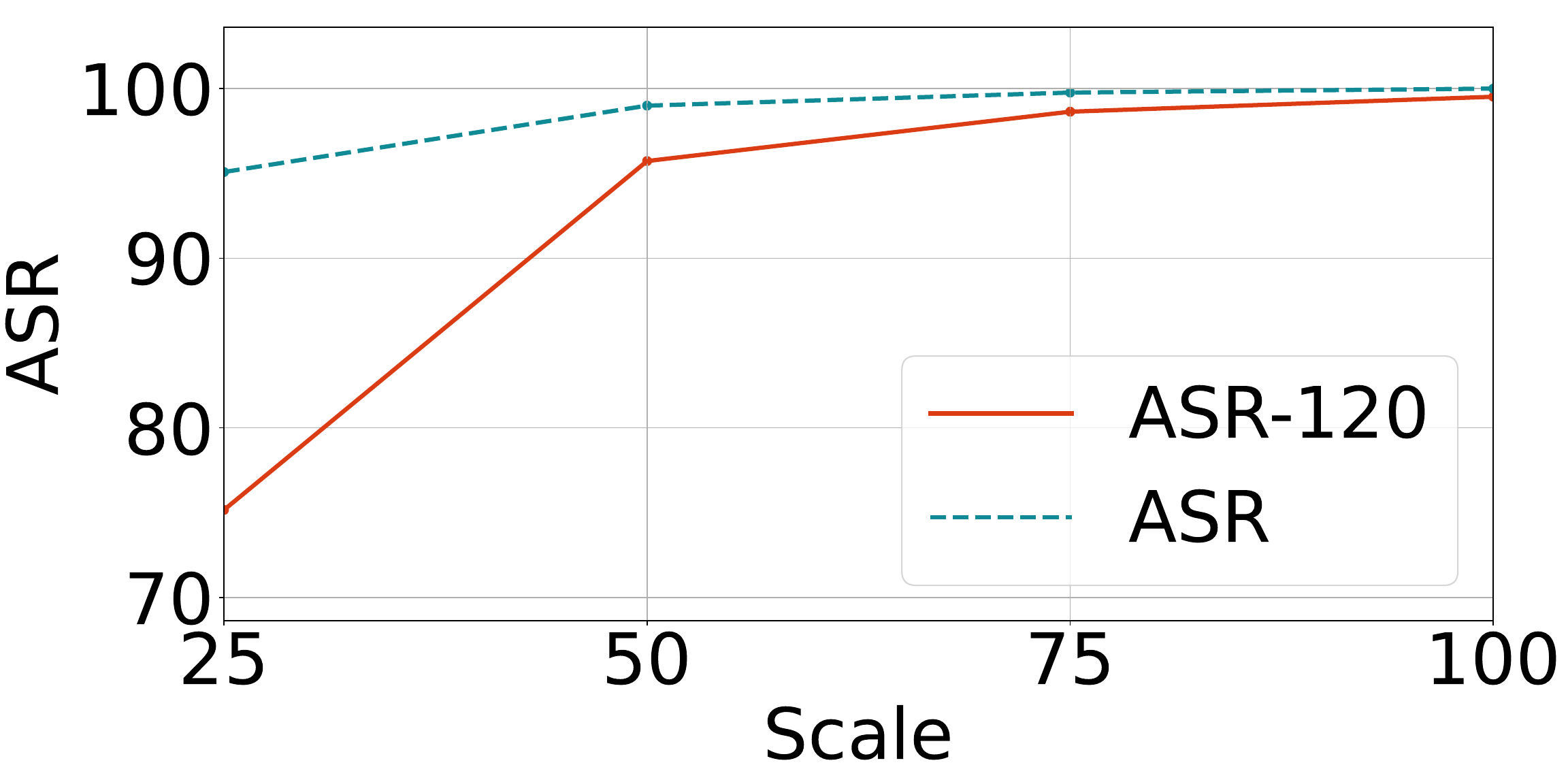}
        \caption{MNIST}
    \end{subfigure}
    \begin{subfigure}[b]{0.23\textwidth}
        \includegraphics[width=\linewidth]{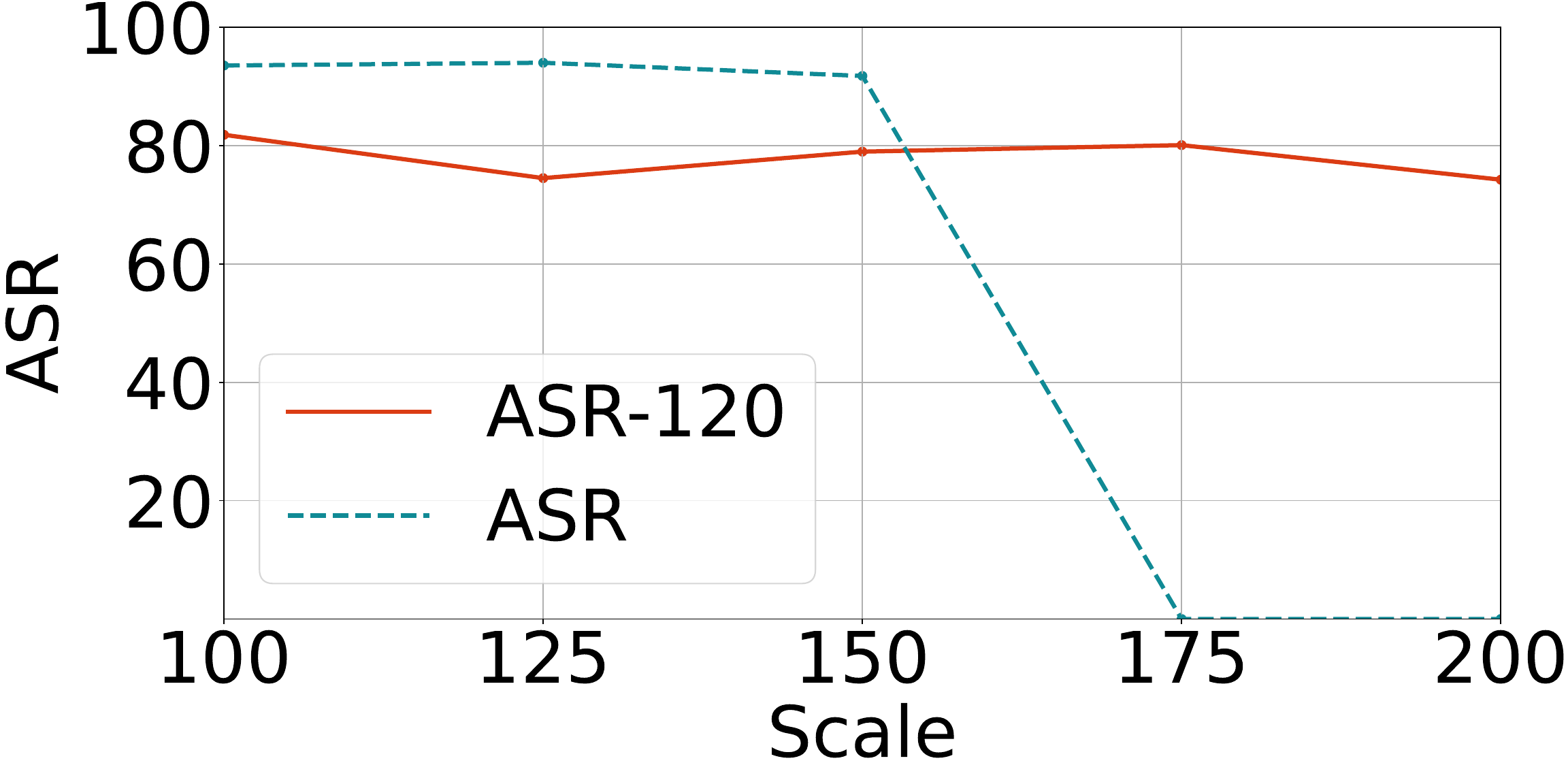}
        \caption{CIFAR-10}
    \end{subfigure}
    
    \caption{Effects of Scale on ASR and ASR-$t$.}
    \label{Effects-of-Scale}
\end{figure}

\subsubsection{Effects of $\alpha$.} FL typically assumes a non-i.i.d. data distribution among participants. We use the Dirichlet distribution~\cite{minka2000estimating} with different $\alpha$ values to shift from i.i.d. to non-i.i.d., where smaller $\alpha$ indicates more data imbalance. See more details in Appendix A.7.

As Fig. 9 illustrates, FCBA maintains high ASR and ASR-$t$, barring $\alpha=0.1$, indicating our attack's robust efficiency and persistence.

At $\alpha=0.1$, FCBA's performance in both backdoor and main tasks declines significantly. This stems from data imbalance, inhibiting optimal model training. Essentially, as data distribution deviates from i.i.d., model performance and attack efficacy deteriorate.

Due to space constraints, see Appendix A.8 for further discussion about other factors.

\begin{figure}[htbp]
    \centering    
    \begin{subfigure}[b]{0.23\textwidth}
        \includegraphics[width=\linewidth]{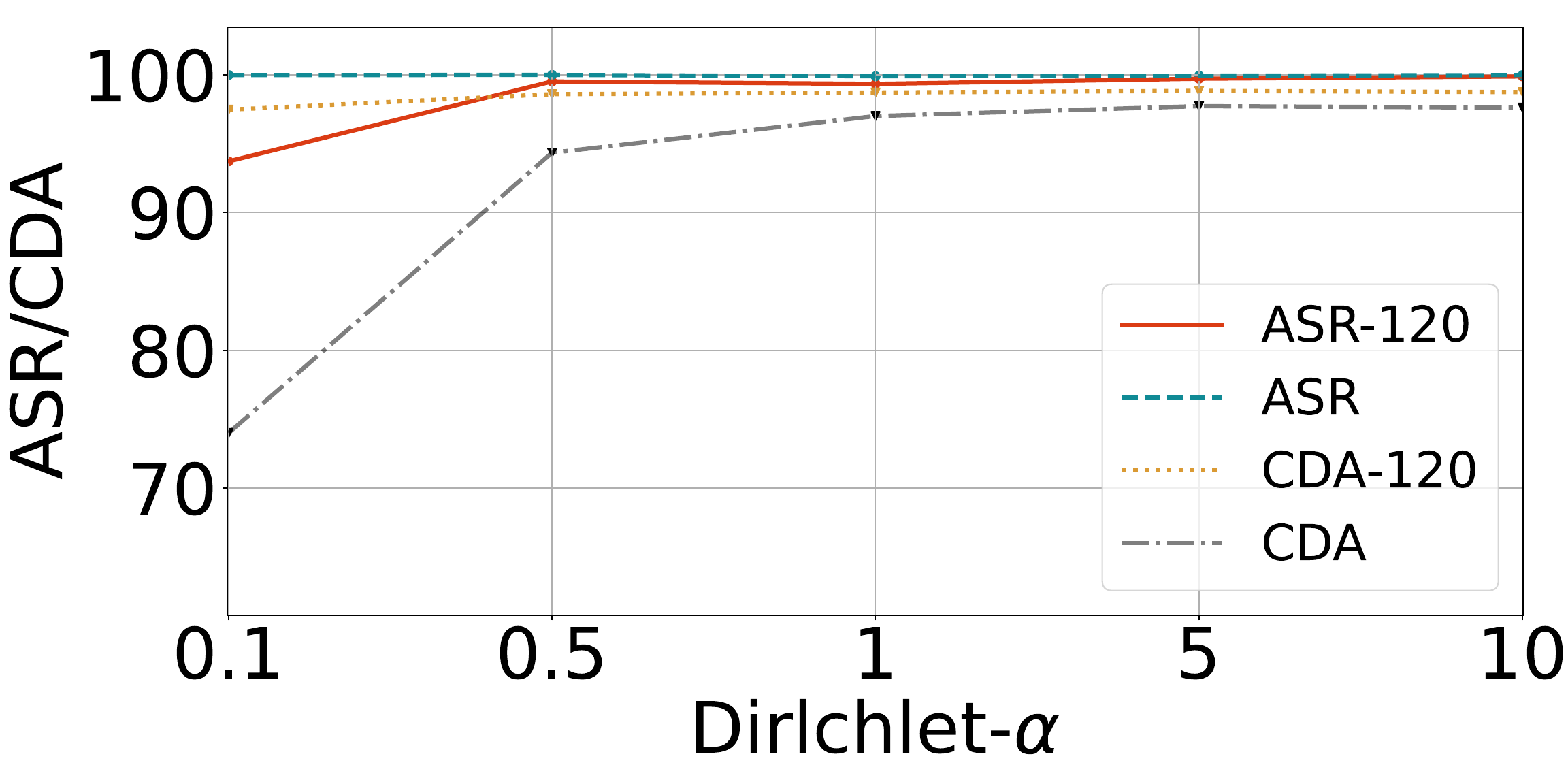}
        \caption{MNIST}
    \end{subfigure}
    \begin{subfigure}[b]{0.23\textwidth}
        \includegraphics[width=\linewidth]{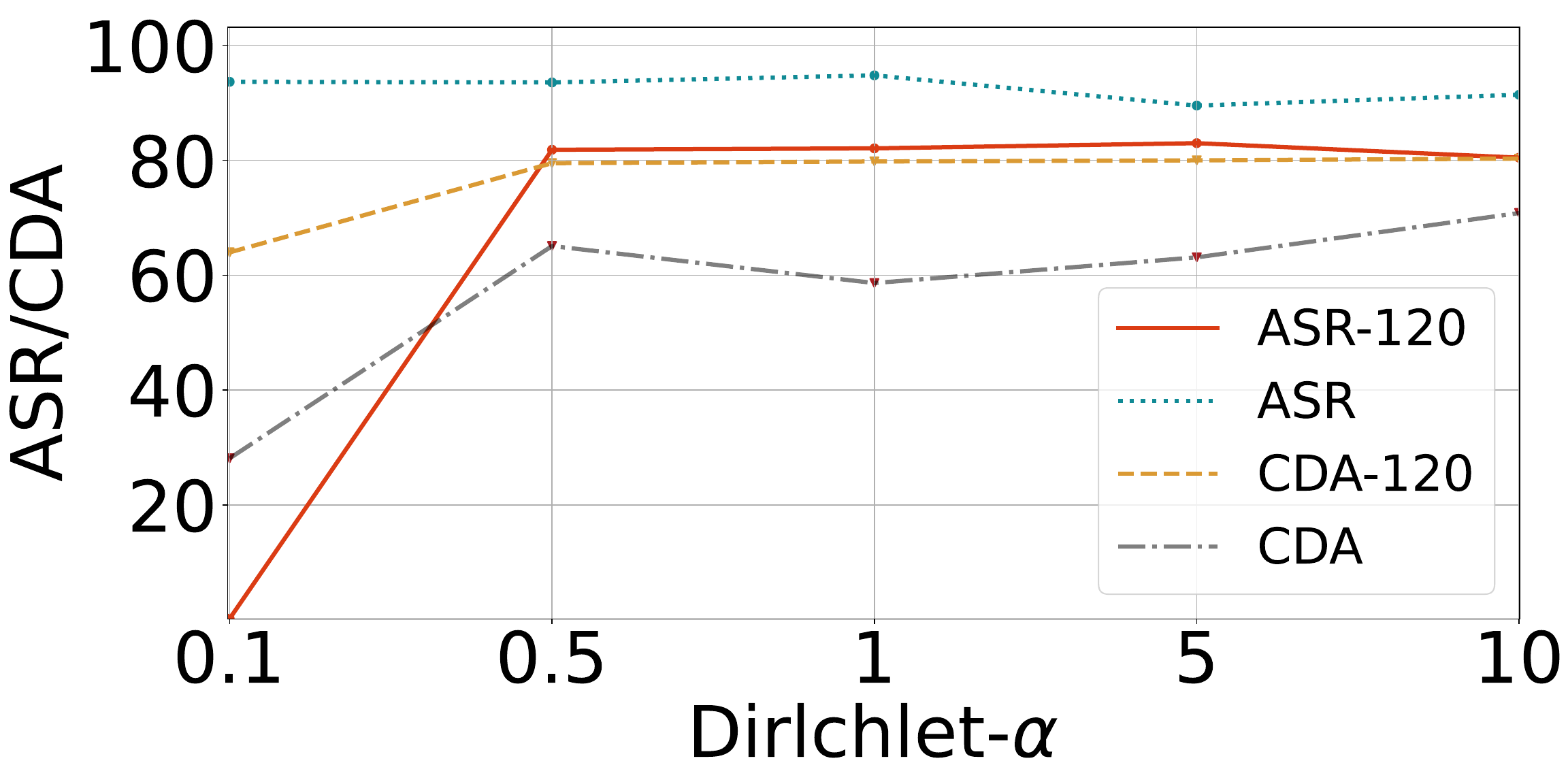}
        \caption{CIFAR-10}
    \end{subfigure}
    
    \caption{Effects of $\alpha$ in Direchlet data distribution on ASR(-$t$) and CDA(-$t$)}
    \label{Effects-of-Direchlet}
\end{figure}

\subsection{Robustness of FCBA}

As previously stated, defenses leveraging anomaly detection and Byzantine aggregation are infeasible in FL. Given BaFFle's high overhead, we omit its detailed discussion and focus primarily on participant-level differential privacy techniques. 

In participant-level differential privacy training~\cite{sun2019can}, two crucial stages might curtail the potency and durability of backdoor intrusions. (1) Participant updates are clipped to constrain the sensitivity of model updates, multiplied by $\min (1,\frac{S}{\left \| L_{i}^{t+1}-G^{t}   \right \| }_2 )$, with $S$ as the clipping threshold. While attackers avoid local clipping, they calibrate updates within this limit. (2) Gaussian noise $N(0,\sigma)$ is added to the weighted average of the updates. While low clipping thresholds and high noise variance mitigate backdoor impacts, they compromise the model's primary performance. We study the effects of varying $S$ and $\sigma$ on model efficacy, as presented in Fig. 10 and Fig. 11. (For a detailed comparison with DBA's performance, see Appendix A.9.)
\subsubsection{Effects of $S$.} Fig. 10 shows that a reduced clipping boundary $S$ lowers the ASR-$t$ curve, weakening the backdoor effect. The stable CDA-$t$ curve indicates minimal influence on the main task by $S$ variations.
\begin{figure}[htbp]
    \centering    
    \begin{subfigure}[b]{0.23\textwidth}
        \includegraphics[width=\linewidth]{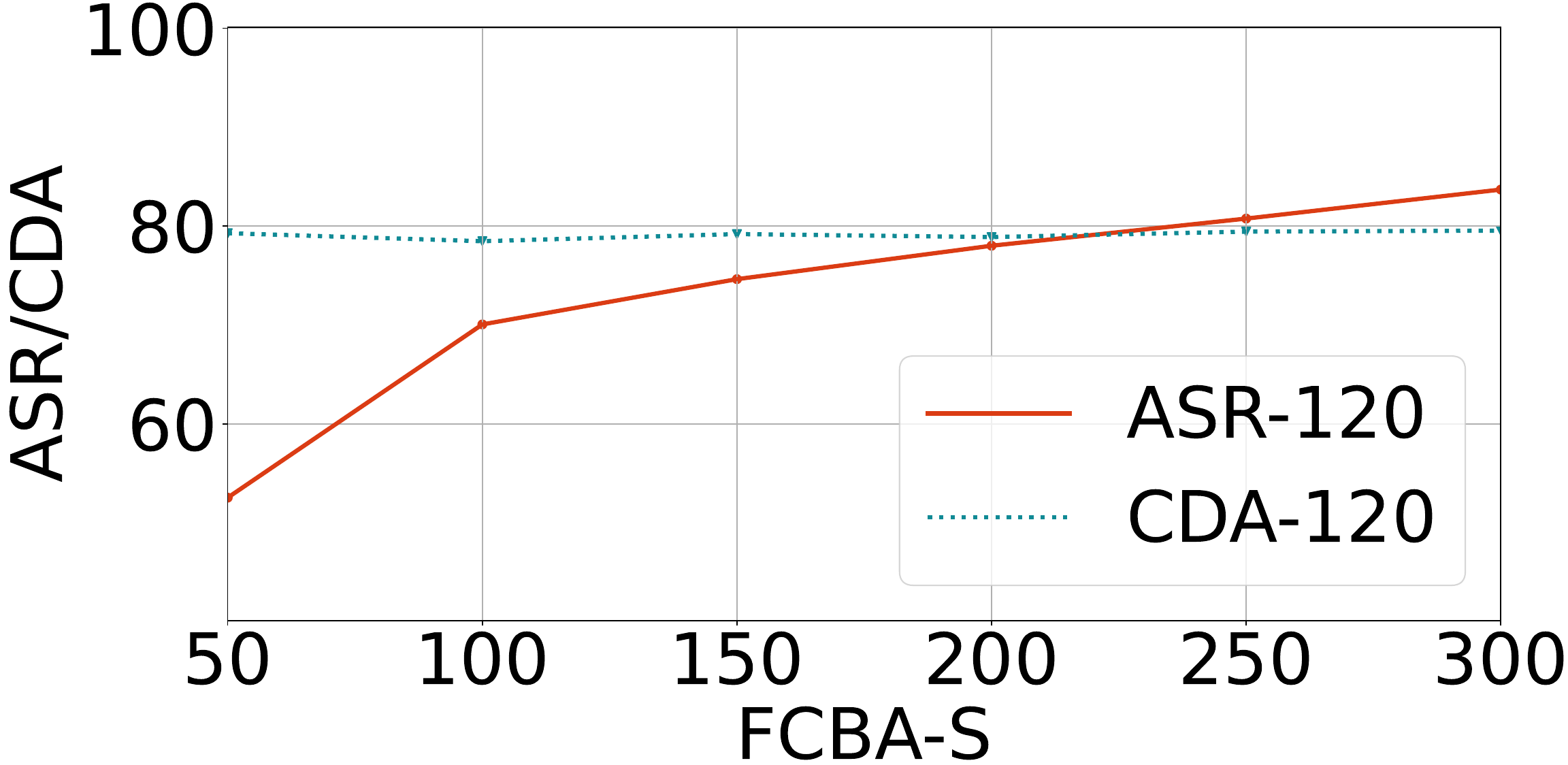}
        \caption{FCBA on CIFAR-10}
    \end{subfigure}
    \begin{subfigure}[b]{0.23\textwidth}
        \includegraphics[width=\linewidth]{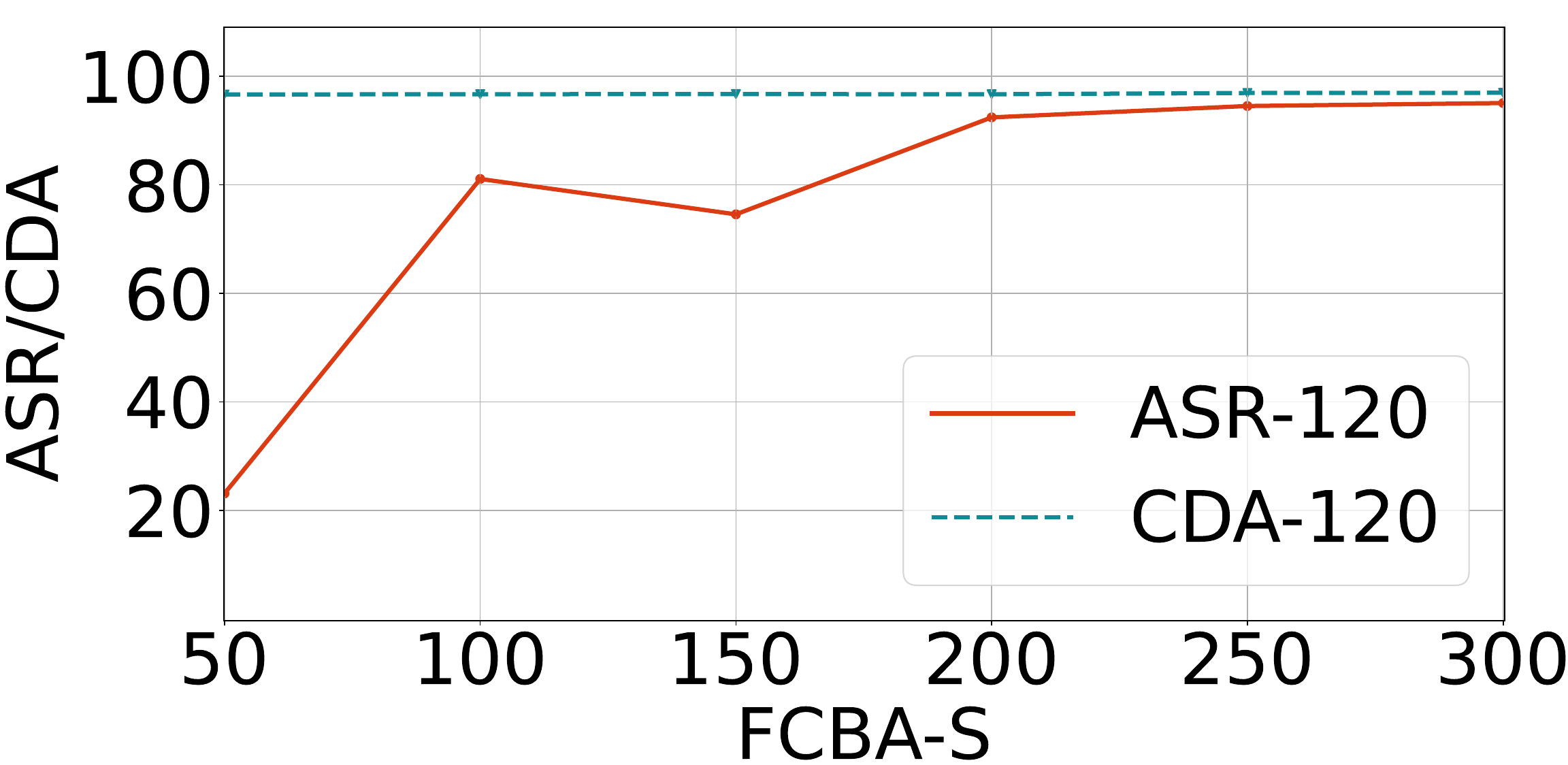}
        \caption{FCBA-on-MNIST}
    \end{subfigure}
    
    \caption{Effects of Clipping Boundary $S$ on ASR-$t$ and CDA-$t$}
    \label{Effects-of-Clipping-Boundary-S}
\end{figure}
\subsubsection{Effects of $\sigma$.} In Fig. 11, as noise variance $\sigma$ rises, both ASR-$t$ and CDA-$t$ curves decline, suggesting increased disturbances negatively affect main and backdoor task performances.

Amplifying noise reduces both primary and backdoor accuracies. This disturbance destabilizes model performance (see Appendix A.10), emphasizing the method's inefficacy against our attack.

Our analysis reveals that participant-level differential privacy in FL is vulnerable to FCBA. Further, existing defense mechanisms prove insufficient against its robustness.

\begin{figure}[htbp]
    \centering    
    \begin{subfigure}[b]{0.23\textwidth}
        \includegraphics[width=\linewidth]{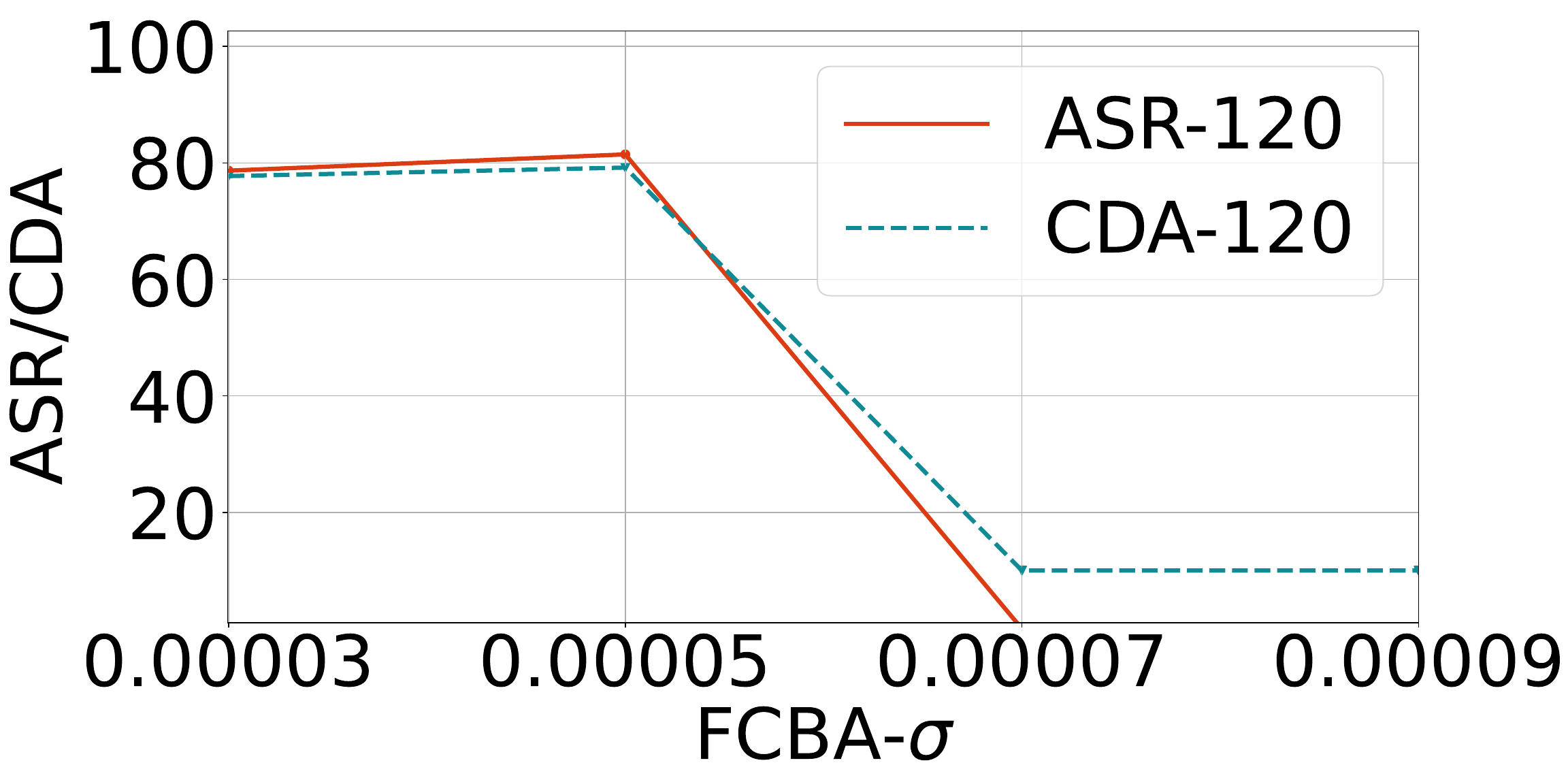}
        \caption{FCBA on CIFAR-10}
    \end{subfigure}
    \begin{subfigure}[b]{0.23\textwidth}
        \includegraphics[width=\linewidth]{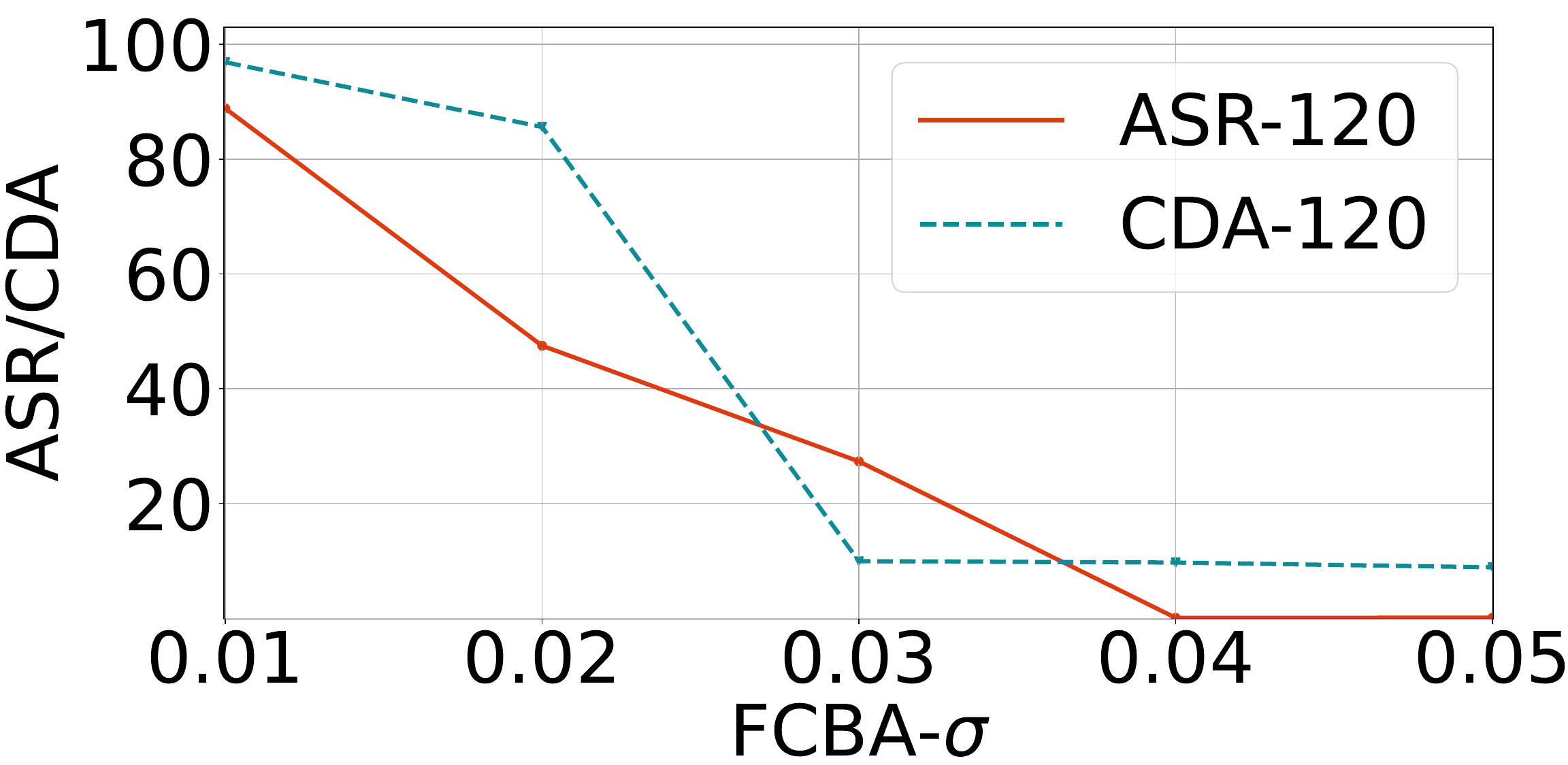}
        \caption{FCBA on MNIST}
    \end{subfigure}
    
    \caption{Effects of Noise Variance $\sigma$ on ASR-$t$ and CDA-$t$}
    \label{Effects-of-Noise-Variance}
\end{figure}

\section {Conclusion}
This paper reports on a new backdoor attack against FL, called FCBA, that has excellent persistence even in a single-shot attack setting. Extensive experiments have shown that FCBA outperforms the SOTA method in terms of attack persistence and stealth across multiple tasks. We use combinatorics theory for the first time to design trigger strategies that reinforce the backdoor effect of the global model, and demonstrate this through data distribution plots. We perform an ablation analysis of the factors influencing FCBA and found that FCBA has a relatively stable persistence in various settings. We show that FCBA is robust and that existing backdoor defense methods struggle to effectively defend against FCBA in practice. Our analysis and findings can provide new threat assessment tools and insights for evaluating the adversarial robustness of FL.

\section{Acknowledgments}
This project was sponsored by National Key R\&D Program of China (grant no.2021YFB3101401), NSFC-Xinjiang Joint Fund Key Program (grant no.U2003206), key research project supported National Natural Science Foundation of China (grant no.61831007), NSFC-Regional Joint Fund Key Program (grant no.U22A2036),  National Natural Science Foundation of China (grant no.62272127), research team project supported by Natural Science Foundation of Heilongjiang (grant no.TD2022F001), National Natural Science Foundation of China (grant no.61971154) and National Natural Science Foundation of China
(grant no.U21B2019).



\thispagestyle{empty}
\mbox{}

\bibliography{aaai24}

\newpage
\newpage
\newpage
\newpage

\thispagestyle{empty}
\mbox{}

\thispagestyle{empty}
\mbox{}

\appendix
\renewcommand{\thefigure}{A.\arabic{figure}}
\renewcommand{\thetable}{A.\arabic{table}}

\setcounter{secnumdepth}{2}
\newpage
\newpage
\section{Appendix}

\subsection{More details on the poison pixel alignment method}
To equitably compare FCBA and DBA's adversarial capabilities, we must standardize the poisoning payload. This is achieved by ensuring congruent pixel modifications in trigger regions, a methodology we term \textit{Poison Pixel Alignment}.

To ensure consistency in our analysis, we standardize the number of pixel alterations across both DBA and FCBA backdoor methods. Two key assumptions guide this: (1) It is presumed that the alteration in the value of each individual pixel is precisely 128, thereby rendering the median pixel value to be equivalent to a white pixel. (2) 
The four pixel sub-blocks in Fig. 2 are uniform in size and shape. With these assumptions, the total pixel modifications for each local trigger can be deduced from the number of its sub-blocks. In our experiments, we detail the frequency of trigger samples in each poisoned batch (i.e., poison rate, $r$) when comparing FCBA and DBA. The results are presented in the subsequent table.
\begin{table}[hbt]
\centering
\begin{tabular}{c|cc}
\toprule
\multirow{2}{*}{Dataset} & Poison Ratio & Poison Ratio\\
&(FCBA)&(DBA) \\
\midrule
MNIST & 3 & 21\\
CIFAR-10 & 2 & 14\\
GTSRB & 4 & 28\\
\bottomrule
\end{tabular}
\caption{Poison Pixel Alignment}
\end{table}

The detailed calculation process is shown in the equation below. $ps_{FCBA}*pb_{FCBA}*pr_{FCBA}=ps_{DBA}*pb_{DBA}*pr_{DBA}$.

Here, $ps$ denotes the content of poisoned samples in each poisoned batch, and $ps$ takes the values of 3, 2, and 4 in the three datasets in FCBA. $pb$ denotes the average of the number of pixel blocks in each poisoned sample, $pb=(1*4+2*6+3*4)/14=2$ in FCBA, and $pb=1$ in DBA. $pr$ denotes the total number of poisoned rounds, which is equal to the total number of malicious clients $M$, and is 14 in FCBA, 4 in DBA. Finally, the values of $ps$ for each dataset in DBA are found to be 21, 14, and 28 based on this equation and the five variables mentioned above.

\subsection{A brief overview of task for each dataset}
\begin{itemize}
    \item \textbf{MNIST:} This task aims to identify digits in handwritten images using MNIST, comprising 10 grayscale classes representing digits 0-9~\cite{lecun1998gradient}.
    \item \textbf{CIFAR-10:} The task was to classify 10 categories of color physics pictures, such as airplanes, horses, cats, and so on~\cite{krizhevsky2009learning}.
    \item \textbf{GTSRB:} This task emulates intelligent recognition for autonomous driving with 43 traffic sign categories~\cite{houben2013detection}. For optimal model performance, we focus on the 16 most sampled categories: 6 speed limits, 3 prohibitions, 2 dangers, and 5 mandates.
\end{itemize}

\subsection{Hardware details for the experiment}
Using the PyTorch framework, we execute the federated learning experiments on a server equipped with dual Intel(R) Xeon(R) Silver 4214R CPUs and two NVIDIA Corporation Device 2204 GPUs, each with 12 GB RAM, running CentOS Linux 7.6.1810(Core) OS.
\subsection{Better to attack late}
In our primary research, FCBA allows backdoor injection both before and after the global model's convergence. While our main focus is on post-convergence injection, we also explore the implications of early backdoor introduction.
\begin{figure}[H] 
\centering 
\includegraphics[width=0.45\textwidth]{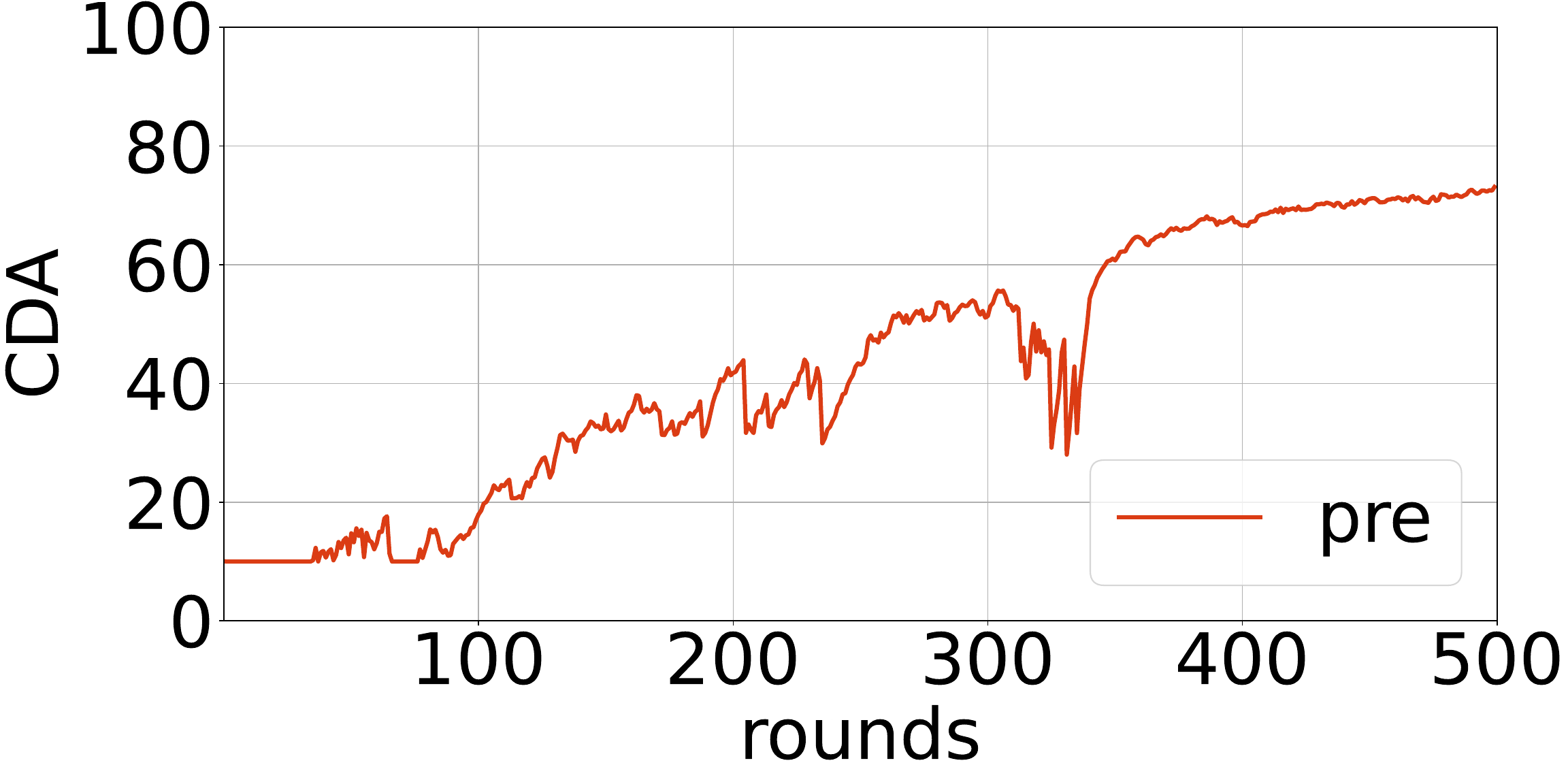} 
\caption{CDA on CIFAR-10, pre-inject FCBA.} 
\label{FCBA-describe} 
\end{figure}
Empirically, on datasets like CIFAR-10, early backdoor introduction results in suboptimal and slow-growing accuracy (refer to Fig. A.1). Additionally, premature injection adversely impacts ASR and ASR-$t$ values, particularly on CIFAR-10 and GTSRB datasets (see Fig. A.2 and Fig. A.3). This suggests early injections might compromise FCBA attack's effectiveness and longevity. Thus, we opt for post-convergence attacks: round 2 in MNIST, round 314 in CIFAR-10, and round 154 in GTSRB. As mentioned by \citet{CBA}, a delayed backdoor injection enhances attack performance. As the global model converges, benign client updates, with their unique traits, are more likely neutralized, minimizing backdoor impact.


\begin{figure}[H]
\centering
\begin{minipage}{0.23\textwidth}
\centering
\includegraphics[width=\textwidth]{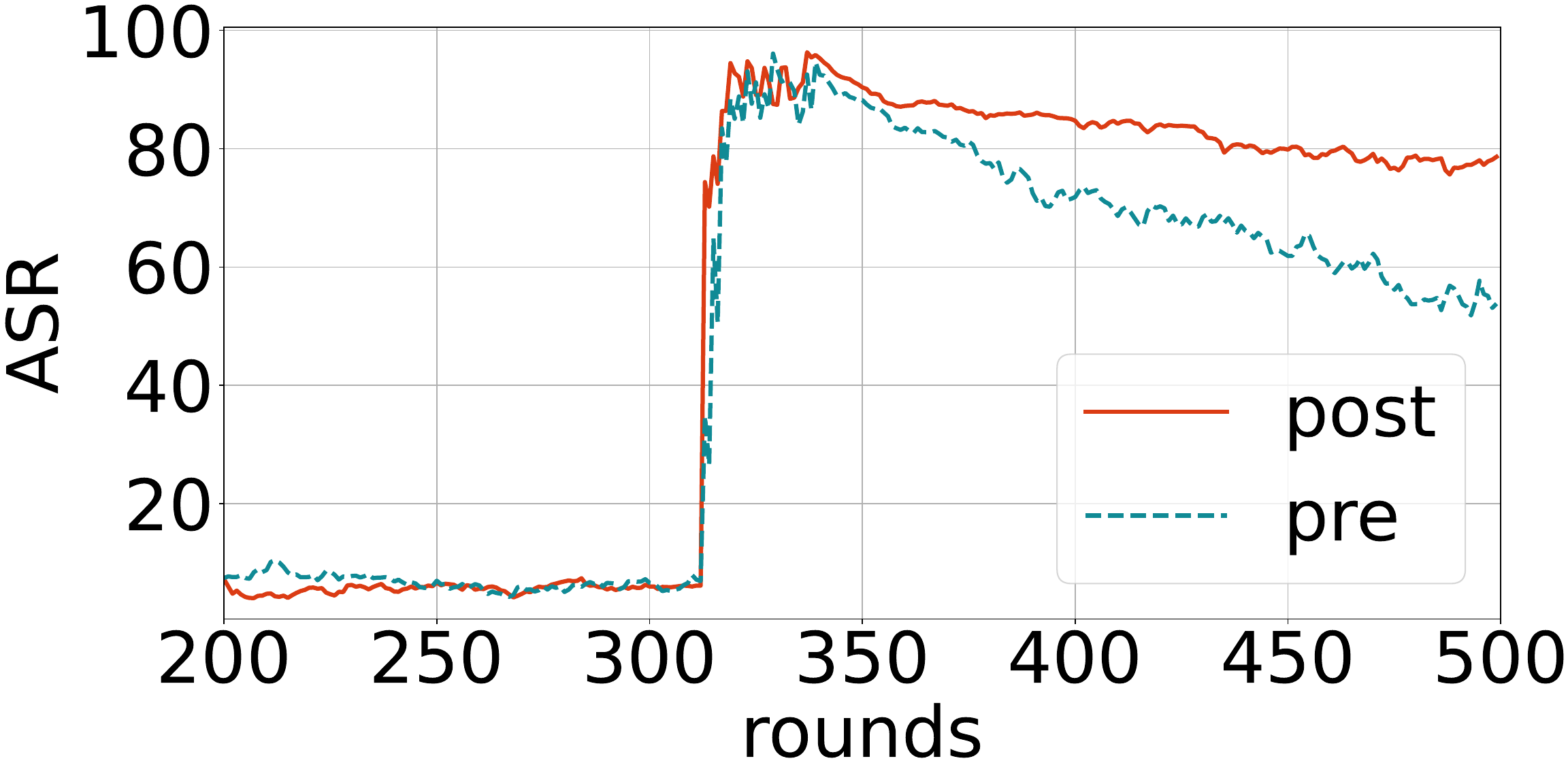}
\caption{ASR on CIFAR-10, pre-injection and post-injection FCBA.}
\label{FCBA-describe1}
\end{minipage}
\hfill
\begin{minipage}{0.23\textwidth}
\centering
\includegraphics[width=\textwidth]{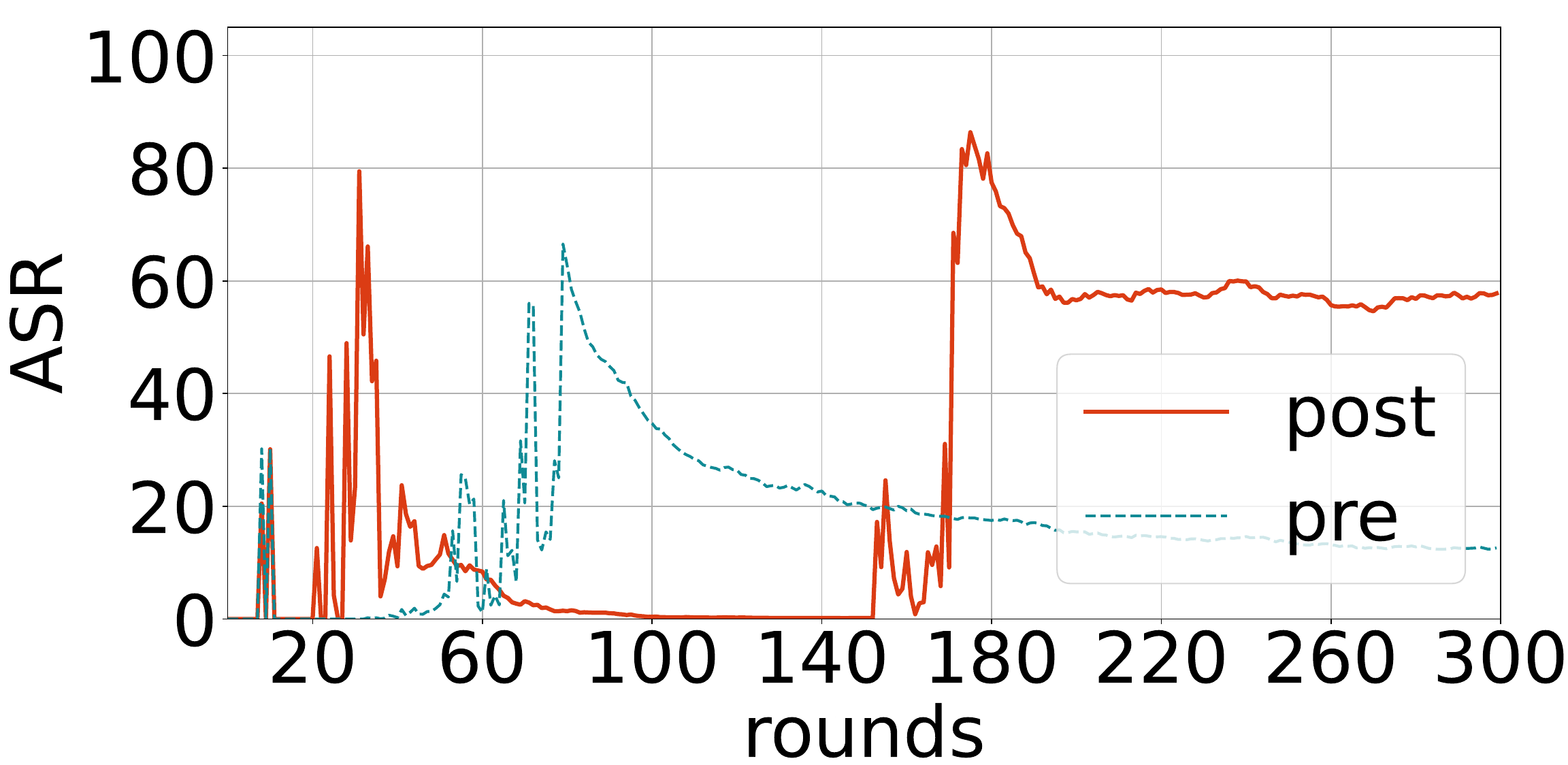}
\caption{ASR on GTSRB, pre-injection and post-injection FCBA.}
\label{FCBA-describe2}
\end{minipage}
\end{figure}

\subsection{Further exploration into the reasons for FCBA's high persistence}
Referencing \citet{DBA}, our FCBA employs data poisoning with local triggers. However, Fig. A.4 (CIFAR-10) shows that its global triggers are more enduring. This indicates FCBA's enhanced resilience to benign updates, possibly contributing to its high persistence.
\begin{figure}[H]
\centering
\begin{minipage}{0.23\textwidth}
\centering
\includegraphics[width=\textwidth]{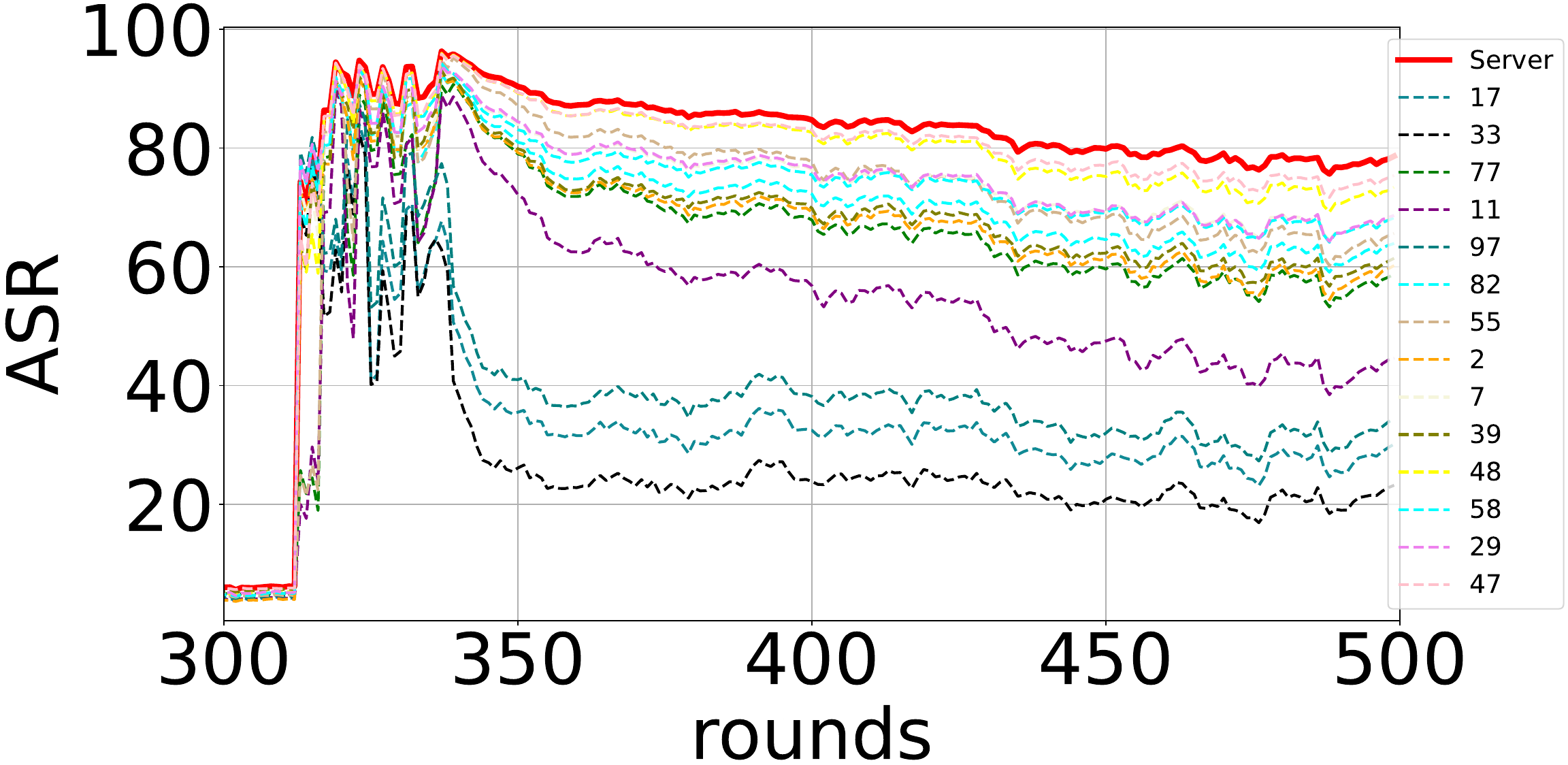}
\caption{ASR for all triggers in FCBA.}
\label{FCBA-describe1}
\end{minipage}
\hfill
\begin{minipage}{0.23\textwidth}
\centering
\includegraphics[width=\textwidth]{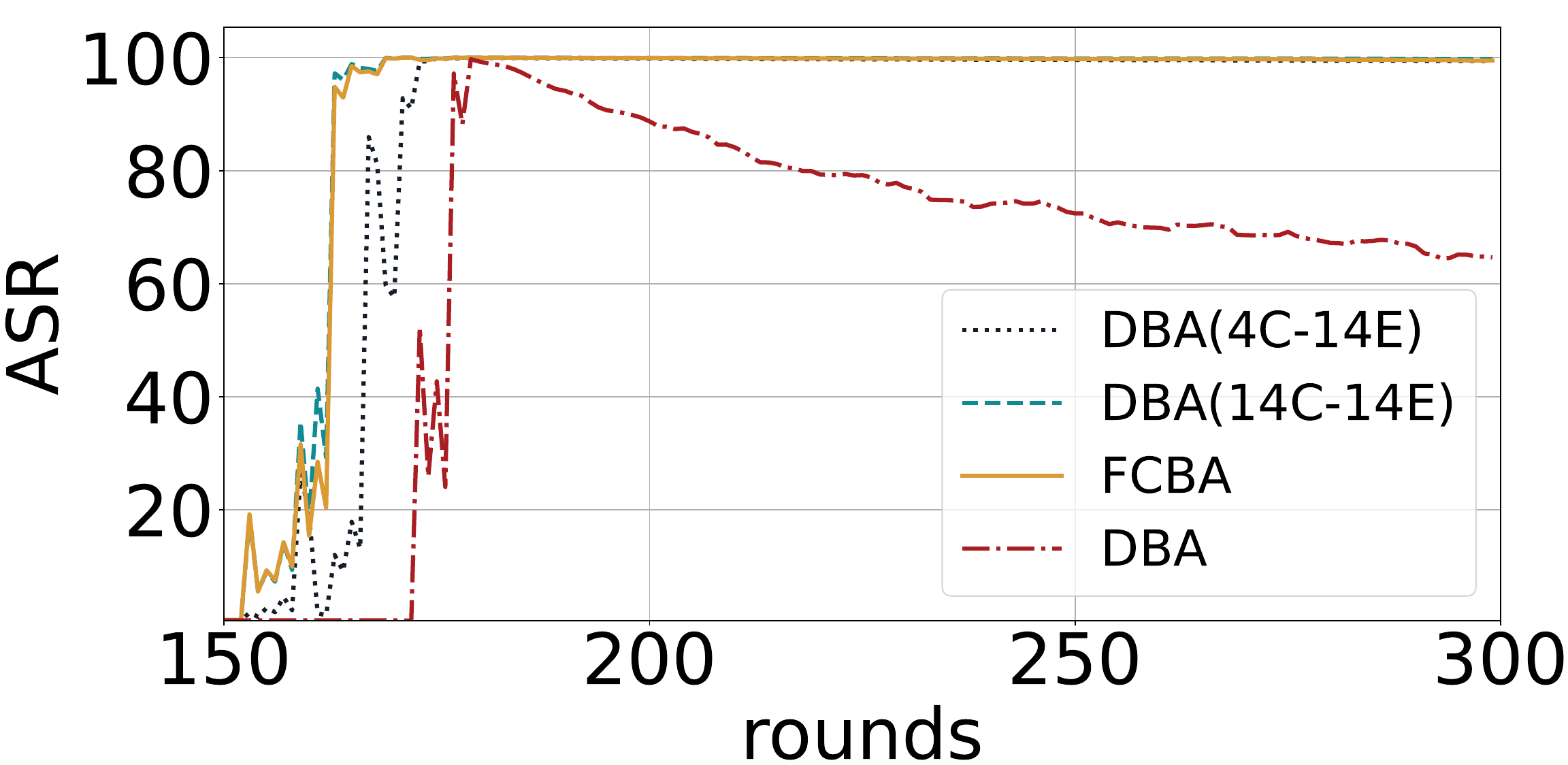}
\caption{ASR for 4 types of attacks on MNIST}
\label{FCBA-describe2}
\end{minipage}
\end{figure}
The added rounds from the full combination strategy might boost persistence. We examined this using control groups $DBA_{4C}^{14E}$ and $DBA_{14C}^{14E}$, ensuring equal poisoned pixels across three image tasks and aligned parameters. Attack details are:

\begin{itemize}
    \item $DBA$: A single sub-block local trigger uses a single-shot attack with 4 malicious clients over 4 poison injection epochs.
    \item $DBA_{4C}^{14E}$: $DBA_{4C}^{14E}$ modifies the $DBA$ approach, using 4 clients over 14 poison epochs. It disperses the same poison pixels count across more epochs, termed as \textit{Temporally Dispersed Poisoning} (TDP).
    \item $DBA_{14C}^{14E}$: $DBA_{14C}^{14E}$ adapts the $DBA$ model, using 14 clients for 14 poison epochs. Unlike $DBA_{4C}^{14E}$, it distributes poison pixels across more clients, termed \textit{Spatially Dispersed Poisoning} (SDP).
    \item $FCBA$: Uses full combination local triggers in a single-shot attack with 14 malicious clients over 14 poison epochs.
\end{itemize}

Fig. A.5-A.7 displays attack performance results. Aside from the blue curve ($DBA_{14C}^{14E}$), curve positions remain consistent across classifications. Full combination triggers and TDP enhance persistence, but SDP's effects are uncertain.

\begin{figure}[H]
\centering
\begin{minipage}{0.23\textwidth}
\centering
\includegraphics[width=\textwidth]{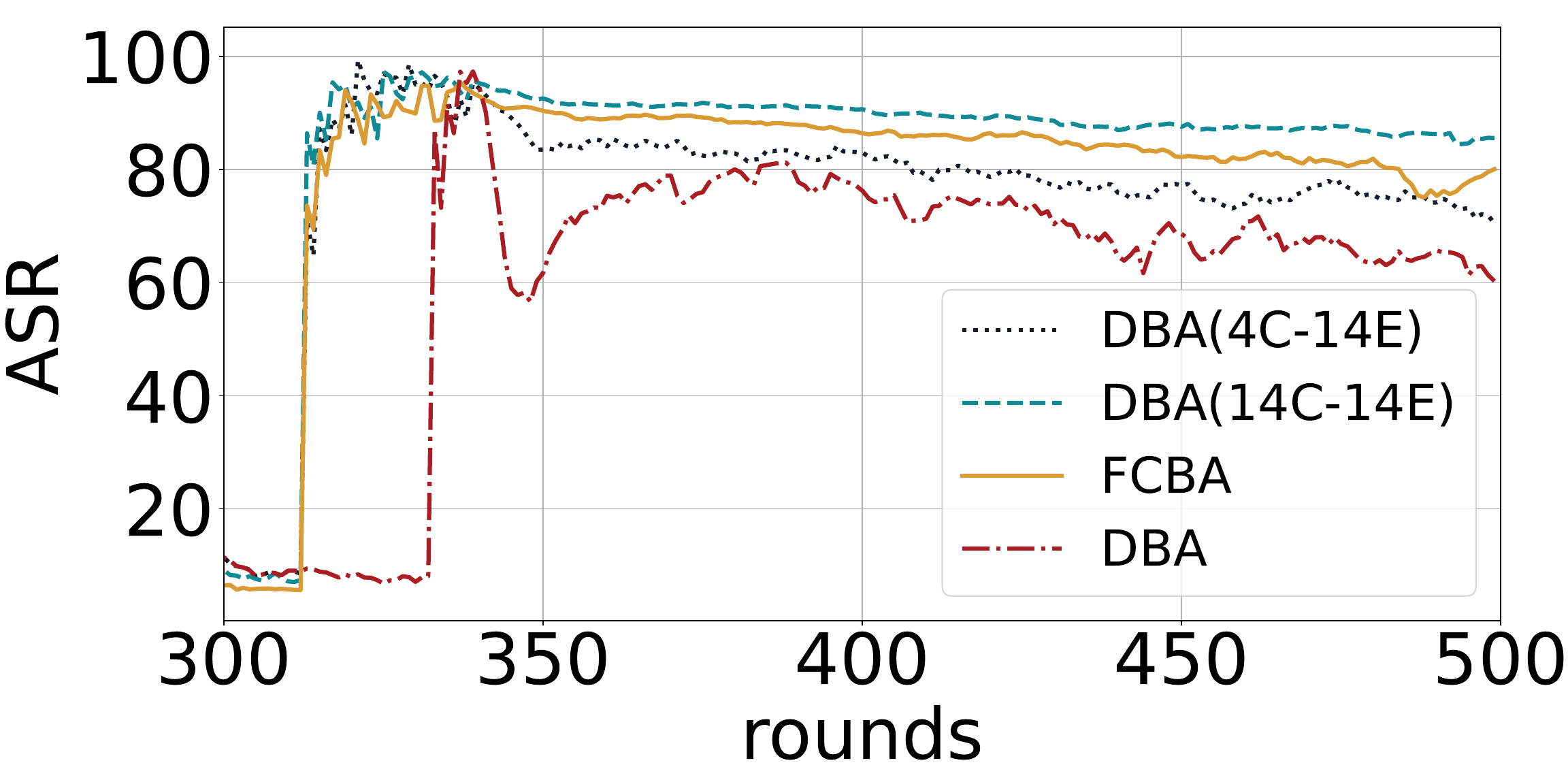}
\caption{ASR for 4 types of attacks on CIFAR-10}
\label{FCBA-describe1}
\end{minipage}
\hfill
\begin{minipage}{0.23\textwidth}
\centering
\includegraphics[width=\textwidth]{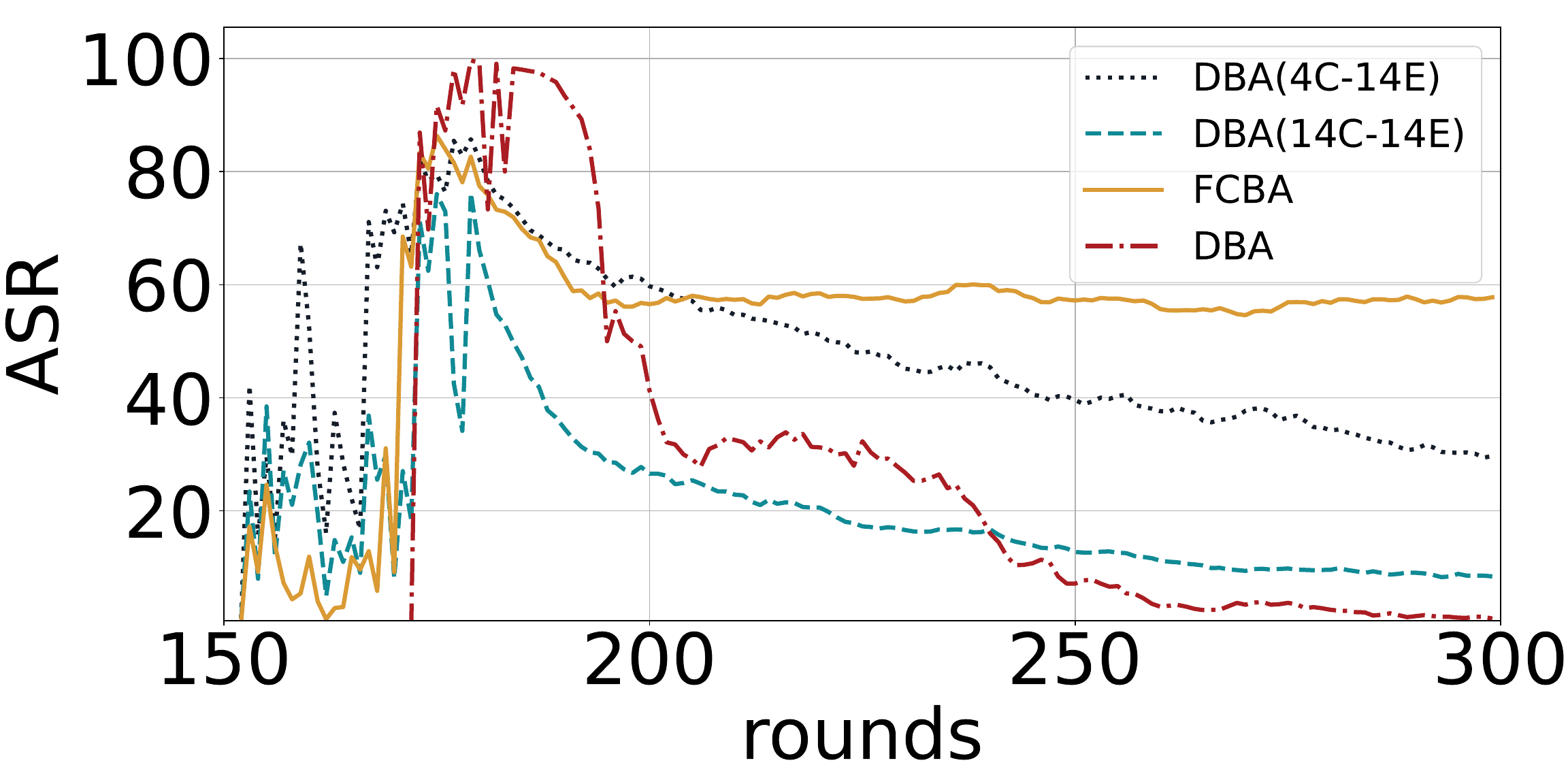}
\caption{ASR for 4 types of attacks on GTSRB}
\label{FCBA-describe2}
\end{minipage}
\end{figure}


The full combination trigger strategy aims to enhance global model's backdoor performance by maximizing trigger information. Using combinatorics, all local trigger patterns are considered, enabling malicious participants to understand sub-pixel block relationships. Using combinatorics, it encompasses all local triggers. Malicious updates sent to the central server enrich the global model with a detailed backdoor pattern, potentially enhancing attack persistence.

TDP disperses poison pixels across more epochs. While findings in \textbf{A.8} indicate minimal impact on backdoor performance by reducing poison ratio in a batch, TDP increases the model's exposure to malicious updates in federated aggregations, enhancing backdoor learning. Subsequent training with only clean samples induces \textbf{backdoor forgetting}, similar to multi-task learning's catastrophic forgetting. TDP preserves more backdoor information, making the model resilient to this forgetting, likely explaining the improved attack persistence.

The conclusions offer insightful approaches for designing potent backdoor attacks. In federated learning, to enhance backdoor attacks: (1) Disperse toxic pixels over extended epochs. (2) Use information-rich local triggers. These strategies enrich the global model's backdoor knowledge and boost attack persistence.


\subsection{More details on the re-aggregation}
As shown in Fig. A.8 and Fig. A.9, there is a clear U-shaped portion in the middle of the CDA and ASR curves with the increase of the scale factor $\gamma$. It reflects a model's temporary failure due to an outlier malicious update from excessive scaling. This disrupts the global model's steady state, leading to a \textbf{re-convergence}. During this phase, both the backdoor and main tasks fail, with the main task having a 10\% success prediction chance. As rounds progress, the model stabilizes, performing well on both tasks. The larger the scale, the longer re-convergence takes, suggesting a sufficiently large scale might permanently derail the model. Excessive scaling can detrimentally affect both main and backdoor task performances.

\begin{figure}[H]
\centering
\begin{minipage}{0.23\textwidth}
\centering
\includegraphics[width=\textwidth]{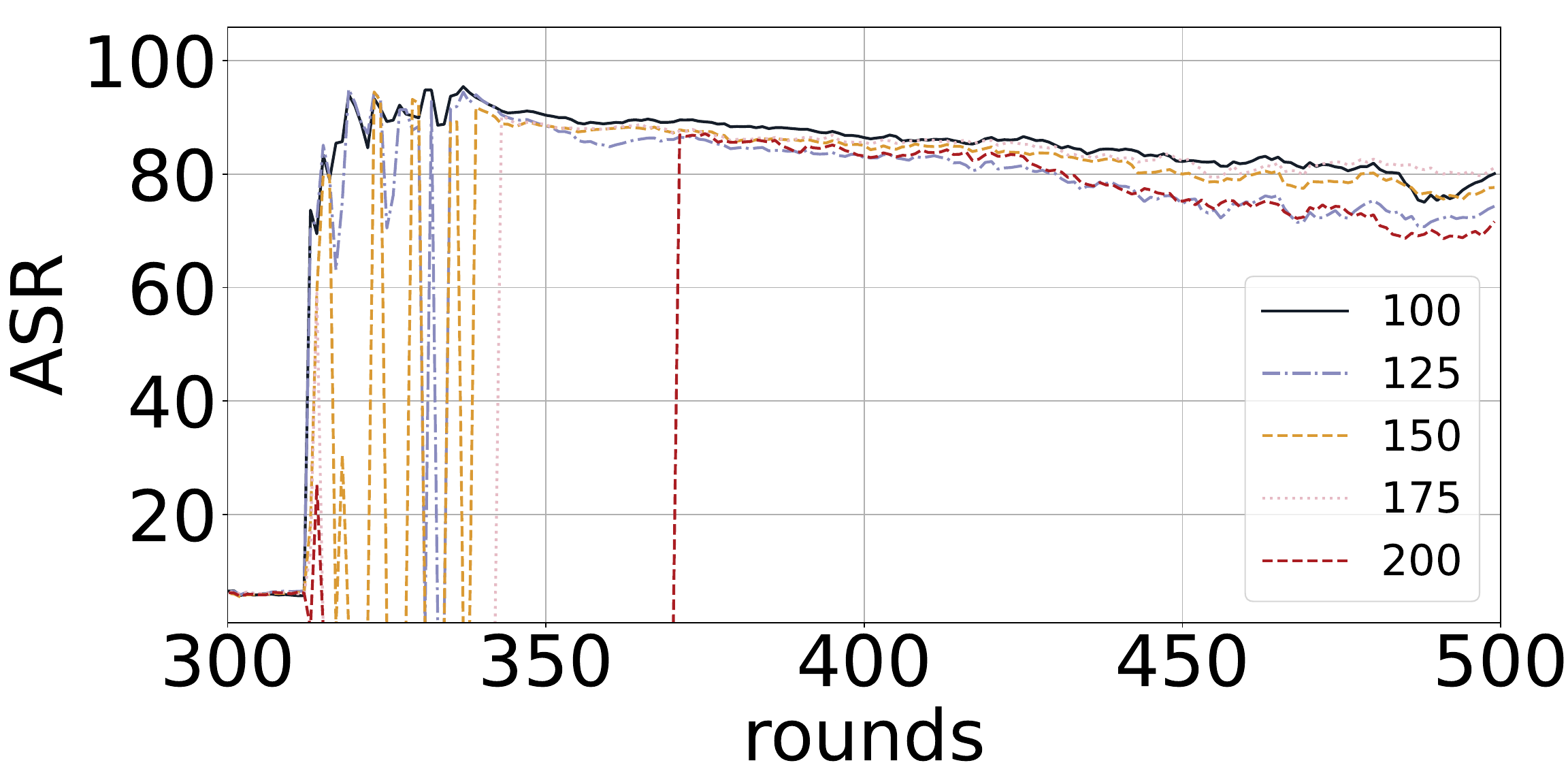}
\caption{FCBA-ASR on CIFAR-10 at different scales.}
\label{FCBA-describe1}
\end{minipage}
\hfill
\begin{minipage}{0.23\textwidth}
\centering
\includegraphics[width=\textwidth]{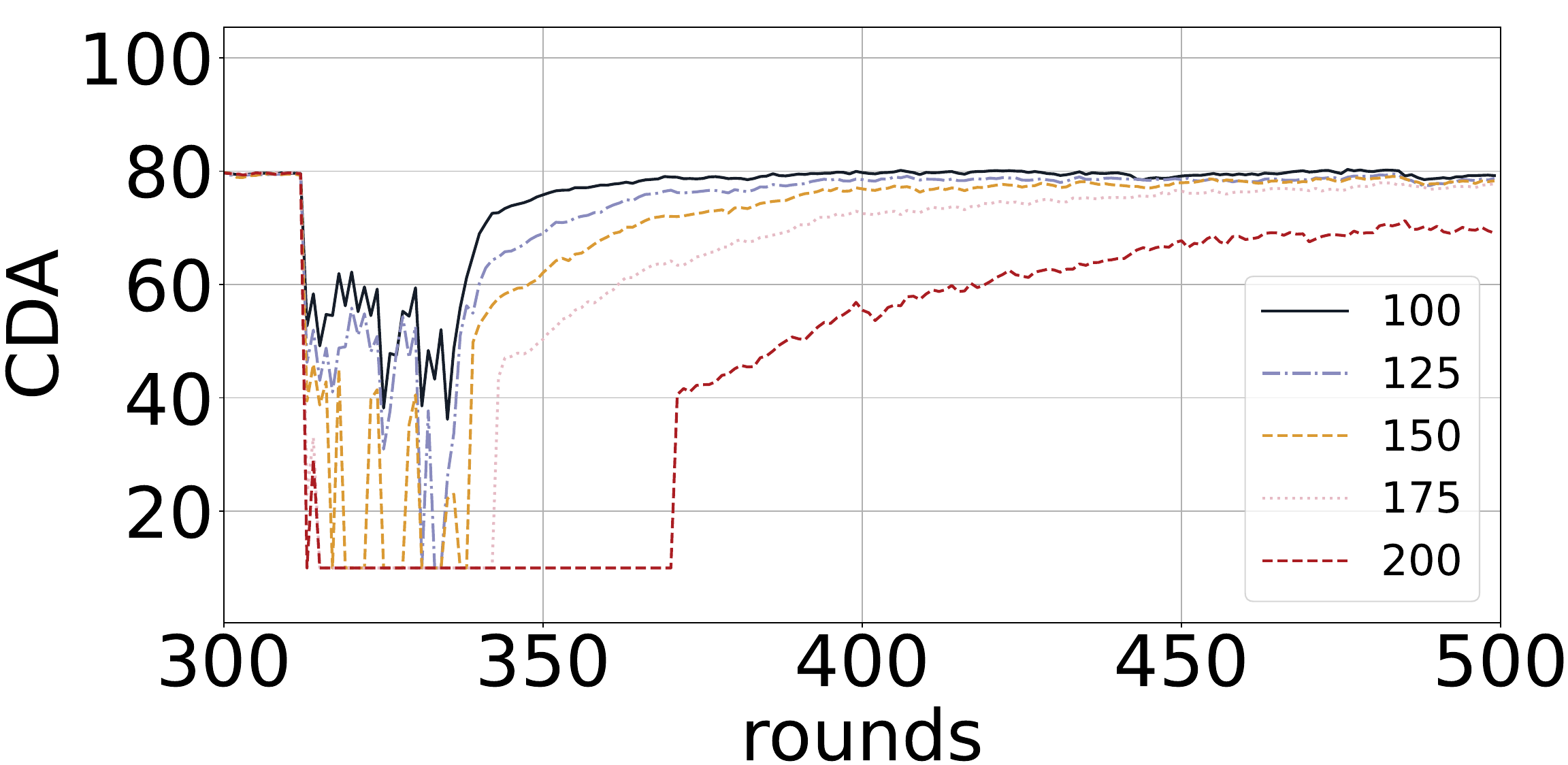}
\caption{FCBA-CDA on CIFAR-10 at different scales.}
\label{FCBA-describe2}
\end{minipage}
\end{figure}


\subsection{The impact of $\alpha$ on data distribution}
In the Dirichlet distribution, a smaller $\alpha$ signifies greater data imbalance. We examine non-i.i.d. data distribution for $\alpha=0.1$ and 0.5, focusing on data size and class imbalances. In federated learning, unequal client data sizes can impact the global model's effectiveness. Clients with limited data hinder local model performance and degrade the aggregated global model. Conversely, data-rich clients can disproportionately skew the global model, adversely impacting federated learning. Figures A.10 and A.11 depict client data distributions, revealing more pronounced imbalances at $\alpha=0.1$ than at $\alpha = 0.5$. Specifically, at $\alpha=0.5$, the clients with different total amount of data basically satisfy the normal distribution. At $\alpha=0.1$, 17 clients possess under 100 data pieces, while 13 clients have over 1,000. This imbalance correlates with decreased main precision and backdoor performance.

Class imbalance refers to notable sample disparities between classes within a client's local data, influencing model training and evaluation. Here, we randomly selecte five clients and count the class distribution of their local data, as shown in Fig. A.12 and Fig. A.13 (CIFAR-10). Comparing the two figures we find that the class imbalance is more serious as $\alpha$ decreases. At $\alpha= 0.5$, sample distribution is fairly even; however, at $\alpha=0.1$, there's pronounced disparity. Notably, in Fig. A.13, clients \#5 and \#25 have data labeled ``1'' exceeding the sum of other labels. Such imbalances impede both main and backdoor task performance.

\begin{figure}[H]
\centering
\begin{minipage}{0.23\textwidth}
\centering
\includegraphics[width=\textwidth]{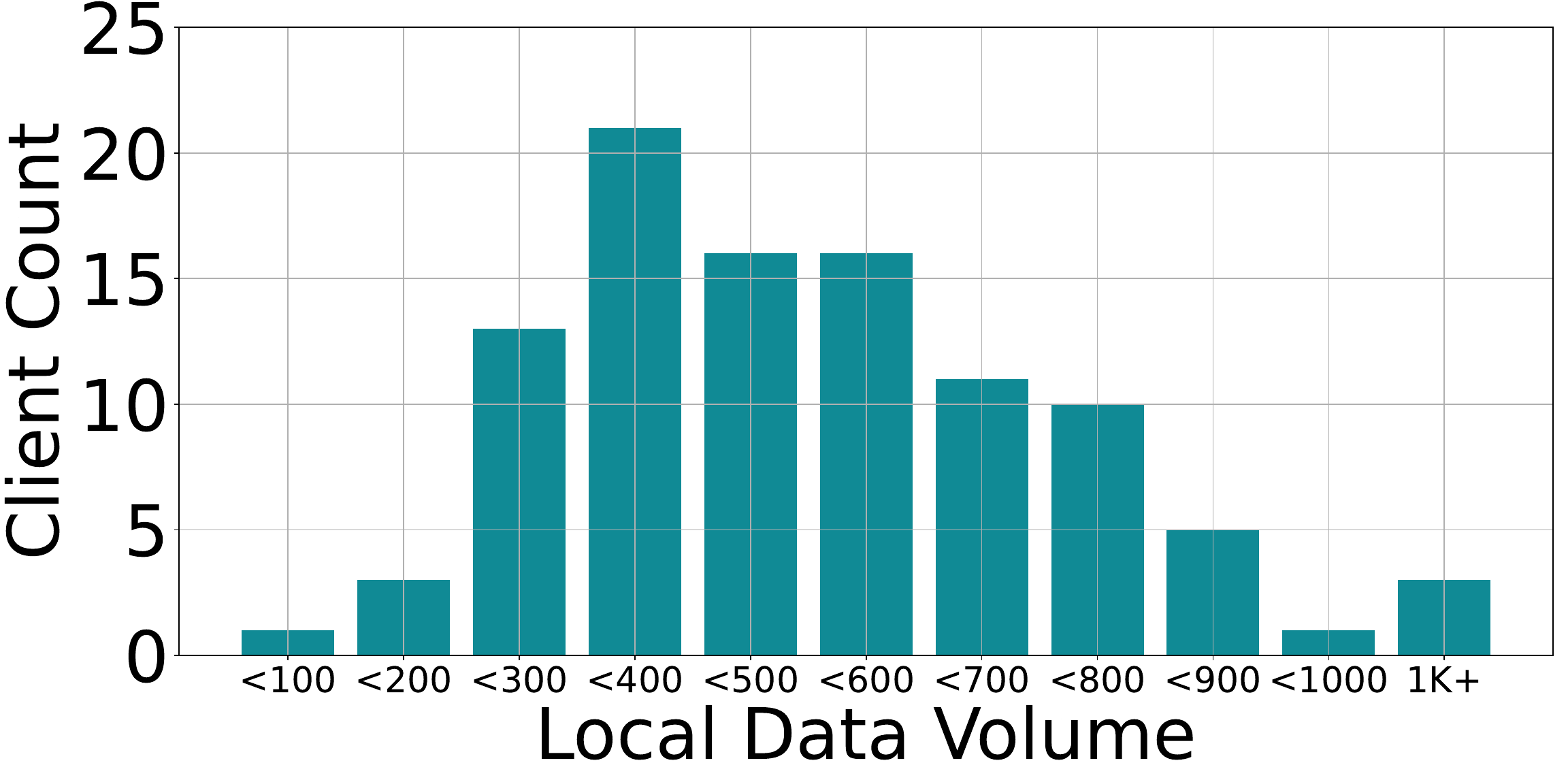}
\caption{Client count by local data volume, $\alpha=0.5$.}
\label{FCBA-describe1}
\end{minipage}
\hfill
\begin{minipage}{0.23\textwidth}
\centering
\includegraphics[width=\textwidth]{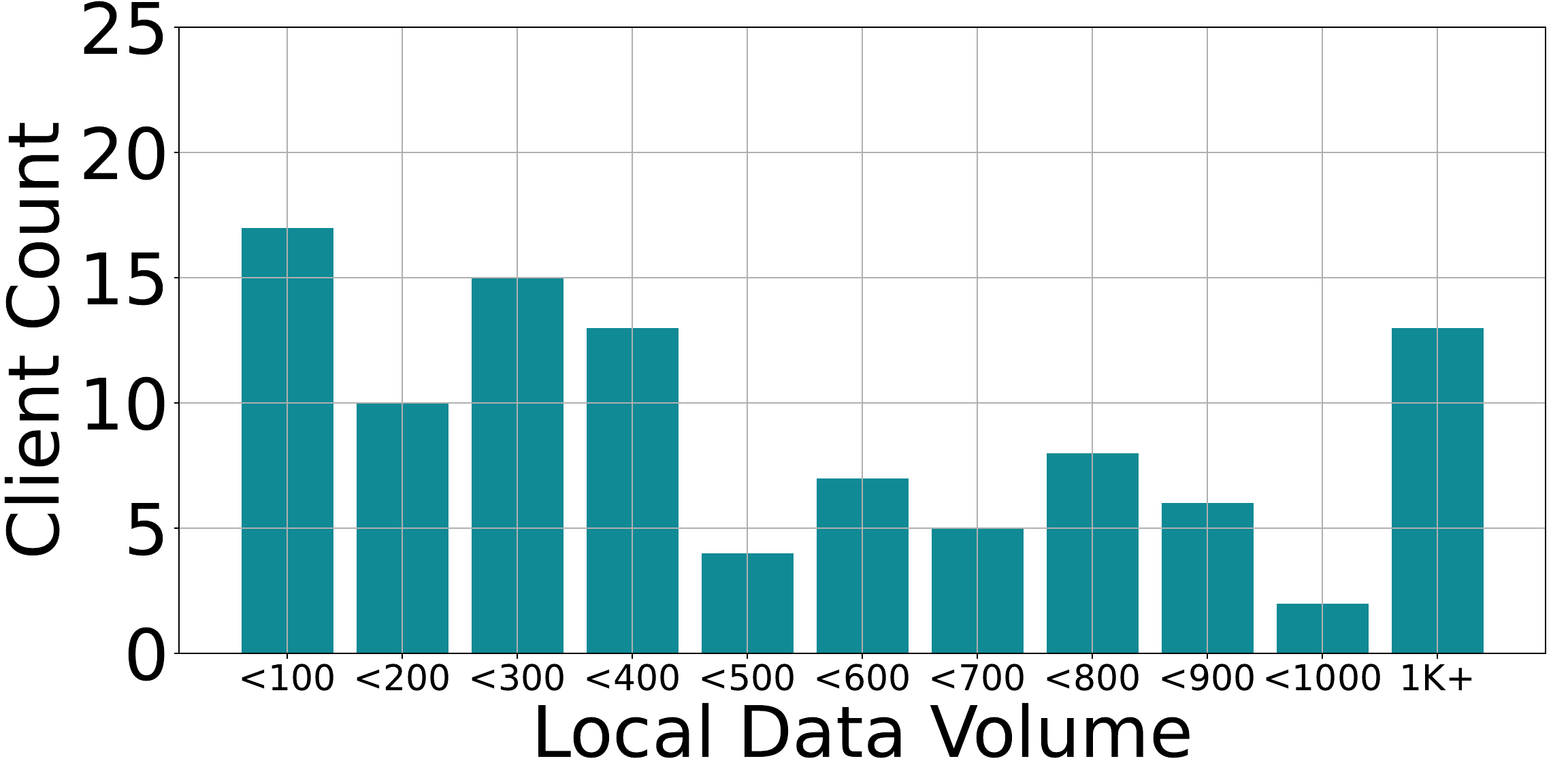}
\caption{Client count by local data volume, $\alpha=0.1$.}
\label{FCBA-describe2}
\end{minipage}
\end{figure}



\begin{figure}[H]
\centering
\begin{minipage}{0.23\textwidth}
\centering
\includegraphics[width=\textwidth]{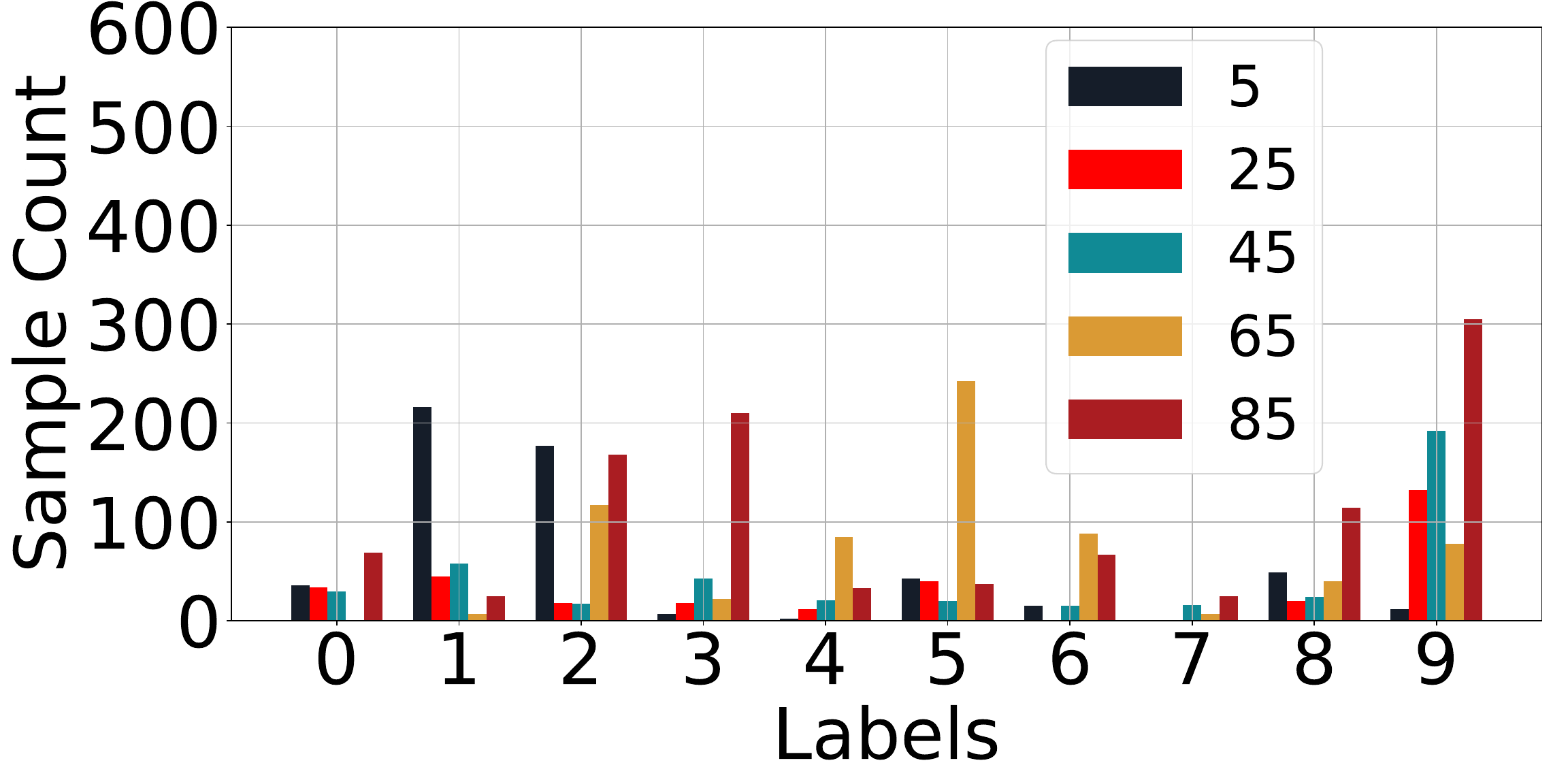}
\caption{Sample count by labels, $\alpha=0.5$.}
\label{FCBA-describe1}
\end{minipage}
\hfill
\begin{minipage}{0.23\textwidth}
\centering
\includegraphics[width=\textwidth]{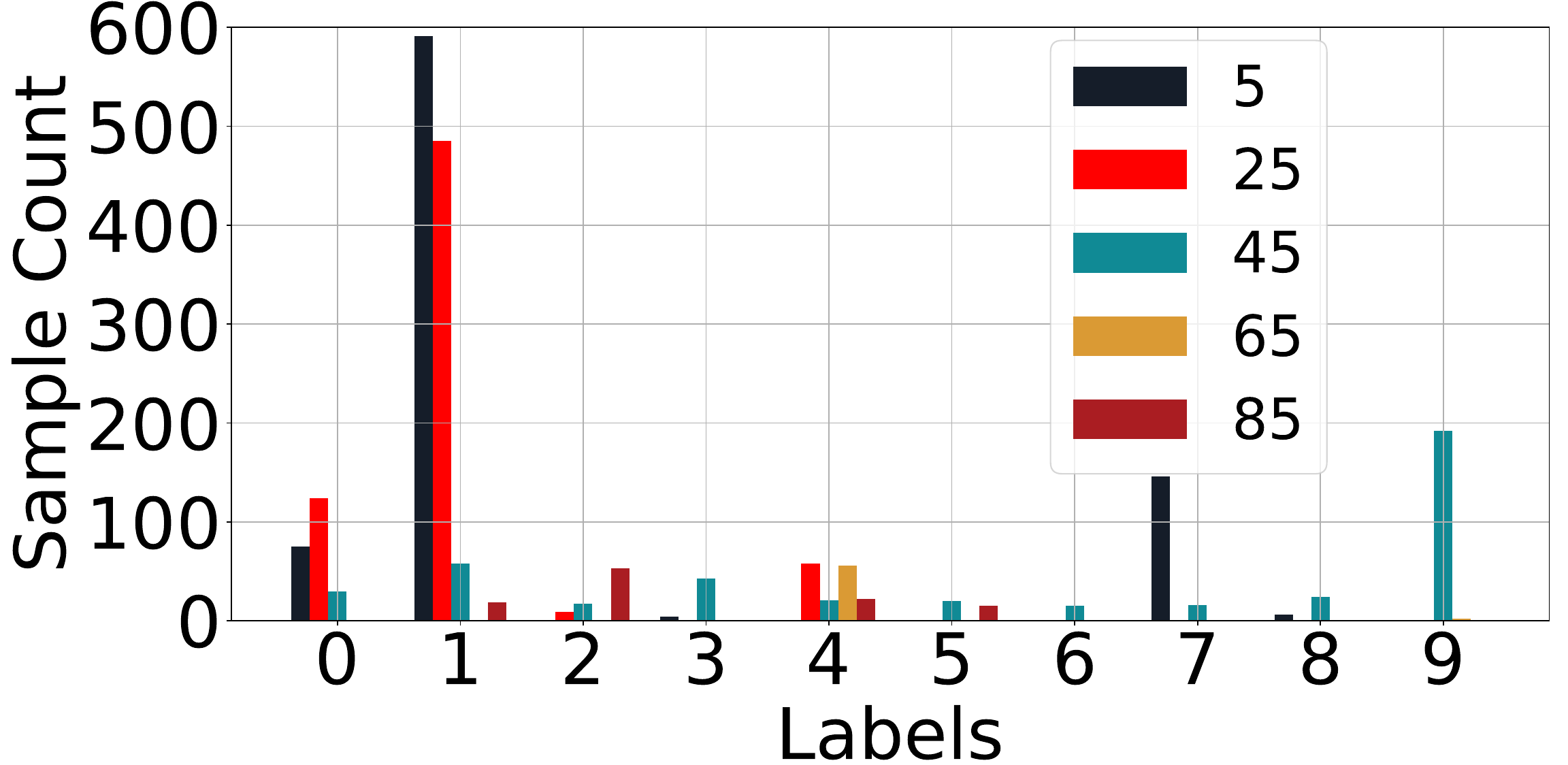}
\caption{Sample count by labels, $\alpha=0.1$.}
\label{FCBA-describe2}
\end{minipage}
\end{figure}



\subsection{Analysis of other factors in FCBA}

\subsubsection{Effects of $TS$.} $TS$ denotes the pixel columns in a sub-block, with rows fixed at 1 due to the image's size.

Fig. A.14 indicates that an initial rise in $TS$ boosts ASR and ASR-$t$, but then stabilizes. Small $TS$ values make triggers harder for the model to recognize, resulting in weaker backdoor performance. Thus, $TS$ minimally impacts attack effectiveness and persistence.

\begin{figure}[htbp]
    \centering    
    
    \begin{subfigure}[b]{0.23\textwidth}
        \includegraphics[width=\linewidth]{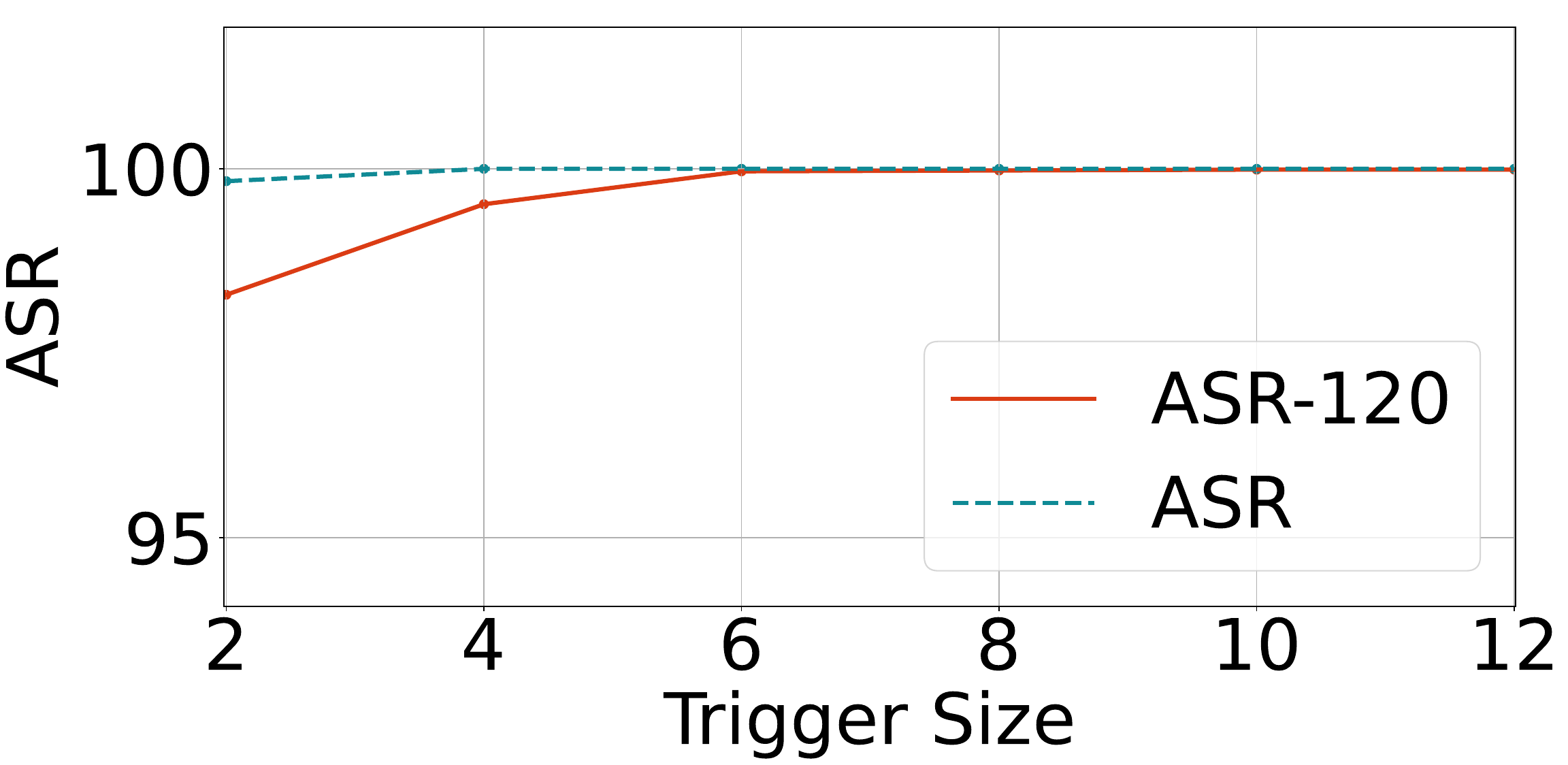}
        \caption{MNIST}
    \end{subfigure}
    \begin{subfigure}[b]{0.23\textwidth}
        \includegraphics[width=\linewidth]{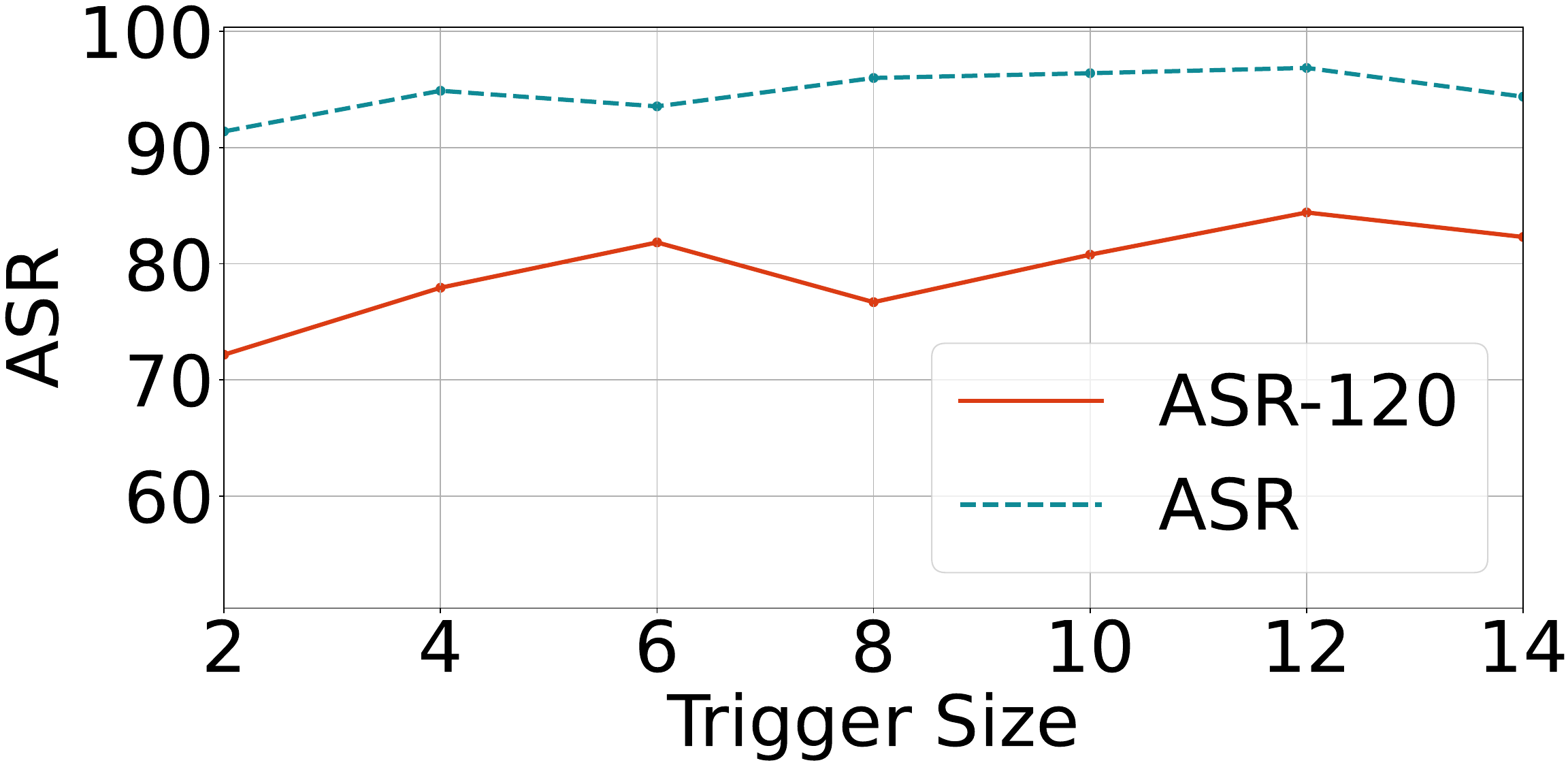}
        \caption{CIFAR-10}
    \end{subfigure}
    \caption{Effects of Trigger Size on ASR and ASR-$t$.}
    \label{Effects-of-Trigger-Size}
\end{figure}

\subsubsection{Effects of $TL$.}  $Shift_x$ and $Shift_y$ represent the trigger's horizontal and vertical offsets, with $TL$ being the larger of the two.

Fig. A.15 shows that $TL$ changes have minimal impact on ASR and ASR-$t$, except in a V-shaped segment. FCBA is less effective in MNIST's central region where primary objects reside. In essence, $TL$ affects attack performance only if the trigger significantly overlaps the main object.

\begin{figure}[htbp]
    \centering    
    \begin{subfigure}[b]{0.23\textwidth}
        \includegraphics[width=\linewidth]{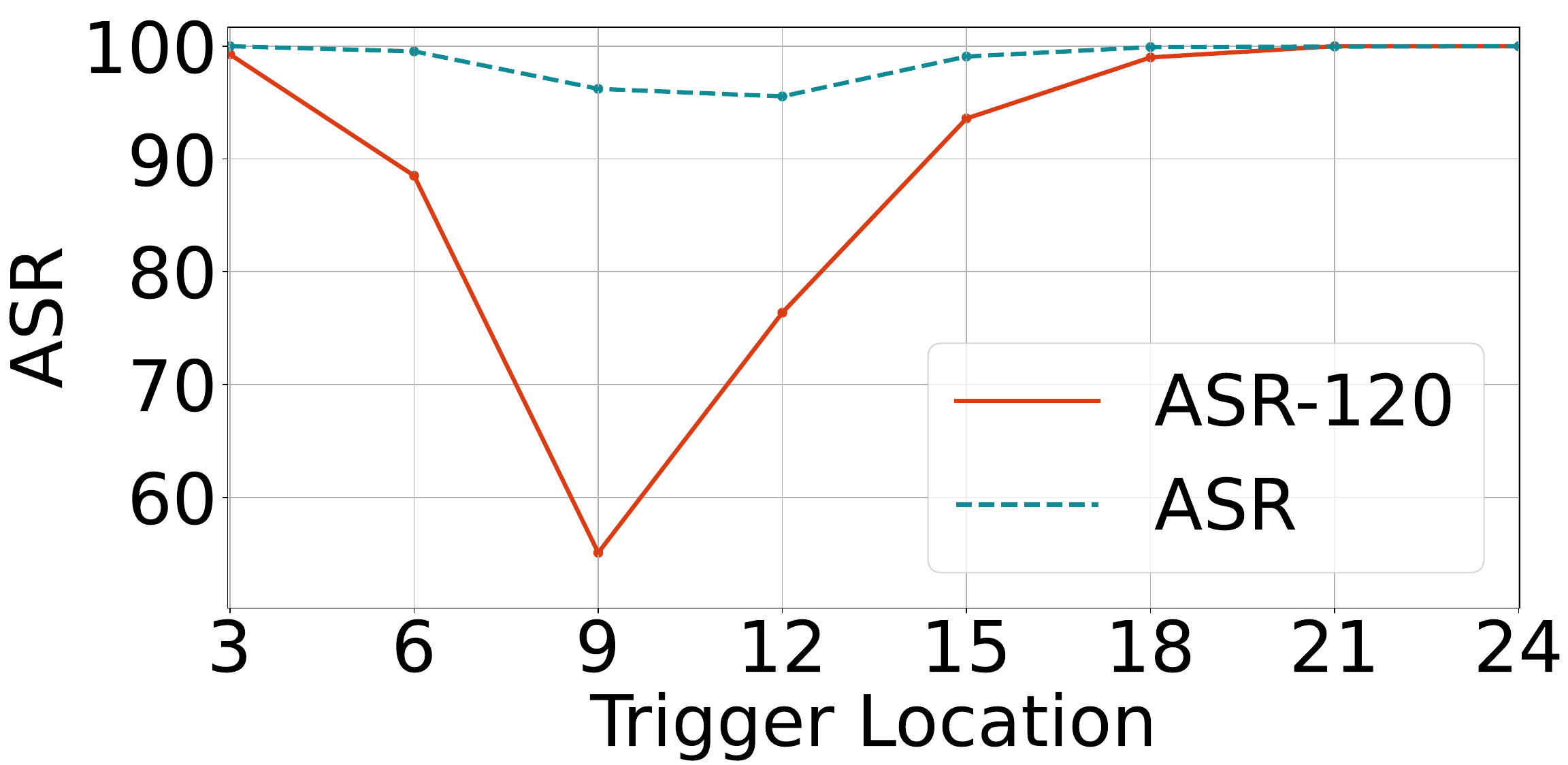}
        \caption{MNIST}
    \end{subfigure}
    \begin{subfigure}[b]{0.23\textwidth}
        \includegraphics[width=\linewidth]{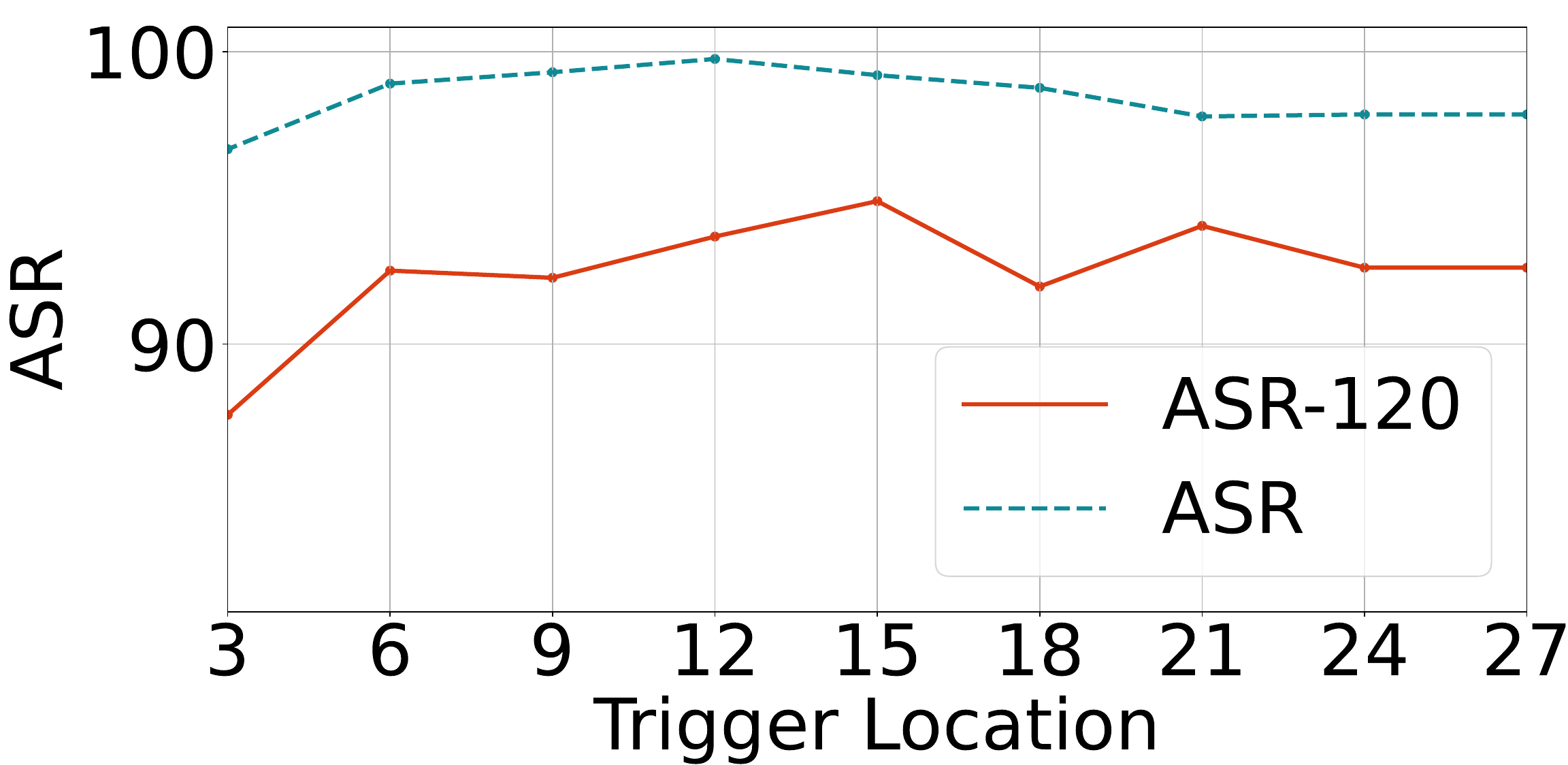}
        \caption{CIFAR-10}
    \end{subfigure}
    
    \caption{Effects of Trigger Location on ASR and ASR-$t$.}
    \label{Effects-of-Trigger-Location}
\end{figure}

\subsubsection{Effects of $TG$.} $TG$ is the spacing between sub-pixel blocks, with $Gap_x$ equating to $Gap_y$ for simplicity.

Findings from $TL$ also pertain to $TG$, though the V-shape in Fig. A.16 is subtler. In $TL$, four pixel blocks overlap the main object; in $TG$, only the lower right block does.

\begin{figure}[htbp!]
    \centering    
    \begin{subfigure}[b]{0.23\textwidth}
        \includegraphics[width=\linewidth]{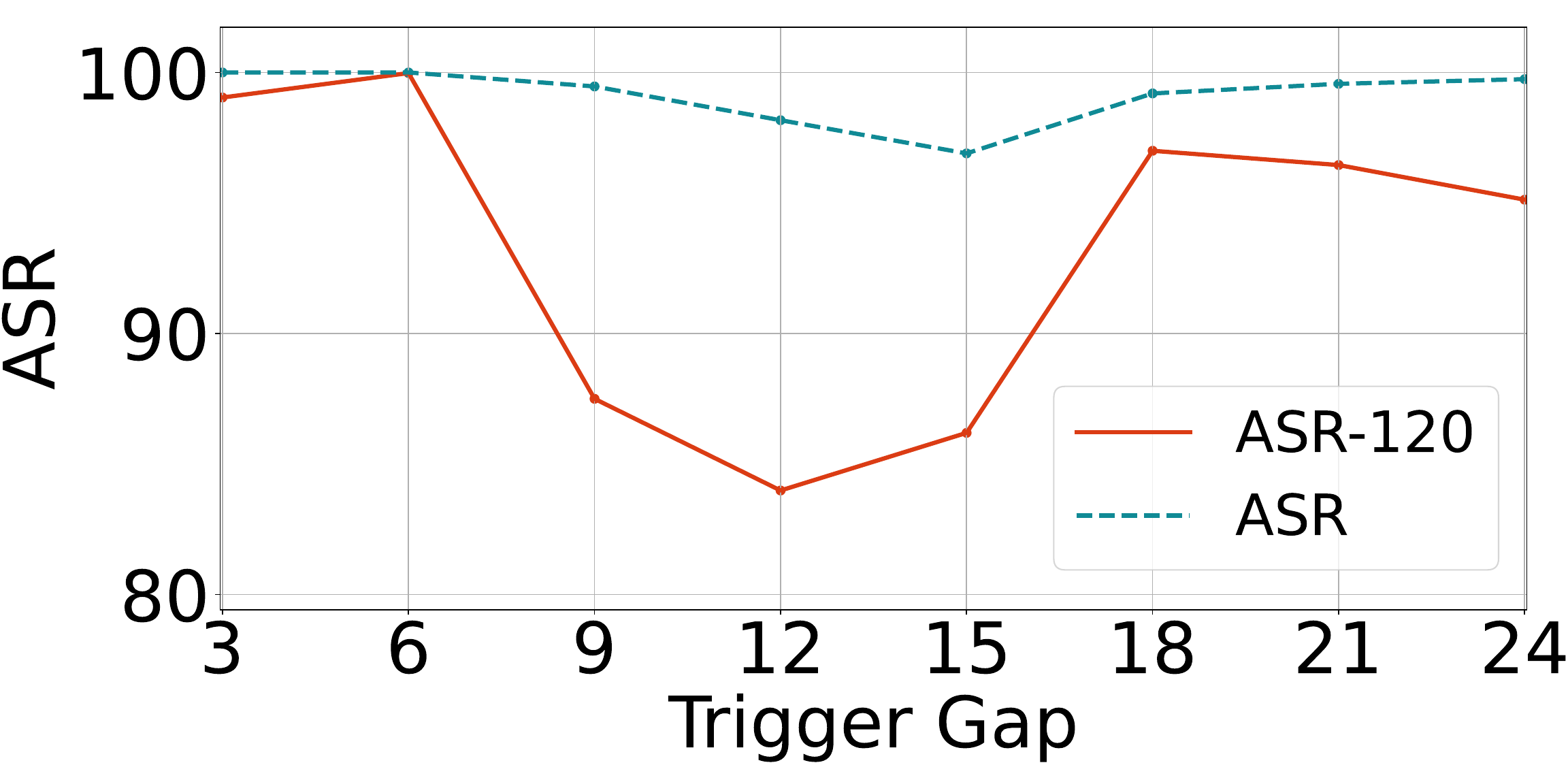}
        \caption{MNIST}
    \end{subfigure}
    \begin{subfigure}[b]{0.23\textwidth}
        \includegraphics[width=\linewidth]{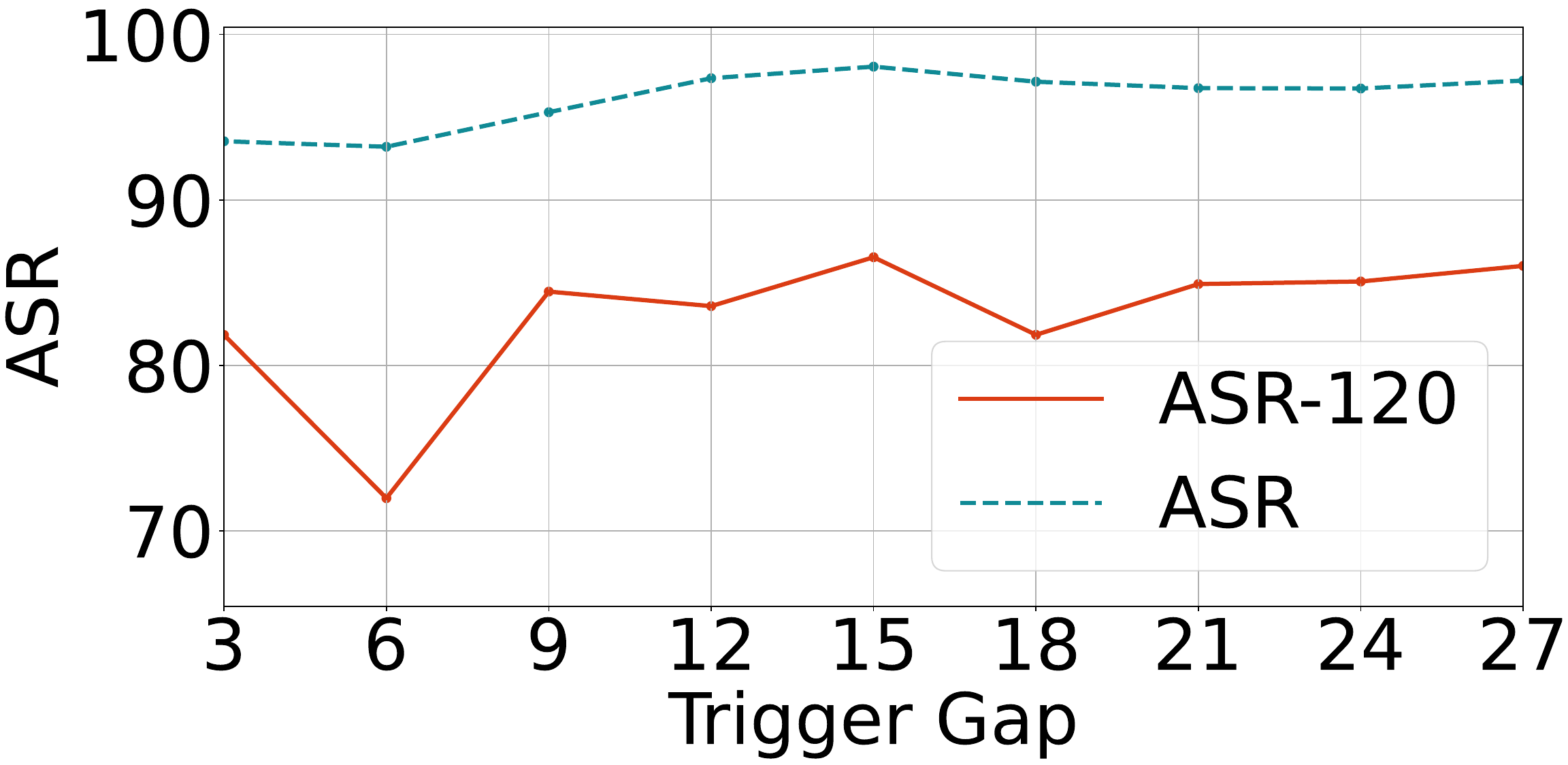}
        \caption{CIFAR-10}
    \end{subfigure}
    
    \caption{Effects of Trigger Gap on ASR and ASR-$t$.}
    \label{Effects-of-Trigger-Gap}
\end{figure}

\subsubsection{Effects of $r$.} $r$ sets the ratio of backdoor samples in training batches. Fig. A.17 reveals that variations in $r$ have minimal impact on ASR and ASR-$t$. 
\begin{figure}[htbp!]
    \centering    
    \begin{subfigure}[b]{0.23\textwidth}
        \includegraphics[width=\linewidth]{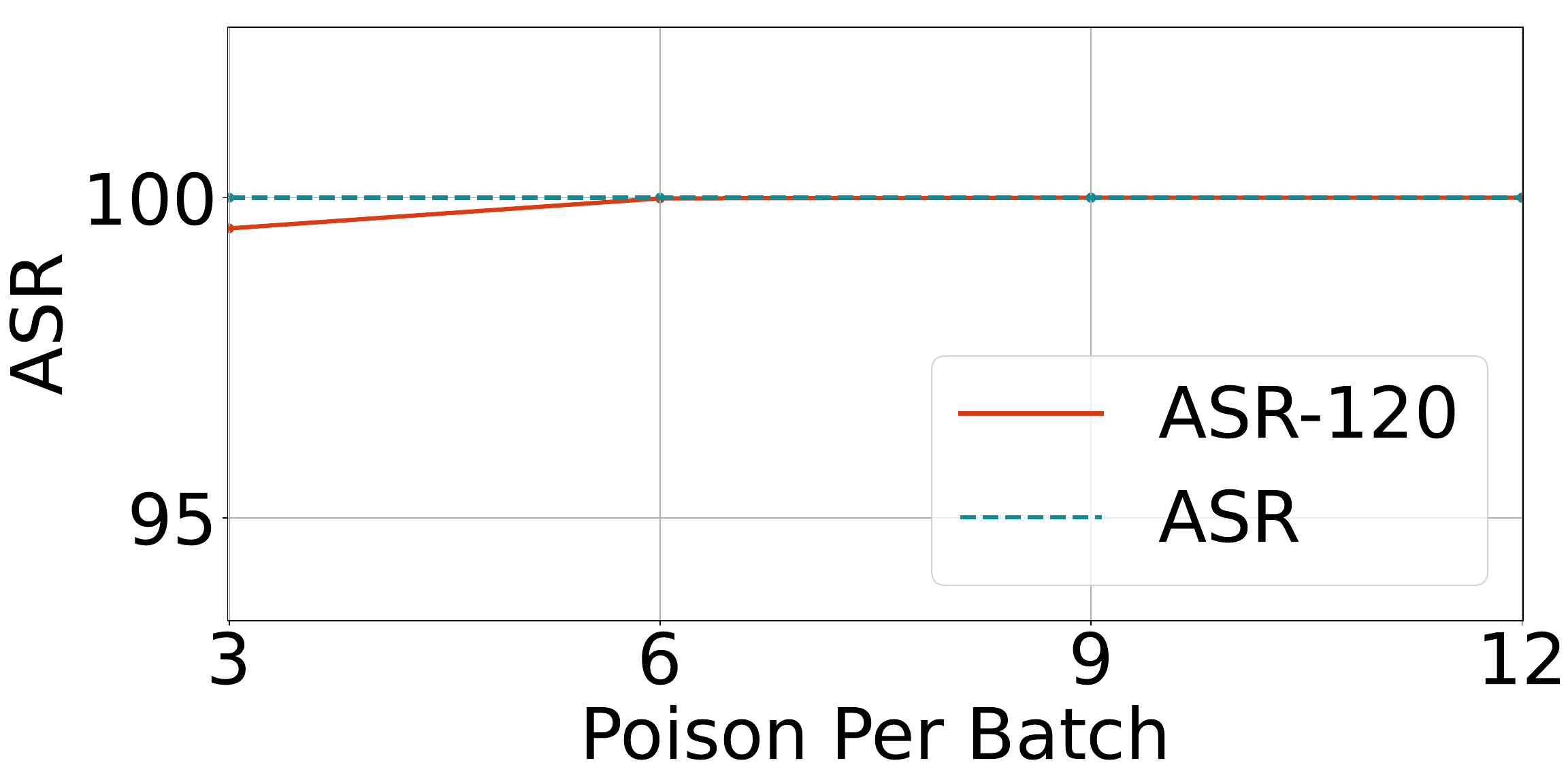}
        \caption{MNIST}
    \end{subfigure}
    \begin{subfigure}[b]{0.23\textwidth}
        \includegraphics[width=\linewidth]{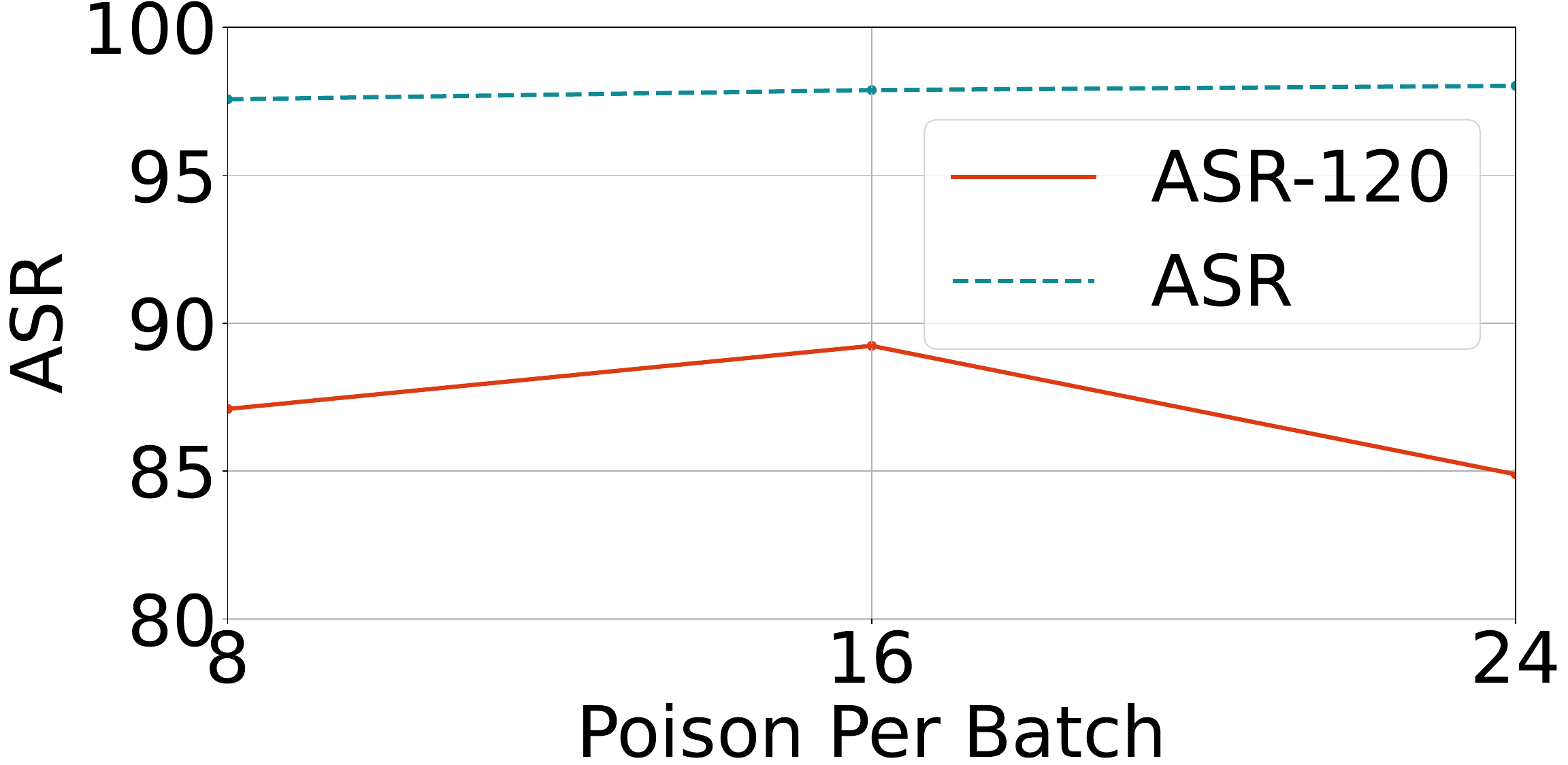}
        \caption{CIFAR-10}
    \end{subfigure}

    \caption{Effects of Poison Ratio on ASR and ASR-$t$.}
    \label{Effects-of-Poison-Ratio}
\end{figure}

\subsubsection{Effects of $I$.} $I$ is the gap between backdoor injections. Fig. A.18 suggests $I$ has little effect on ASR and ASR-$t$.
\begin{figure}[htbp!]
    \centering    
    \begin{subfigure}[b]{0.23\textwidth}
        \includegraphics[width=\linewidth]{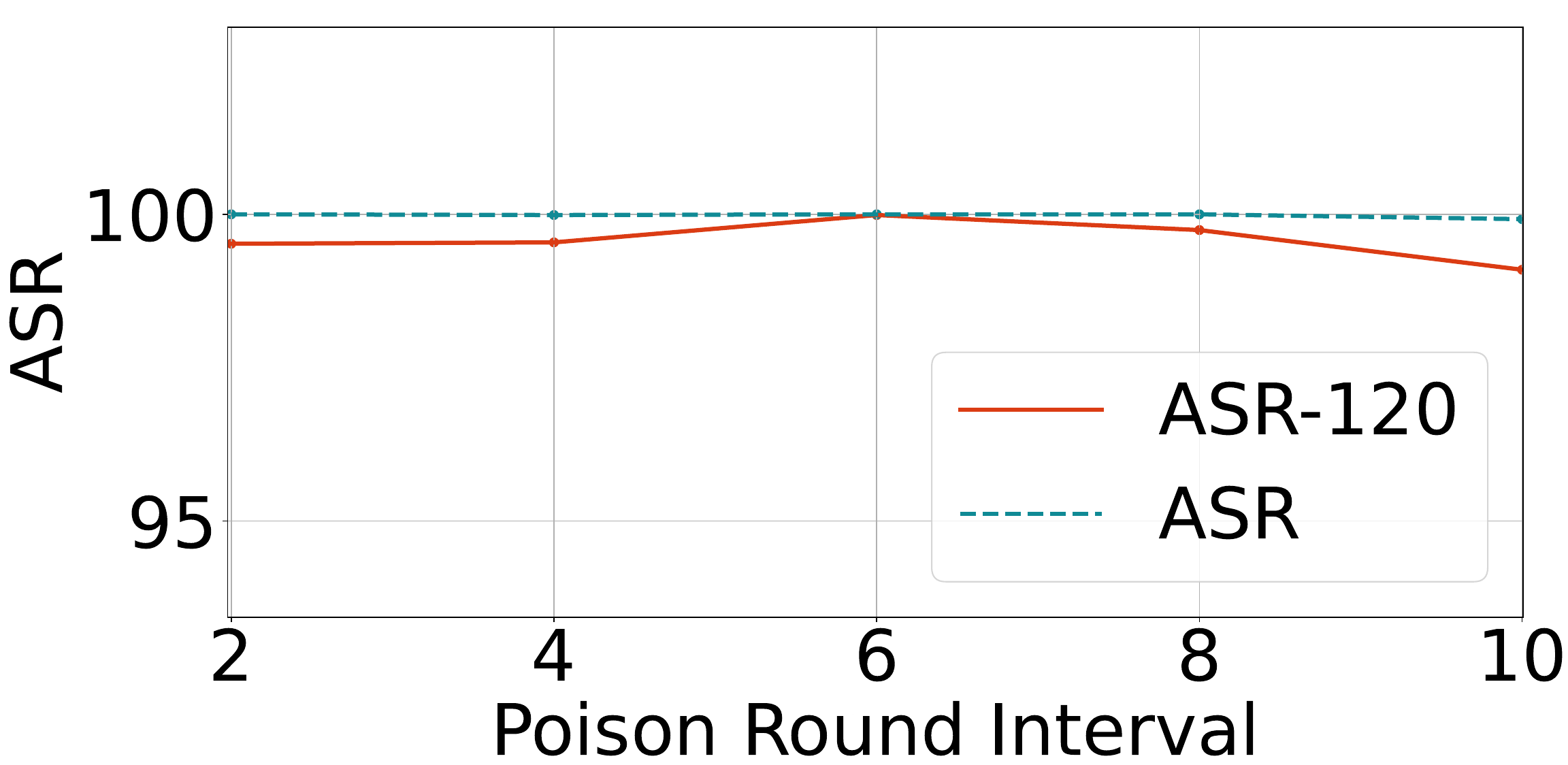}
        \caption{MNIST}
    \end{subfigure}
    \begin{subfigure}[b]{0.23\textwidth}
        \includegraphics[width=\linewidth]{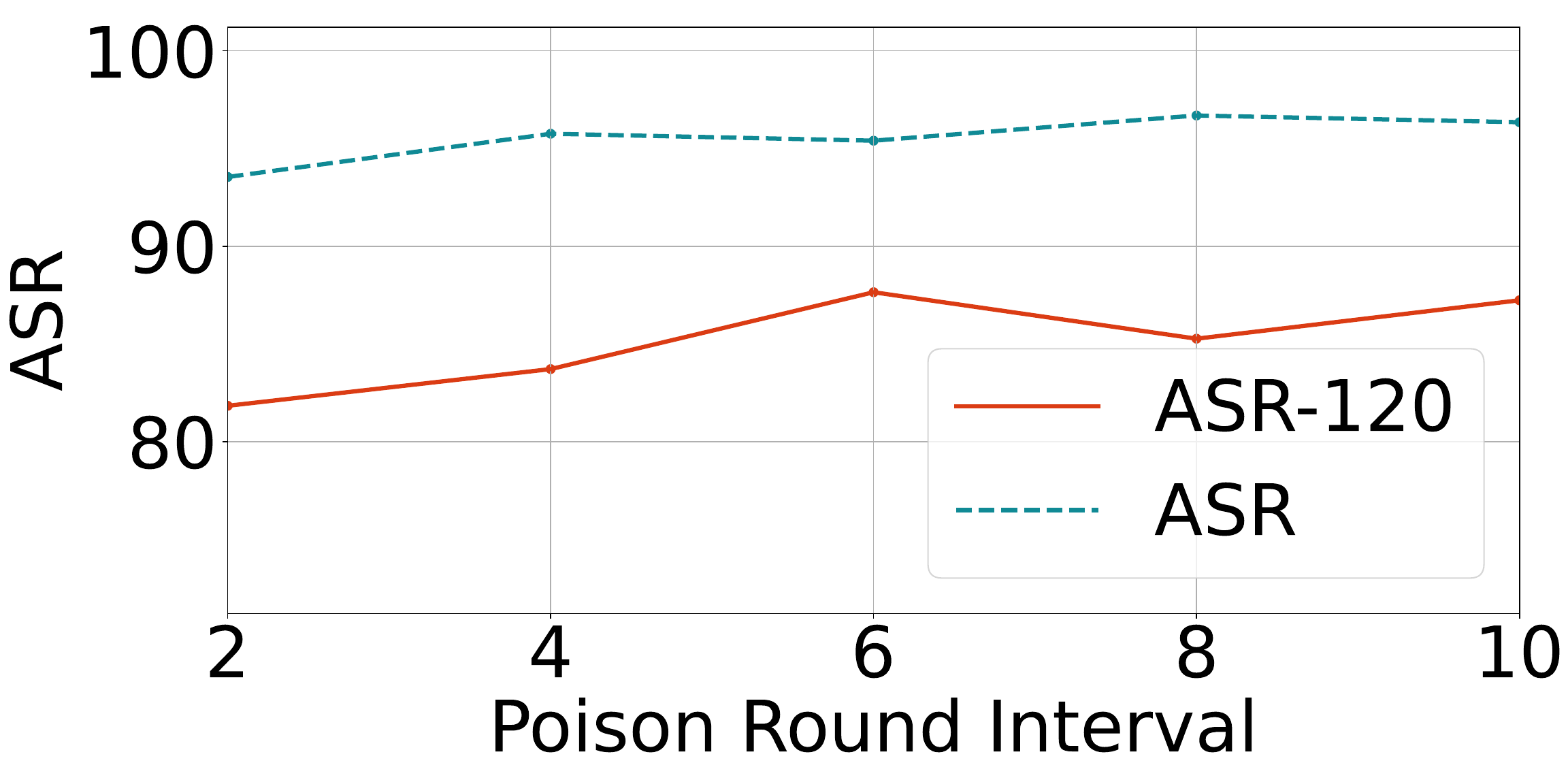}
        \caption{CIFAR-10}
    \end{subfigure}
    
    \caption{Effects of Poison Round Interval on ASR and ASR-$t$.}
    \label{Effects-of-Poison-Round}
\end{figure}

\subsection{Robustness comparison between FCBA and DBA}
Fig. A.19-A.20 evaluate the impact of clipping boundary $S$ and noise variance $\sigma$ on DBA (MNIST, CIFAR-10). From the comparison, we deduce: (1) FCBA's ASR-$t$ curve consistently surpasses DBA across all $S$ values, showcasing better resilience to low clipping boundaries. (2) For all $\sigma$ values, FCBA again outperforms DBA, indicating superior resistance to high noise variance. In summary, FCBA is better equipped against participant-level differential privacy defense than DBA, achieving enhanced attack success and persistence.
\begin{figure}[htbp!]
    \centering    
    \begin{subfigure}[b]{0.23\textwidth}
        \includegraphics[width=\linewidth]{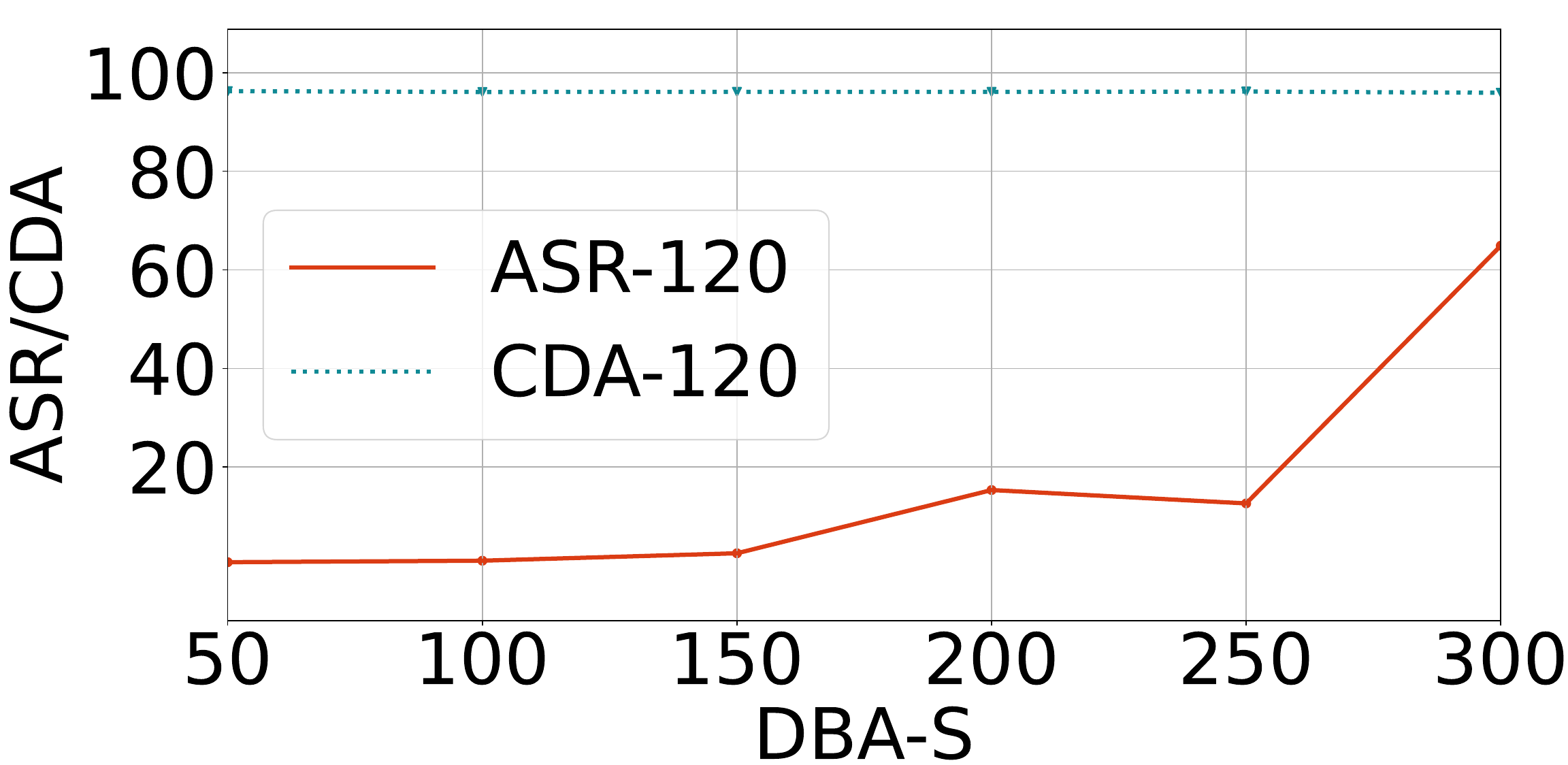}
        \caption{MNIST}
    \end{subfigure}
    \begin{subfigure}[b]{0.23\textwidth}
        \includegraphics[width=\linewidth]{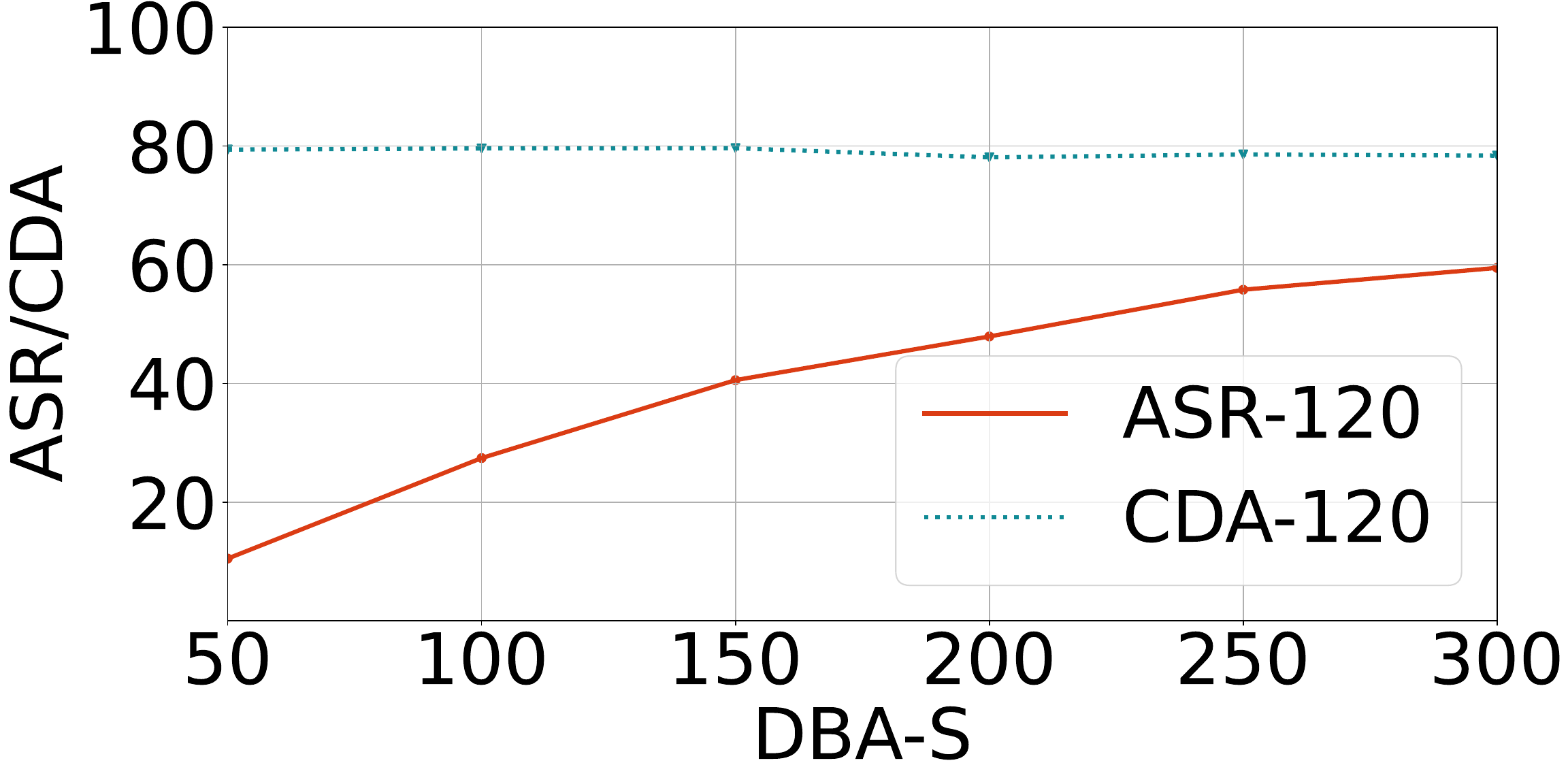}
        \caption{CIFAR-10}
    \end{subfigure}
    
    \caption{Effects of Clipping Boundary $S$ on ASR-$t$ and CDA-$t$ (DBA)}
    \label{Effects-of-Poison-Round}
\end{figure}

\begin{figure}[htbp!]
    \centering    
    \begin{subfigure}[b]{0.23\textwidth}
        \includegraphics[width=\linewidth]{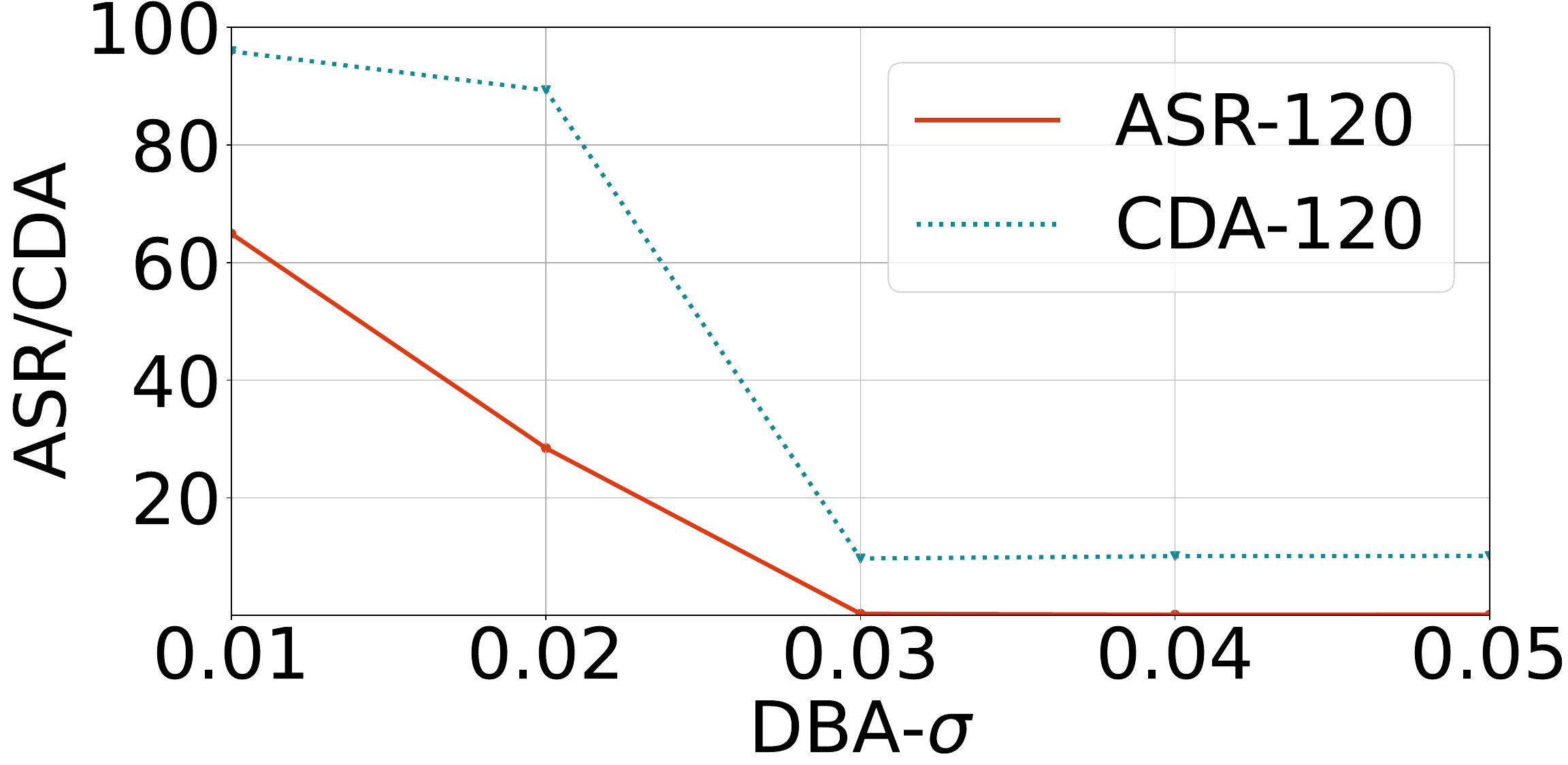}
        \caption{MNIST}
    \end{subfigure}
    \begin{subfigure}[b]{0.23\textwidth}
        \includegraphics[width=\linewidth]{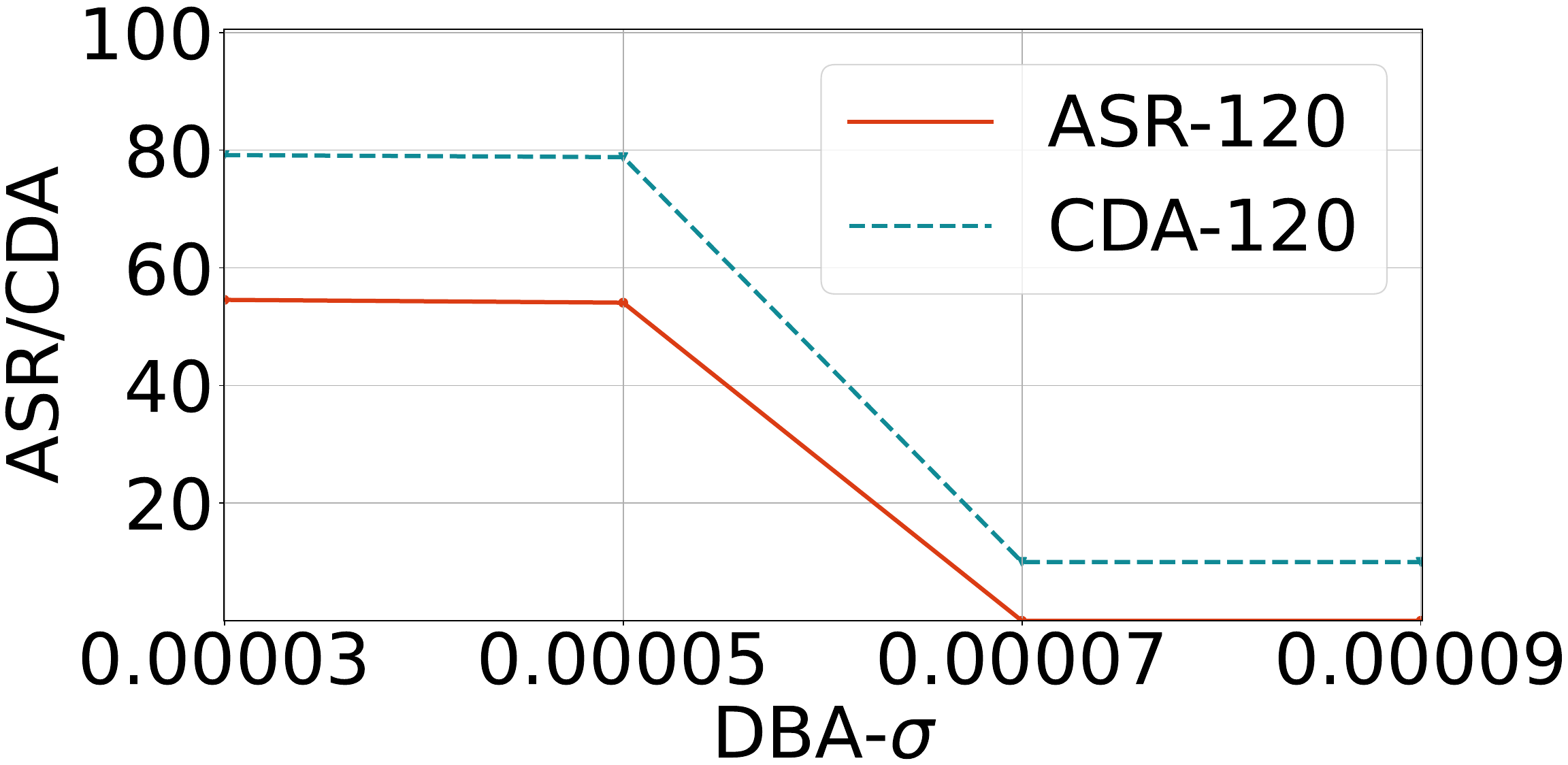}
        \caption{CIFAR-10}
    \end{subfigure}
    
    \caption{Effects of Noise Variance $\sigma$ on ASR-$t$ and CDA-$t$ (DBA).}
    \label{Effects-of-Poison-Round}
\end{figure}





\subsection{More details on $\sigma$}
Fig. A.21 and Fig. A.22 (CIFAR-10) reveal irregular ASR curve oscillations for FCBA and DBA due to the unpredictability of noise, causing unstable model performance. The uncontrollable Gaussian noise perturbation affects the global model's efficiency in main and backdoor tasks.

\begin{figure}[H]   
\centering
\begin{minipage}{0.23\textwidth}
\centering
\includegraphics[width=\textwidth]{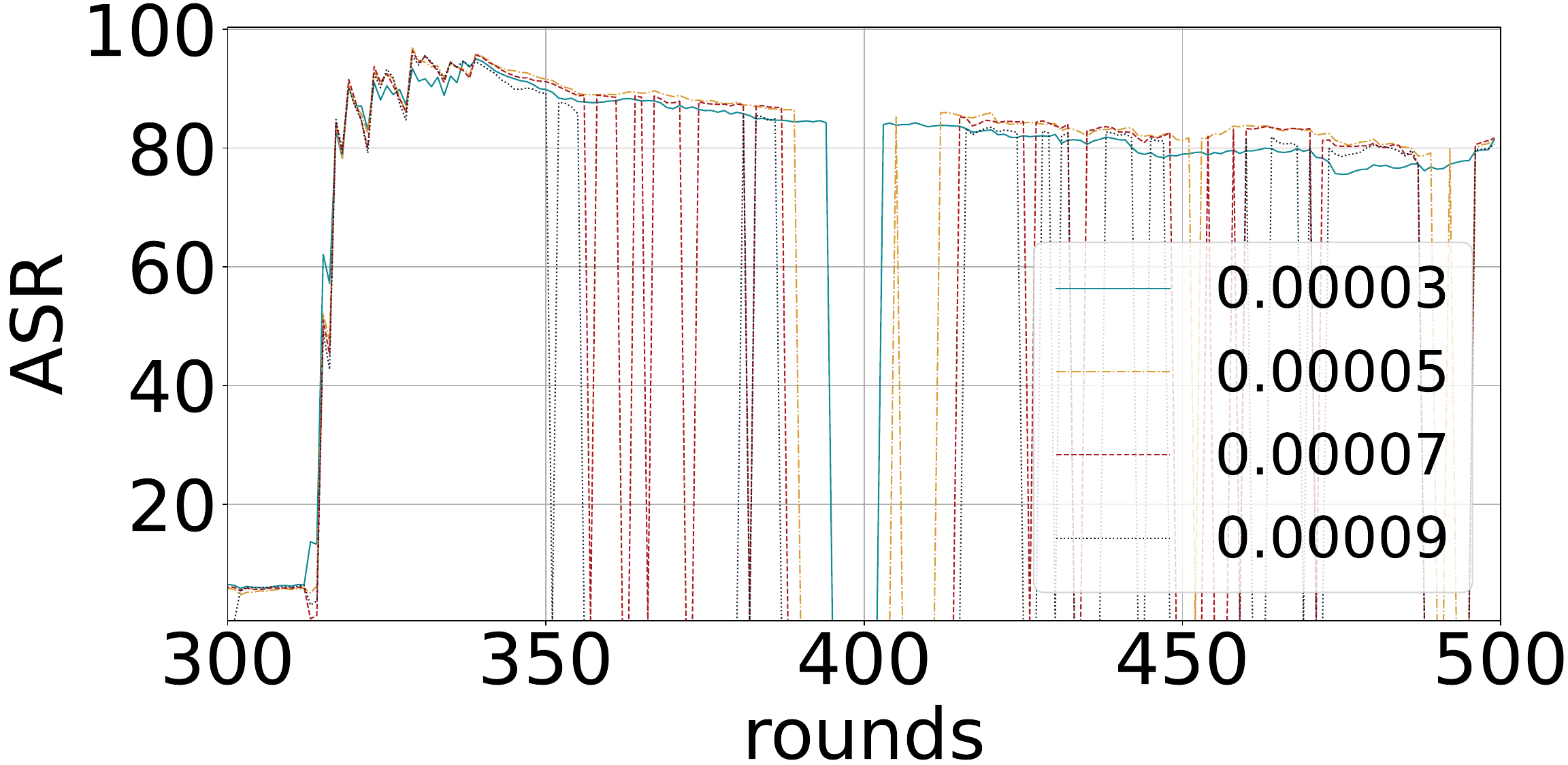}
\caption{FCBA-ASR on CIFAR-10 at different $\sigma$.}
\label{FCBA-describe1}
\end{minipage}
\hfill
\begin{minipage}{0.23\textwidth}
\centering
\includegraphics[width=\textwidth]{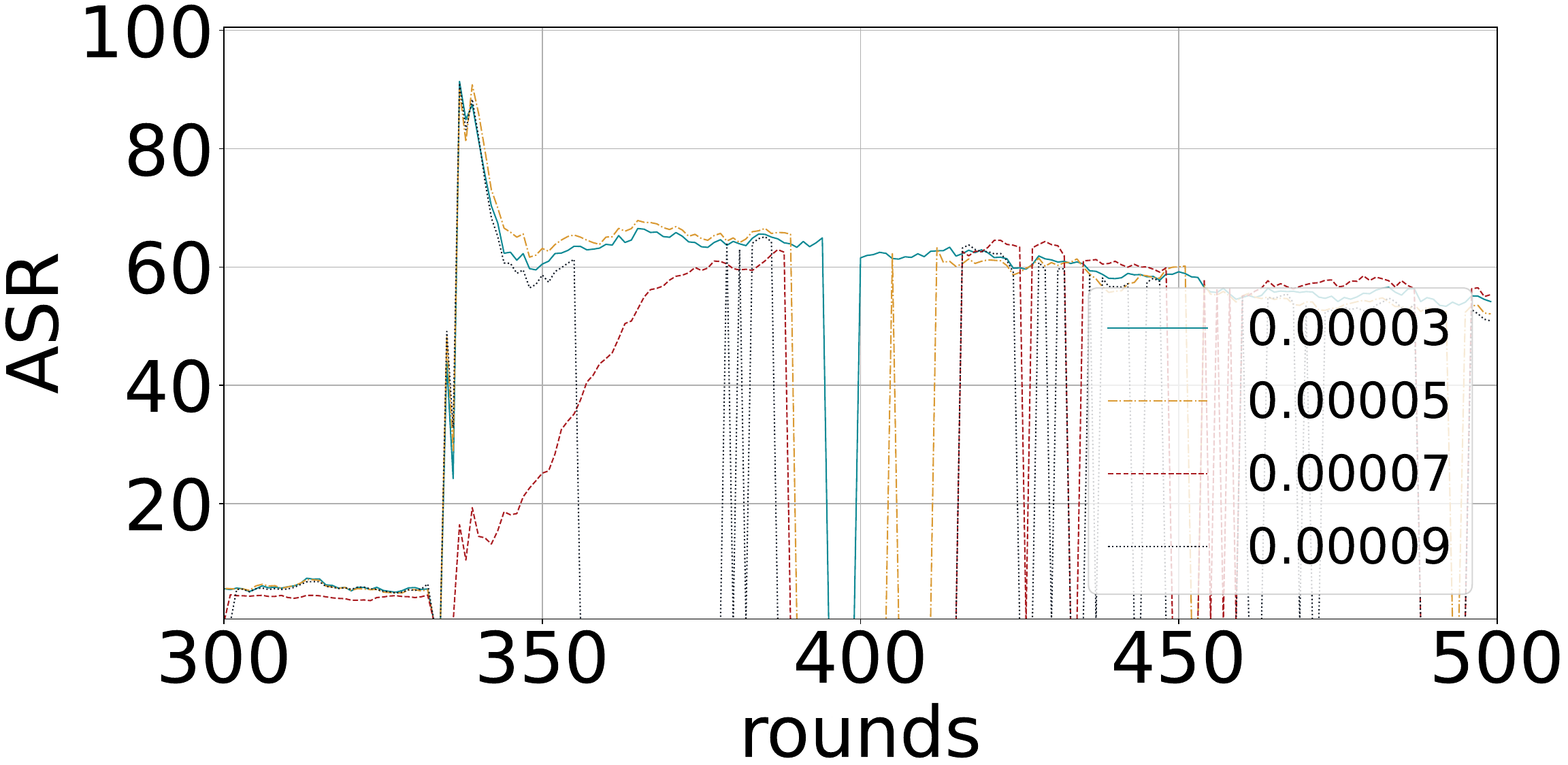}
\caption{DBA-ASR on CIFAR-10 at different $\sigma$.}
\label{FCBA-describe2}
\end{minipage}
\end{figure}

\end{document}